\documentclass{aa}

\usepackage{graphicx}
\usepackage{subcaption}
\usepackage{placeins}
\usepackage{float}
\usepackage{stfloats}
\usepackage{subcaption}
\usepackage[varg]{txfonts}
\mathchardef\mhyphen="2D 
 \usepackage[colorlinks=true,linkcolor=blue, citecolor=blue]{hyperref}
\usepackage{array}
\newcommand{\PreserveBackslash}[1]{\let\temp=\\#1\let\\=\temp}
\newcolumntype{C}[1]{>{\PreserveBackslash\centering}p{#1}}
\newcolumntype{R}[1]{>{\PreserveBackslash\raggedleft}p{#1}}
\newcolumntype{L}[1]{>{\PreserveBackslash\raggedright}p{#1}}

\begin{document}
	\title{Jet formation in post-AGB binaries\thanks{Based on observations made with the \textit{Mercator} Telescope, operated on the island of La Palma by the Flemish Community, at the Spanish Observatorio del Roque de los Muchachos of the Instituto de Astrofísica de Canarias.}}
	
	\subtitle{Confronting cold magnetohydrodynamic disc wind models with observations}
	
	\author{
		T.\ De Prins\inst{1}
		\and
		H.\ Van Winckel\inst{1}
		\and
		J.\ Ferreira\inst{2}
		\and
		O.\ Verhamme\inst{1}
		\and
		D.\ Kamath\inst{3}\fnmsep\inst{4}
		\and
		N.\ Zimniak\inst{2}
		\and
		J.\ Jacquemin-Ide\inst{5}
	}
	
	\institute{Institute of Astronomy, KU Leuven,
		Celestijnenlaan 200D, 3001 Leuven, Belgium\\
		\email{toon.deprins@kuleuven.be}
		\and
		Univ. Grenoble Alpes, CNRS, IPAG, 38000 Grenoble, France
		\and
		Department of Physics \& Astronomy, School of Mathematical and Physical Sciences, Macquarie University, Sydney, NSW 2109, Australia
		\and
		Astronomy, Astrophysics and Astrophotonics Research Centre, Macquarie University, Sydney, NSW 2109, Australia
		\and
		Center for Interdisciplinary Exploration \& Research in Astrophysics (CIERA), Physics and Astronomy, Northwestern University, Evanston, IL 60202, USA
	}
	
	\date{Received 16 February 2024 / Accepted 12 June 2024}
	
	
	\abstract
	{Jets are launched from many classes of astrophysical objects, including post-asymptotic giant branch (post-AGB) binaries with a circumbinary disc. Despite dozens of detections, the formation of these post-AGB binary jets and their connection to the inter-component interactions in their host systems remains poorly understood.}
	{Building upon the previous paper in this series, we consider cold self-similar magnetohydrodynamic (MHD) disc wind solutions to describe jets that are launched from the circumcompanion accretion discs in post-AGB binaries. Resulting predictions are matched to observations. This both tests the physical validity of the MHD disc wind paradigm and reveals the accretion disc properties.}
	{Five MHD solutions are used as input to synthesise spectral time-series of the $\mathrm{H_{\alpha}}$ line for five different post-AGB binaries. A fitting routine over the remaining model parameters is developed to find the disc wind models that best fit the observed time-series.}
	{Many of the time-series' properties are reproduced well by the models, though systematic mismatches, such as overestimated rotation, remain. Four targets imply accretion discs that reach close to the secondary's stellar surface, while one is fitted with an unrealistically large inner radius at $\gtrsim 20$ stellar radii. Some fits imply inner disc temperatures over $10\,000\,\mathrm{K}$, seemingly discrepant with a previous observational estimate from H band interferometry. This estimate is, however, shown to be biased. Fitted mass-accretion rates range from $\sim 10^{-6}-10^{-3}\,\mathrm{M_\odot/yr}$. Relative to the jets launched from young stellar objects (YSOs), all targets prefer winds with higher ejection efficiencies, lower magnetizations and thicker discs.}
	{Our models show that current cold MHD disc wind solutions can explain many of the jet-related $\mathrm{H_{\alpha}}$ features seen in post-AGB binaries, though systematic discrepancies remain. This includes, but is not limited to, overestimated rotation and underestimated post-AGB circumbinary disc lifetimes. The consideration of thicker discs and the inclusion of irradiation from the post-AGB primary, leading to warm magnetothermal wind launching, might alleviate these.}
	
	
	\maketitle
	%
	
	\section{Introduction}
	{{
			The post-asymptotic giant branch (post-AGB) is a short phase of contraction after the envelope of a star has been stripped, either by a dusty wind on the asymptotic giant branch \citep[e.g.][]{vassiliadis1993, habing2004, hofner2018} or, as is more likely for binary systems, through strong binary interactions \citep[e.g.][]{Decin2021}. A post-AGB star continually contracts and heats up at a constant luminosity of $\sim 1000 - 10\,000\,\mathrm{L_\odot}$ \citep{VanWinckel2003}, eventually reaching the white dwarf phase in $\sim 10\,000\,\mathrm{yr}$ \citep{MillerBertolami2016}. In this paper, we focus on a specific class of post-AGB stars, namely post-AGB binaries surrounded by a stable, dusty circumbinary disc \citep[for a comprehensive review, see][]{vanWinckel2019}. These systems show strong infrared excesses in their spectral energy distributions, indicative of a large dusty disc \citep{Kluska2022}. The binarity of these disc sources' central stars is confirmed through radial velocity (RV) measurements \citep[e.g.][]{Oomen2018}. This revealed orbital periods in the range of a few hundred to a few thousand days and eccentricities up to $\approx 0.65$, while also showing that the secondary companions are likely main sequence (MS) stars. Meanwhile, Galactic sources resolved through both infrared and (molecular) sub-mm interferometry \citep[e.g.][]{Corporaal2023, GallardoCava2021}, as well as single-telescope imaging \citep{Ertel2019, Andrych2023}, confirm the circumbinary nature of the dusty discs.
	}}
	
	{{
			This type of circumbinary disc is likely formed from envelope material, stripped from the evolved primary and ejected during a poorly understood phase of binary interaction during the AGB or, even earlier, on the red giant branch (RGB) \citep{vanWinckel2019}. Examples of the latter population, typically distinguished from their post-AGB analogues by their systematically lower luminosities \citep[e.g.][]{Kamath2016}, are called post-RGB binaries. Due to the lack of reliable luminosity estimates (see the note under Table \ref{table:target_star_params}) and for the sake of brevity, we will hereafter refer to circumbinary disc-bearing post-AGB/RGB systems simply as post-AGB binaries, despite the possibility that some of our sample stars might be post-RGB in nature. The physics described in this paper is applicable to both types of systems.
	}}
	
	There is strong observational evidence for ongoing re-accretion streams from the dusty post-AGB circumbinary discs to the central stars. First, the primaries' atmospheres often appear depleted in refractory elements, which is explained through the trapping of dust in the circumbinary disc and the subsequent infall of refractory-depleted gas on the primary. This dilutes its atmosphere, leaving behind a specific chemical signature \citep[][and references therein]{Oomen2019}. Second, thermal emission from a smaller accretion disc around the secondary star (i.e.\ a circumcompanion accretion disc) has been detected in near-infrared interferometry for several systems \citep{Hillen2016, Anugu2023}. Several more such detections are as of yet unpublished (N.\ Anugu, private communication). Current estimates of the accretion rates in the circumcompanion discs indicate that re-accretion from the circumbinary disc is the most likely mechanism feeding them \citep{Bollen2022}. {{To be succinct during the rest of this paper, we will use the term accretion disc to refer to a post-AGB binary system's circumcompanion accretion disc, while circumbinary disc is used explicitly to refer to the larger, dusty disc surrounding the entire system}}.
	
	Using high-resolution spectra from a more than a decade long monitoring campaign \citep{vanWinckel2009} with the HERMES spectrograph \citep{Raskin2011}, it was shown that the accretion discs in post-AGB binaries can launch fast, collimated outflows \citep[e.g.][]{Gorlova2012, gorlova2015}. Such outflows, colloquially called jets, are ubiquitously observed around a large variety of accreting astrophysical objects. Jet energetics range from extremely relativistic, as when launched from active galactic nuclei, to non-relativistic in both evolved and non-evolved systems. Examples of the latter include those launched from (proto)planetary nebulae ((P)PNe) and young stellar objects (YSOs). These jets are suspected to affect the properties of both their launching object and the ambient surrounding medium. Indeed, jets in (P)PNe are being studied as crucial nebular shaping agents \citep[e.g.][]{Lopez1993, Clairmont2022, lora2023}, while jets in YSOs are thought to be removing a significant fraction of the angular momentum in their accretion discs, promoting stellar growth and affecting the underlying disc structure \citep[e.g.][]{Ray2021}. {{The jets in post-AGB binaries are also non-relativistic, and thus share similarities to these systems in terms of jet-launching physics.}}
	
	Post-AGB binary jets are detected through variable $\mathrm{H_\alpha}$ line profiles. Due to the high contrast between the very luminous primary and the secondary in post-AGB binaries, the latter's light cannot be detected. Outside of superior conjunction, the $\mathrm{H_\alpha}$ profile consists of the primary's photospheric absorption profile, superposed with two emission wings, which either follow the primary's RV or stay fixed on the centre of mass (COM). The origin of these wings are as of yet unclear. At superior conjunction, a jet launched from the accretion disc can obscure the primary, causing a deep and extended absorption feature as background photons are scattered out of the line-of-sight (LOS). Parametric modelling of such $\mathrm{H_\alpha}$ profiles in a series of papers constrained the geometry and kinematics of post-AGB binary jets, indeed revealing many similarities to YSO jets and again suggesting a MS nature for the secondaries \citep{Bollen2017, Bollen2019, Bollen2021, Bollen2022}. Examples of typical jet-related $\mathrm{H_\alpha}$ variations can be seen in the left column of Fig.\ \ref{fig:spectra_main_results}.
	
	Still, parametric prescriptions provide little connection to the actual underlying physics of the accretion disc and the accretion-ejection process. For this reason, \citet{verhamme2024} (from here on out Paper I) introduced the use of self-similar magnetohydrodynamic (MHD) disc wind solutions \citep{Jacquemin2019} to describe post-AGB binary jets. These solutions are based on the extended disc wind paradigm, which is deemed one of the main contenders for the dominant jet formation mechanism in YSOs, through both theoretical and observational considerations \citep{Ferreira2006, Launhardt2023, moscadelli2022}. They consistently describe the accretion-ejection process, allowing for a direct connection to be made between the disc wind properties and the accretion disc. Nevertheless, Paper I only considered a parameter study using one MHD solution, which was heavily tailored to YSO jets, and applied it to one target system without a minimisation routine.
	
	In this paper, we introduce an MHD disc wind solution fitting routine to match predictions for the $\mathrm{H_\alpha}$ line profile, throughout the binary orbit, to the observed HERMES time-series. We applied this fitting routine to five different post-AGB binary targets using five different self-similar disc wind solutions as input to describe the jet geometry, density and kinematics. We aim to test the validity of the MHD disc wind paradigm for jet formation, and to reveal properties of the circumcompanion accretion discs and the re-accretion streams from the circumbinary discs feeding them. The self-similar MHD solutions are introduced in Sect.\ \ref{sect:self_similar_sol}. Our methodology is described in Sect.\ \ref{sect:methods}, while our target systems are introduced in Sect.\ \ref{sect:target_stars}. Results from the fitting routine are presented in Sect.\ \ref{sect:results} and extensively discussed in terms of the underlying physics in Sect.\ \ref{sect:discussion}. Our summary and conclusions are presented in Sect.\ \ref{sect:conclusions}.
	
	\begin{figure*}
		\centering 
		
		\centering 
		\begin{subfigure}{0.33\textwidth}
			\includegraphics[width=\linewidth]{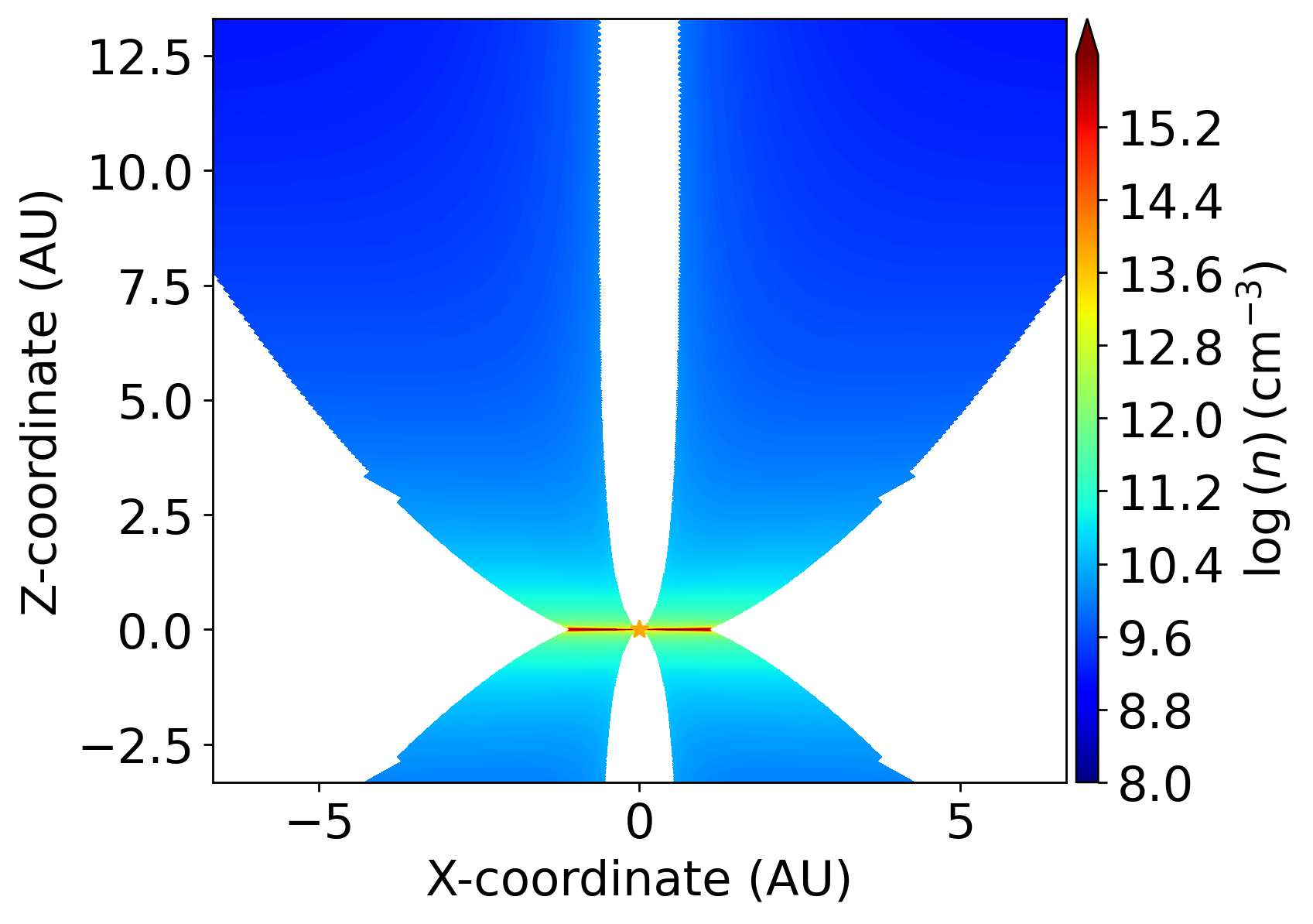}
		\end{subfigure}\hfil 
		\begin{subfigure}{0.33\textwidth}
			\includegraphics[width=\linewidth]{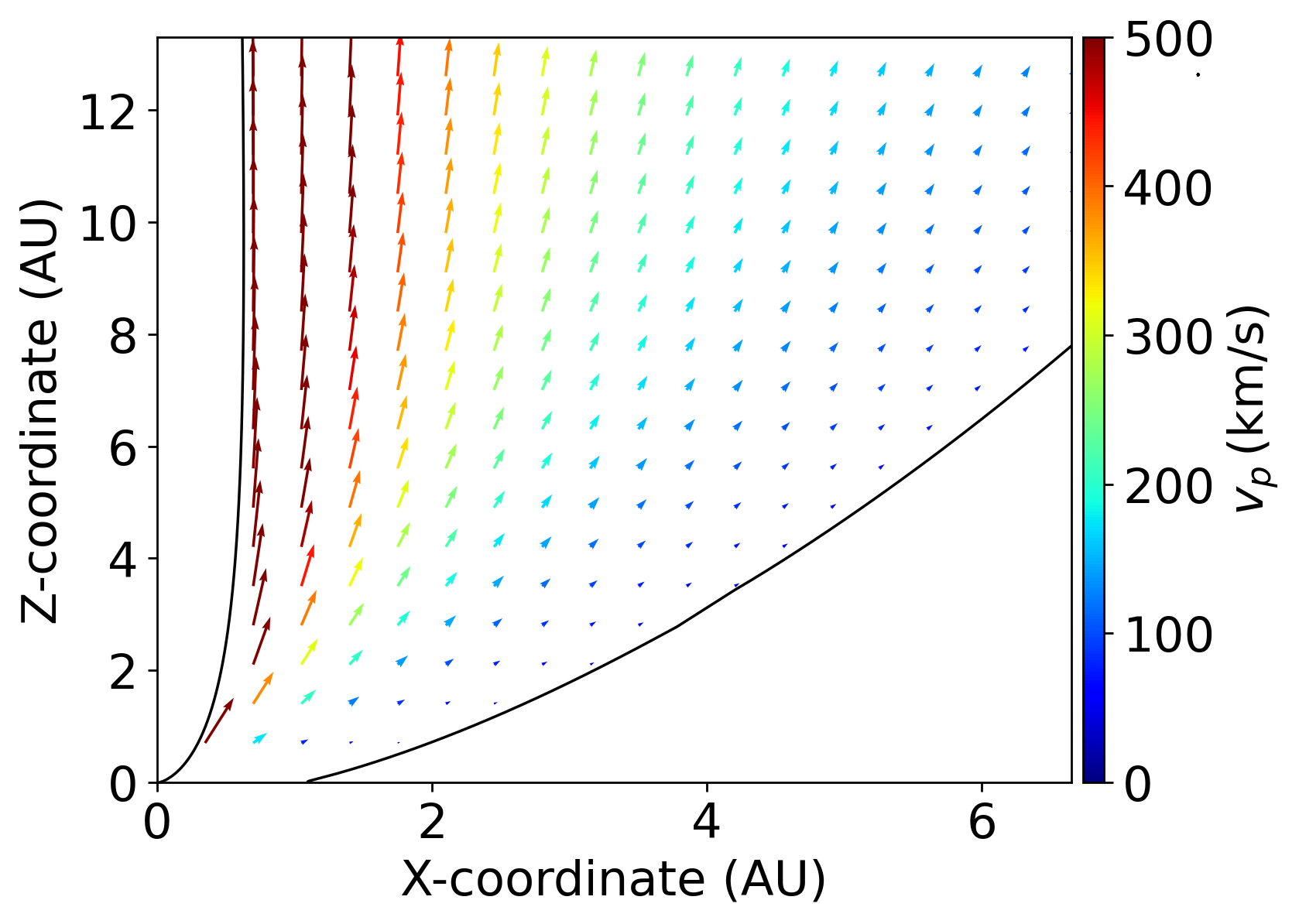}
		\end{subfigure}\hfil
		\begin{subfigure}{0.33\textwidth}
			\includegraphics[width=\linewidth]{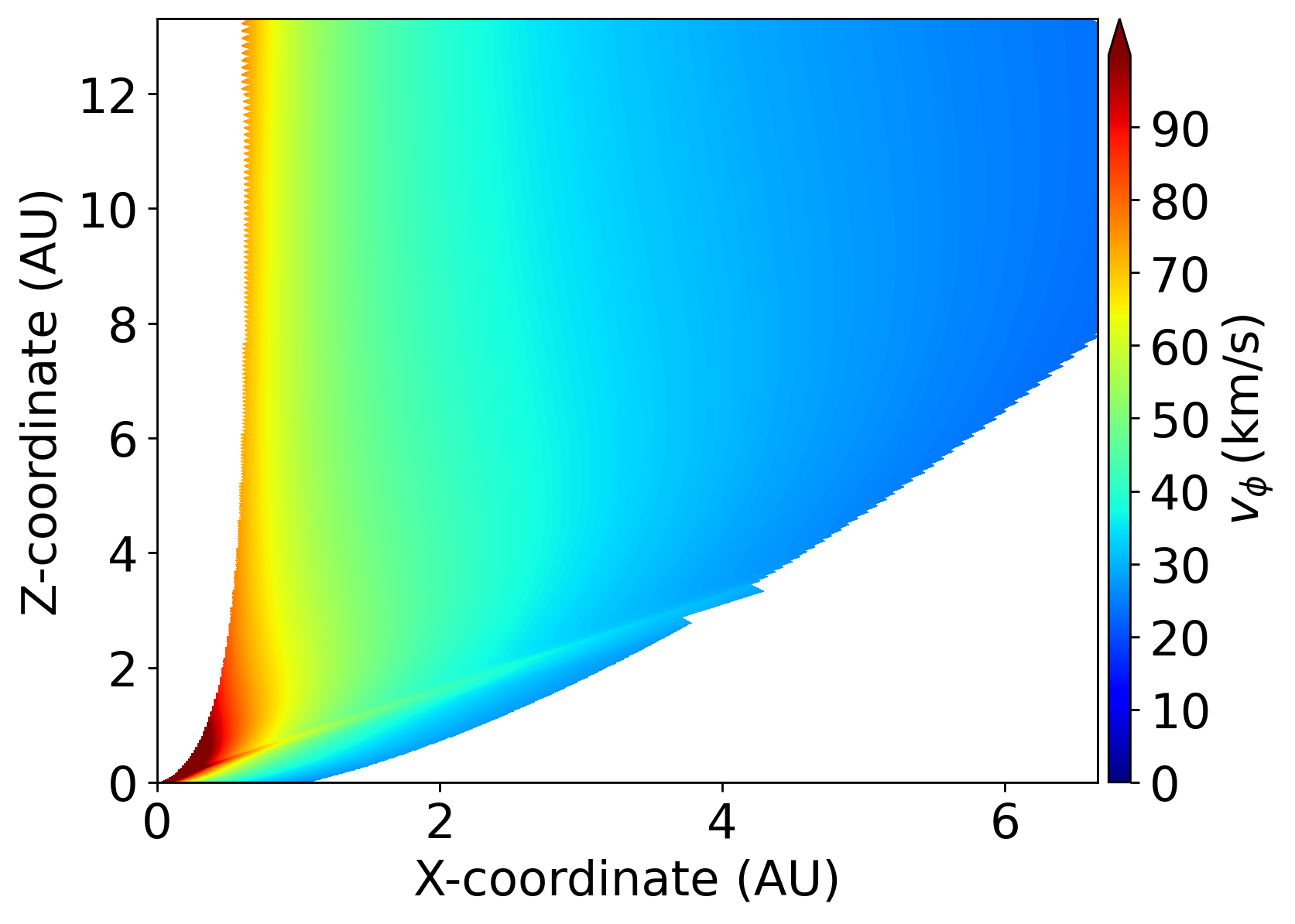}
		\end{subfigure}
		
		\medskip
		\begin{subfigure}{0.33\textwidth}
			\includegraphics[width=\linewidth]{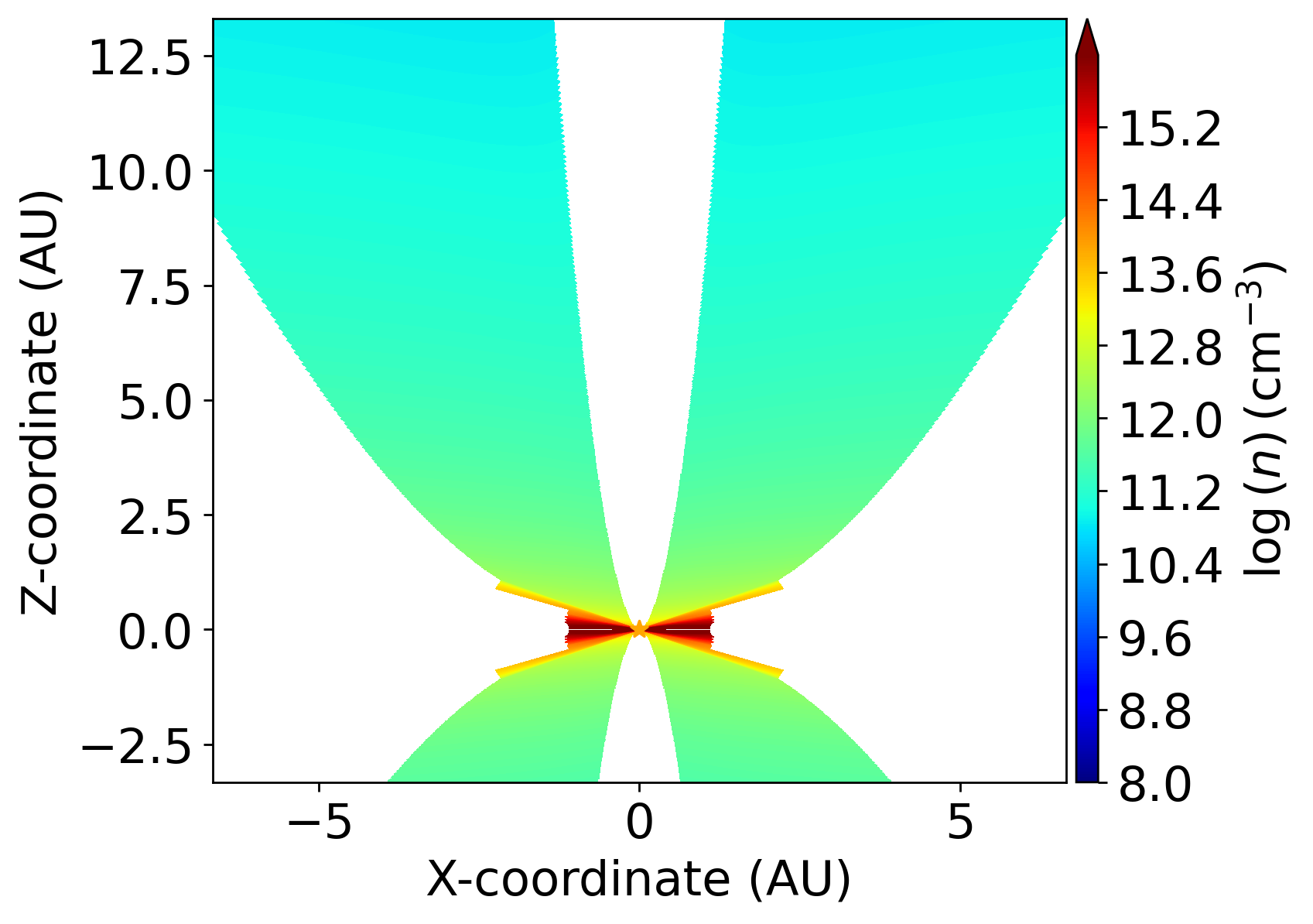}
		\end{subfigure}\hfil 
		\begin{subfigure}{0.33\textwidth}
			\includegraphics[width=\linewidth]{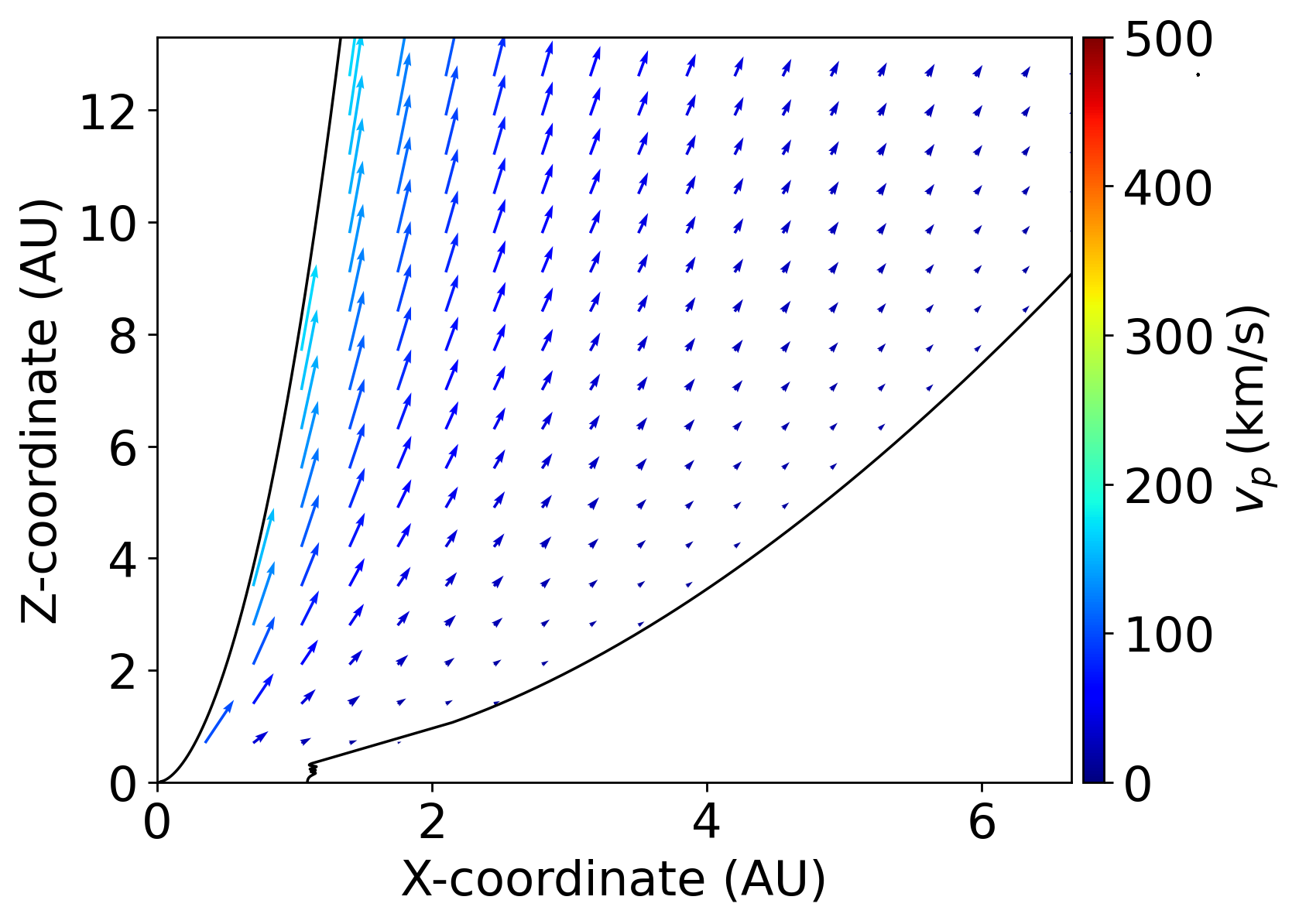}
		\end{subfigure}\hfil
		\begin{subfigure}{0.33\textwidth}
			\includegraphics[width=\linewidth]{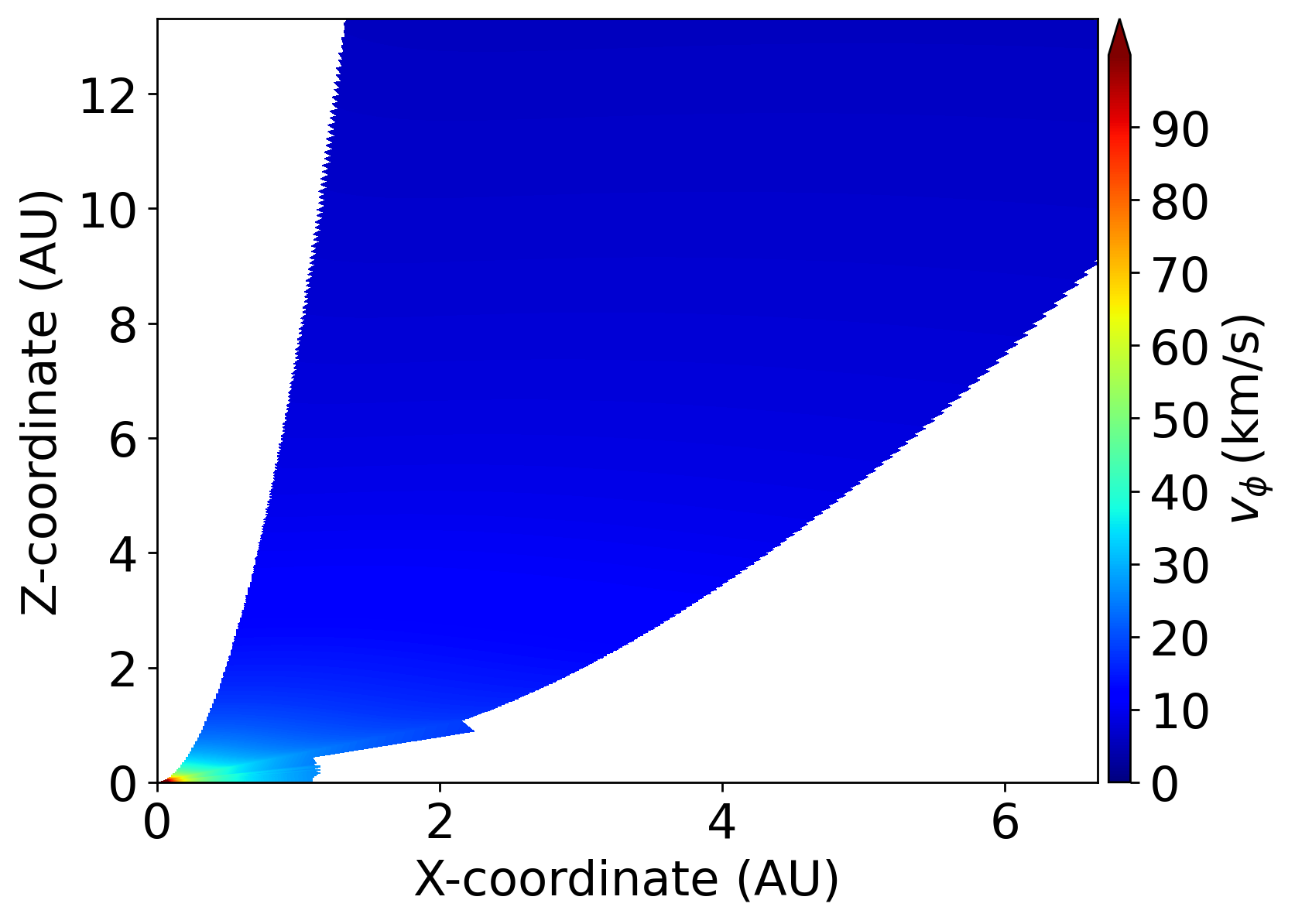}
		\end{subfigure}

		\caption{Example disc wind structures using two different MDWs. \textit{Top:} using \#MDW 1. \textit{Bottom:} using \#MDW 5. \textit{Left:} hydrogen number density $n$. \textit{Middle:} poloidal velocity $v_p$. \textit{Right:} toroidal/rotational velocity $v_\phi$. Calculated assuming the HD 52961 system and $i=70^\circ$, $r_{in} = 5\, R_2$, $r_{out} = 0.99\, R_{RL,2}$ and $\dot{M}_{in} = 10^{-3}\,\mathrm{M_{\odot}/yr}$ (see Sect.\ \ref{sect:methods}). We note the slight discontinuity at a fixed polar angle in the $n$ and $v_\phi$ profile. This is an artefact from the leapfrog integration performed over the Alfv\'en surface in the MDW solutions \citep{Jacquemin2019}, in combination with our projection onto Cartesian coordinates. This introduces no uncertainty on our results.}
		\label{fig:example_jed_structures}
	\end{figure*}
	
	\section{Self-similar disc wind solutions}\label{sect:self_similar_sol}
	
	To describe our disc wind models, we use the super-Alfv\'enic, self-similar MHD solutions of \citet{Jacquemin2019}. These accretion-ejection solutions, originally intended to describe YSO jets, are henceforth termed magnetised disc winds (MDWs). They assume a vertical magnetic field threading the midplane and driving the wind. Using a self-similar ansatz, they consistently solve the equations of resistive, single-fluid 2D MHD for the combined disc-jet system \citep[e.g.][]{Ferreira1997, Casse2000}. This gives the relative scaling of the wind geometry, density and kinematics. Absolute values for these quantities can then be calculated for any assumed combination of accretion rate and stellar mass (as is done in Sect.\ \ref{sect:methods}). We describe MDWs in general, as well as the specific MDWs used in this work, below.
	
	\begin{table}
		\caption{Dimensionless MDW parameters.}
		\label{table:MDW_solution_parameters}
		\setlength\tabcolsep{0mm}
		\centering
		\begin{tabular}{L{0.13333333333 \columnwidth} C{0.15666666666 \columnwidth} C{0.24\columnwidth} C{0.15666666666\columnwidth} C{0.15666666666\columnwidth} C{0.15666666666\columnwidth}}
			\hline\hline & \\[-1.6ex]
			\#MDW & $\epsilon$ & $\mu$ & $\alpha_m$ & $\xi$ & $\eta_{rad}$\\[1.0ex]
			\hline & \\[-1.7ex] 
			$1$ & $0.03$ & $0.351$ & $2$ & $0.04$ & $0.118$\\[0.5ex]
			$2$ & $0.1$ & $2.85\cdot 10^{-3}$ & $1$ & $0.103$ & $0.545$\\[0.5ex]
			$3$ & $0.1$ & $1.53\cdot 10^{-2}$ & $1$ & $0.104$ & $0.445$\\[0.5ex]
			$4$ & $0.1$ & $2.3\cdot 10^{-3}$ & $1$ & $0.215$ & $0.364$\\[0.5ex]
			$5$ & $0.1$ & $2.2\cdot 10^{-3}$ & $2$ & $0.305$ & $0.325$\\[0.8ex]
			\hline
		\end{tabular}
		\tablefoot{Relevant dimensionless parameters of the self-similar MDW solutions considered in this work. The solutions are labelled by number (\#MDW) and ordered by ascending ejection efficiency $\xi$. See text Sect.\ \ref{sect:MDW_parameters_description} for these parameters' definitions.}
	\end{table}
	
	\subsection{MDW parameters}\label{sect:MDW_parameters_description}
	MDWs are fully described by a complex set of dimensionless parameters. Our MDW models refer only to the volume associated with the material outflowing from the disc, but their dynamics are always computed self-consistently with the underlying disc physics. Below, we give a brief, simplified description of the MDW parameters most relevant to us, all related to the disc structure. This includes the disc aspect ratio $\epsilon$, disc magnetization $\mu$, disc turbulence parameter $\alpha_m$, the ejection efficiency $\xi$ and the radiative efficiency $\eta_{rad}$. For more rigorous details, we refer the interested reader to Paper I and \citet{Jacquemin2019}. Table \ref{table:MDW_solution_parameters} summarises these parameters for the five MDWs considered in this work. Fig.\ \ref{fig:example_jed_structures} shows an example wind structure calculated using \#MDW 1 and 5, which we use to showcase the general effects of the parameters on the wind. An analogous figure for all five MDWs is presented in Appendix \ref{sect:jed_example_structures}.

	The scale height of the disc is parametrised by $h(r) = \epsilon r$, with $r$ the radius along the midplane and $\epsilon$ the disc aspect ratio. Higher $\epsilon$ values imply thicker accretion discs. \#MDW 1 is a very thin disc at $\epsilon = 0.03$, while all other MDWs we consider are considerably thicker at $\epsilon = 0.1$. Under magneto-hydrostatic equilibrium, thicker discs rotate significantly slower (see Paper I Eqs.\ 2 \& 16), which weakens rotation of the wind as well. This is seen in the rightmost panels of Fig.\ \ref{fig:example_jed_structures}.
	
	The strength of the vertical magnetic field is parametrised by the disc magnetization $\mu = B^2/(\mu_0 P)$, with $B$ the vertical field strength, $\mu_0$ the permeability of the vacuum and $P$ the total pressure (gas + radiation), all evaluated at the disc midplane. The presence of such a vertical field leads to the generation of both radial and azimuthal field components, the latter providing a torque allowing for accretion and driving ejection \citep{Ferreira1995}. Increasing $\mu$ increases the dynamical importance of the magnetic field. Our MDWs range from $\mu \sim 10^{-1} - 10^{-3}$, covering almost the entire $\mu$-range of super-Alfv\'enic solutions found by \citet{Jacquemin2019}.
	
	Because the disc is magnetised, it is also assumed to be turbulent. This MHD turbulence is described using anomalous transport coefficients, namely a turbulent viscosity $\nu_{\nu}$, and magnetic diffusivity $\nu_m$. These are assumed to follow a vertical Gaussian profile, and have respective midplane amplitudes $\nu_{\nu} = \alpha_\nu C_S h$ and $\nu_{m} = \alpha_m V_A h$, with $\alpha_\nu$ the standard dimensionless Shakura-Sunyaev parameter for viscous angular momentum transport \citep{Shakura1973}, $\alpha_m$ an analogous parameter for the poloidal magnetic diffusivity, $C_S$ the midplane sound speed and $V_A$ the midplane Alfv\'en speed (so that $\mu = (V_A/C_S)^2$). These are related via the magnetic Prandtl number, $\mathcal{P}_m = \nu_\nu / \nu_m$. The diffusivity related to the toroidal magnetic field is then given by $\nu'_m = \nu_m/\chi_m$, where $\chi_m$ is the diffusivity anisotropy parameter. The MDWs we consider here assume $\mathcal{P}_m = 1$ and $\chi_m\approx 1$, consistent with simulations of turbulence induced by the magneto-rotational instability \citep[e.g.][]{Fromang2009, Lesur2009, Zhu2018}. Setting $\alpha_m$ then sets all turbulence-related terms. We note that $\alpha_\nu$ scales as $\alpha_\nu = \alpha_m \mathcal{P}_m \mu^{1/2}$ \citep{salvesen2016}. Increasing $\mu$ thus increases both the turbulent and the laminar (wind) torques. The MDWs considered here assume either $\alpha_m=1$ or $2$.
	
	The disc accretion rate is parametrised as $\dot{M}_{acc}(r) \propto r^\xi$, with $\xi$ the disc ejection efficiency. This is a crucial parameter, directly linking the accretion flow to the wind ejection. Conservation of mass then gives the mass lost to one of the lobes of the wind as $\dot{M}_{wind} = 0.5 \cdot \dot{M}_{in}((r_{out}/r_{in})^\xi-1)$, with $\dot{M}_{in}$ the accretion rate at the inner disc rim, and $r_{out}$ and $r_{in}$ the outer and inner radius of the wind-launching disc (these are fitting parameters, see Sect. \ref{sect:fitting_routine}). Higher $\xi$ values imply that more material is deviated to the wind, making the wind denser and increasing the wind absorption depth in $\mathrm{H_\alpha}$ (see also Paper I Eq.\ (3)). This effect is seen clearly in the leftmost panels of Fig.\ \ref{fig:example_jed_structures}. Throughout the rest of this paper, we will call MDWs with a large disc ejection efficiency $\xi$ 'efficient' solutions.
	
	These MDW solutions assume isothermal magnetic surfaces, meaning that the temperature of a field line remains fixed at the temperature of its anchoring point in the midplane. This simplifying assumption has no effect on the dynamics, since most of the wind power is carried by the magnetic field. For the purely magnetically and rotationally driven MDWs considered here, also known as 'cold' magnetocentrifugal MDWs, the highest efficiencies are achieved at lower magnetization \citep[see Fig.\ 7 in][]{Jacquemin2019}. Assuming that all energy of the magnetic structure is converted to kinetic energy, the asymptotic poloidal and toroidal velocities of the wind, along a field line anchored to the midplane at radius $r_0$, respectively follow $v_p^\infty = r_0 \Omega_{\mathrm{K_0}}(2\lambda -3)^{1/2}$ and $rv_\phi^\infty = r_0^2 \Omega_{\mathrm{K_0}} \lambda$, with $\Omega_{\mathrm{K_0}}$ the Keplerian angular velocity at $r_0$ and $\lambda$ the magnetic lever arm \citep{Blandford1982}. It can be shown under certain assumptions that $\lambda \approx 1 + 1/(2\xi)$ \citep{Ferreira1997}. Deviations from this expression are small for all MDWs considered here \citep[see Fig.\ 5 in][]{Jacquemin2019}. Increasing $\xi$ thus also leads to a decrease in the lever arm $\lambda$, reducing the asymptotic velocities that the disc wind achieves. This effect is seen in the middle and rightmost panels of Fig.\ \ref{fig:example_jed_structures}.
	
	The energy budget of an accretion-ejection structure, settled between outer and inner radii $r_{out}$ and $r_{in}$, is written as $P_{acc} = 2P_{wind} + P_{rad} + P_{adv}$, with $P_{acc}$ the released accretion power, $P_{wind}$ the power fed into one wind lobe, $P_{rad}$ the power released as radiation from the disc and $P_{adv}$ the thermal power carried further inward below $r_{in}$ by the material accreting onto the central object \citep{Ferreira1997}. Dividing this expression by $P_{acc}$ leads to an expression for the disc radiative efficiency $\eta_{rad} = P_{rad}/P_{acc} = 1-2\eta_{wind}-\eta_{adv}$. If we were to consistently solve for the disc thermal equilibrium, namely by taking into account turbulent heating and radiative cooling \citep[e.g.][]{combet2008}, both $\eta_{rad}$ and the disc aspect ratio $\epsilon$ would be an outcome of the calculations. Here, since $\epsilon$ is a prescribed parameter of the MDW solutions, $\eta_{rad}$ is found a posteriori using the expression above, since $\eta_{wind}$ and $\eta_{adv}$ are readily calculated for any MDW solution. The hotter and thicker the disc (the scale height $h$ is related to the midplane temperature), the larger $\eta_{adv}$ ($\eta_{adv} \propto \epsilon^2$). One might thus expect discs with a higher aspect ratio $\epsilon$ to be less bright and have a lower $\eta_{rad}$. While this scaling remains correct for our MDWs, the variations in $\eta_{rad}$ among them (see Table \ref{table:MDW_solution_parameters}) are mostly due to variations in the fraction of power carried away in the wind lobes $\eta_{wind}$. The disc temperature is itself a direct function of the disc accretion rate and $\eta_{rad}$ (see Eq.\ (\ref{eq:disc_temperature_profile})).
	
	\subsection{Considered MDW solutions}
	We consider five different MDW solutions as the basis for our disc wind models. Due to the limited understanding of the properties of post-AGB binary jets, we chose solutions covering a wide range in MDW properties. \#MDW 1 is a strongly magnetised and correspondingly inefficient solution with a thin disc. It also includes a small amount of additional heating at the disc surface \citep{Casse2000}. It has been proven to work well in describing YSO jets, and has been extensively employed to model their observations \citep{Panoglou2012, Yvart2016}. Moreover, it was used in the disc wind parameter study on HD 52961 presented in Paper I. Following Paper I's advice, we also consider various more efficient MDWs with thicker discs. \#MDW 2 and 3 are weakly magnetised, more efficient solutions with a thicker disc. The geometry of \#MDW 2's wind is particularly open compared to the other considered solutions (see Fig.\ \ref{fig:all_jed_example_structures}). \#MDW 4 and 5 are also weakly magnetised, thicker disc solutions, but with even higher ejection efficiencies. They are the most efficient solutions out of the families of MDWs in \citet{Jacquemin2019} assuming either $\alpha_m=1$ or $2$, respectively. 
	
	\section{Methods}\label{sect:methods}
	This section summarises the methodology for modelling and fitting synthetic $\mathrm{H_\alpha}$ time-series using the MDW solutions described in Sect.\ \ref{sect:self_similar_sol}. Sect.\ \ref{sect:spectral_input}  lists the necessary input derived from the observed HERMES spectra. Sect.\ \ref{sect:radiative_transfer} briefly describes the radiative transfer (RT) module. Sect.\ \ref{sect:fitting_routine} then presents our fitting routine, while Sect.\ \ref{sect:no_param_errors} discusses the current omission of error calculations for the fitted parameters. {{Aside from the new fitting routine, our methods are based upon those developed in \citet[development of the spectral input and RT routine]{Bollen2019, Bollen2020} and Paper I (extension of the RT routine to accommodate MDW solutions), and refer the reader to these papers for the details of the associated modelling steps, which are only concisely described here.}}
	
	\subsection{Spectral input}\label{sect:spectral_input}
	{{Our modelling approach requires several inputs from the full spectral time-series of each target. We briefly describe these inputs here, but refer to Appendix \ref{sect:appendix_spectral_inputs} for details on how they are derived.}}
	
	{{A selection of spectra for modelling was made from the full time-series. The resulting dynamic $\mathrm{H_\alpha}$ spectra for our targets are shown in the left column of Fig.\ \ref{fig:spectra_main_results}. A dynamic spectrum plots a time-series of spectra ordered in orbital phase $\phi_{orb}$. The $y$-axis represents the orbital phase, while the $x$-axis represents wavelength in the RV space of the considered line (from the COM frame of reference). A colour gradient interpolated between spectra then represents the normalised flux. In addition, our routine requires the binary orbital parameters, based on SB1 orbital parameters fitted to the HERMES RVs, in order to position the binary stars (Table \ref{table:orbital_params}). A background spectrum was constructed, representing the observed $\mathrm{H_\alpha}$ spectra outside conjunction and unaffected by jet absorption. These background spectra are shown in the middle column of Fig.\ \ref{fig:spectra_main_results}. Finally, an additional error term was calculated to take into account the inherent, non jet-related variability of the spectra (see Fig.\ \ref{fig:extra_background_error_profiles}).}}
	
	\subsection{Radiative transfer}\label{sect:radiative_transfer}
	{{
			After using the orbital solution to place the stars in position according to the considered $\phi_{orb}$, the RT routine draws rays from the primary's surface through the accretion disc wind launched from the secondary and towards the observer. Using the previously constructed $\mathrm{H_\alpha}$ background spectra as the initial radiation leaving the primary, the routine performs 1D RT along the rays to model the effect of disc wind absorption. The wind density and velocity profiles are set by the selected MDW solution, projected onto a Cartesian grid. Furthermore, the disc wind was assumed to be both oriented perpendicular to the orbit and in thermal equilibrium. Spatial resolutions of $\lesssim 0.05\,\mathrm{AU}$ and $0.1\,\mathrm{AU}$, for the directions perpendicular to and along the wind axis respectively, were achieved for all models.
	}}
	
	\begin{table}[t]
		\caption{Fitting routine parameter grid.}
		\label{table:grid_values}
		\setlength\tabcolsep{0mm}
		\centering
		\begin{tabular}{L{0.3333333\columnwidth} C{0.6666666\columnwidth}}
			\hline\hline & \\[-1.6ex]
			Parameter & Grid values\\[1.0ex]
			\hline & \\[-1.7ex] 
			$i\,(^{\circ})$ & $i_0\tablefootmark{a},\,i_0+4,\, ...\,,\,i_0+60$\\[0.5ex]
			$r_{in}\,(R_2)$ & $1,\,5,\,10,\,15,\,20,\,25$\\[0.5ex]
			$r_{out}\,(R_{RL,2})$ & $0.05,\,0.1,\,0.2,\, ...\,,\,0.9$\\[0.5ex]
			$\log{(\dot{M}_{in})}\,(\mathrm{M_{\odot}}/yr)$ & $-8,\,-7,\,-6,\,-5,\,-4,\,-3$\\[0.5ex]
			$T_{wind}\,\mathrm{(K)}$ & $T_{pAGB},\,T_{pAGB}-200,\, ...\,,\,T_{pAGB}-2000$\\[0.5ex]
			$R_1\,(\mathrm{R_\odot})$ & Fixed\tablefootmark{b} \\[0.8ex]
			\hline
		\end{tabular}
		\tablefoot{An ellipsis (...) indicates that the stepsize is maintained for the rest of the grid values. \tablefoottext{a}{$i_0=20^\circ$ for all targets except HP Lyr, for which $i_0=5^\circ$ was chosen due to the low inclination implied from previous modelling (see Table \ref{table:target_star_params}).}\tablefoottext{b}{Fixed to the values derived from previous modelling by \citet{Bollen2020, Bollen2021, Bollen2022} (see Table \ref{table:target_star_params}).}}
	\end{table}
	
	\subsection{Fitting routine}\label{sect:fitting_routine}
\begin{figure*}
		\centering
		\includegraphics[width=15.9cm]{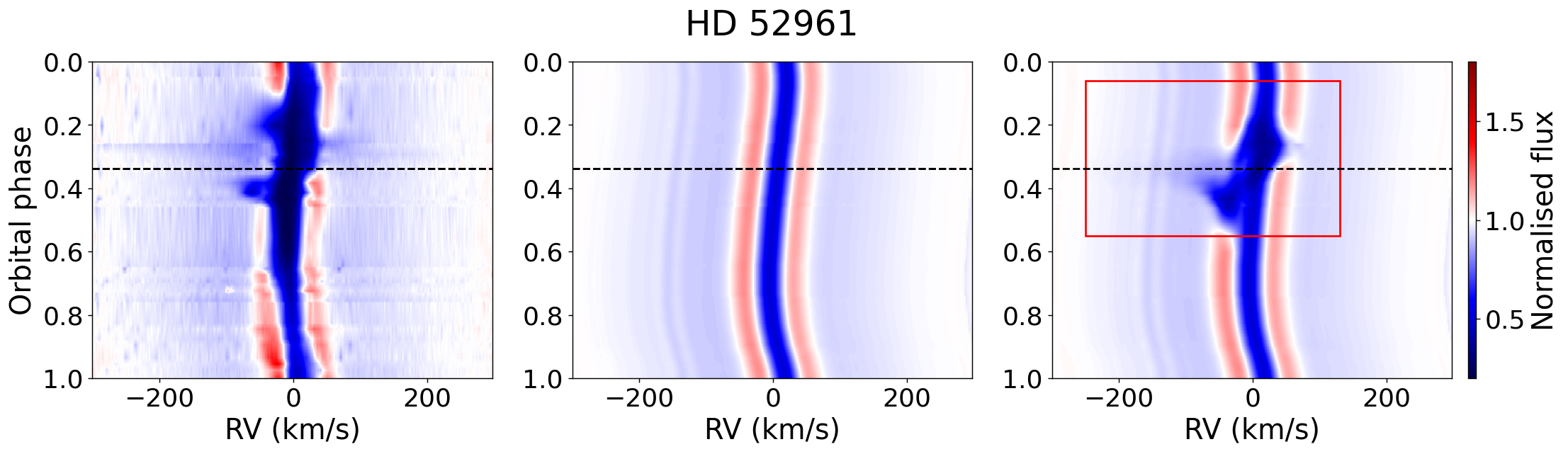}
		\includegraphics[width=15.9cm]{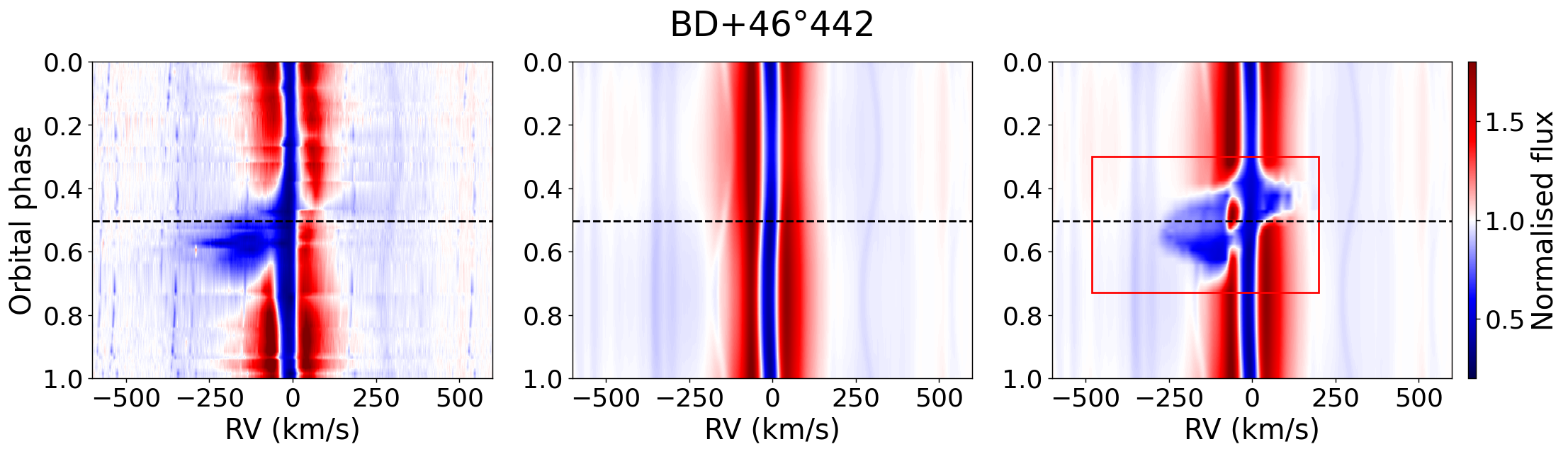}
		\includegraphics[width=15.9cm]{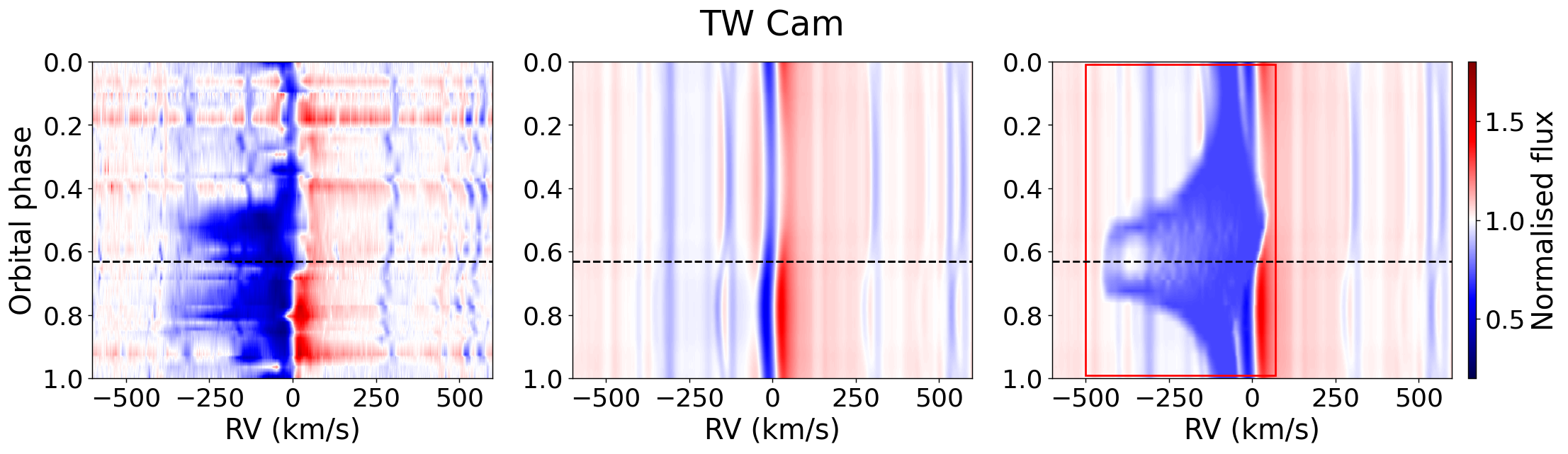}
		\includegraphics[width=15.9cm]{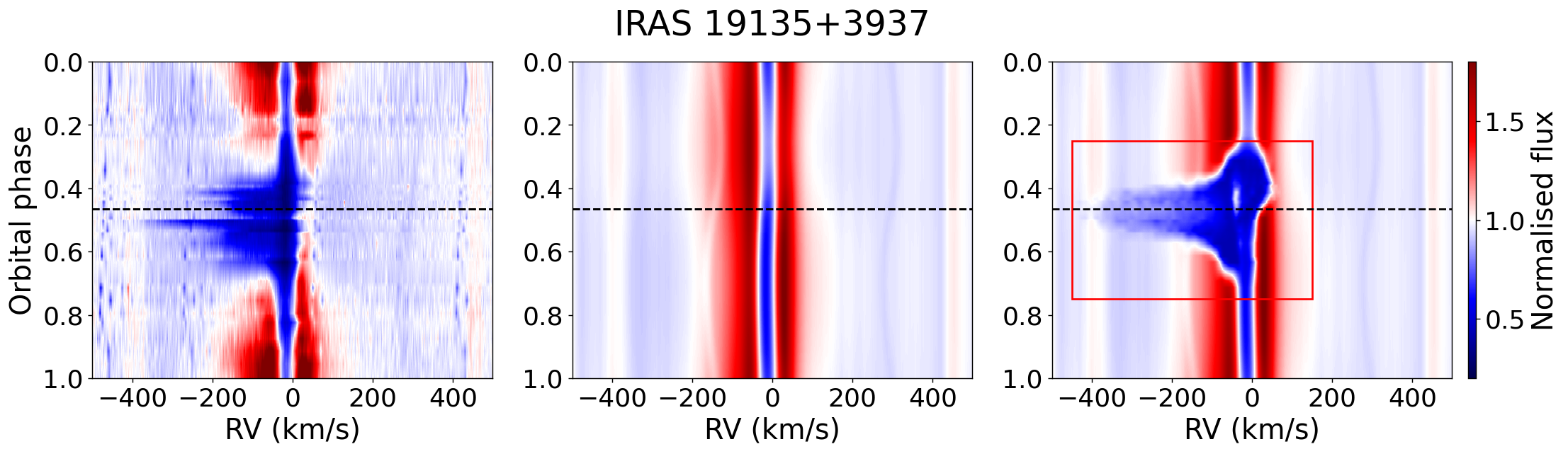}
		\includegraphics[width=15.9cm]{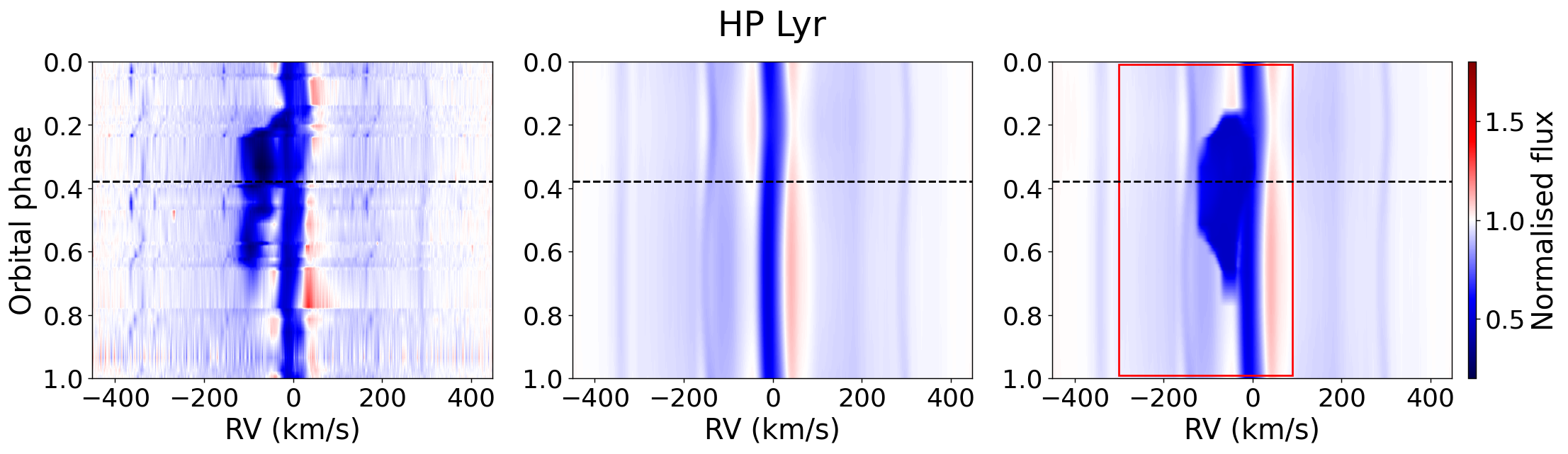}
		\caption{Dynamic $\mathrm{H_\alpha}$ spectra for the target systems. $\textit{Left:}$ observed spectral time-series after spectra selection. $\textit{Middle:}$ background spectrum used as input for the RT module. $\textit{Right:}$ best fitting disc wind model (parameters summarised in Table \ref{table:pAGB_MDW_models_best_fit_params}). A black dashed line marks superior conjunction, while a red selection box denotes the region used for the calculation of $\chi^2_\nu$.}
		\label{fig:spectra_main_results}
	\end{figure*}

	For the sake of robustness and in order not to miss additional degenerate minima, we employed a grid-based approach to fitting the synthetic time-series, iterating over the remaining free parameters. The values assumed in the parameter grid are summarised in Table \ref{table:grid_values}. The best-fitting model for every MDW was found by minimising the reduced chi-squared, $\chi^2_\nu$, within a box in the dynamic spectrum. This box was chosen via a range in $\phi_{orb}$ and RV, denoting the region of points used for the $\chi^2_\nu$ calculation (see Fig.\ \ref{fig:spectra_main_results}). The parameters' physical interpretations, effect on the dynamic spectra, units and limits are briefly summarised below (see Paper I for more details).
	
	\textit{Inclination, $i\,(^{\circ})$}. Sets the angle between the normal to the orbital plane (i.e.\ the disc wind axis) and the LOS. A lower $i$ value implies a wider binary orbit and a higher value of the secondary's mass, $M_2$. This also increases the secondary's radius, $R_2$ \citep[calculated from the MS mass-radius relationship of][]{Demircan1991}, and widens the coverage of the absorption feature in $\phi_{orb}$ as the LOS passes through higher up, geometrically broader regions of the jet. It also decreases the angle between the poloidal velocity and the LOS in well-collimated winds, causing the absorption to extend to higher absolute RV values. It lies between $0^\circ$ and $90^\circ$.
	
	\textit{Inner launching radius, $r_{in}\,(R_2)$}. The inner launching radius of the disc wind at the disc midplane. The field line anchored to this point is used as the inner boundary for the considered wind region. Increasing $r_{in}$ cuts out the high velocity components of the wind close to the secondary. It is expressed in units of the secondary star's radius, $R_2$, and is limited by the secondary's stellar surface, $r_{in} \geq 1 \, R_2$.
	
	\textit{Outer launching radius, $r_{out}\,(R_{RL,2})$}. The outer launching radius of the disc wind. Its field line is used as the outer boundary of the wind. Increasing $r_{out}$ adds material with low velocity to the outer wind regions, deepening the low RV absorption. This also makes the outer regions broader, making the absorption feature in the synthetic spectra wider in $\phi_{orb}$ coverage. It is expressed in units of the secondary's Roche radius, $R_{RL,2}$. It is limited by the inner launching radius and the Roche radius $R_{RL,2} > r_{out} > r_{in}$. $r_{out} \geq R_{RL,2}$ would namely imply that the circumcompanion accretion disc spills over onto the primary.
	
	\textit{Inner rim accretion rate, $\log{(\dot{M}_{in})}\,(\mathrm{M_{\odot}/yr})$}. The rate of material going through the inner edge of the accretion disc at $r_{in}$, and falling further onto the star. Disc wind densities increase proportionally to $\dot{M}_{in}$ \citep[e.g.][]{Jacquemin2019}, and so does the total rate of mass lost to the wind lobes: $\dot{M}_{wind} = 0.5 \cdot \dot{M}_{in}((r_{out}/r_{in})^\xi-1)$. This does not affect the wind geometry or kinematics, but causes deeper absorption signatures. The accretion rate at the outer disc edge then follows $\dot{M}_{out} = \left(r_{out}/r_{in}\right)^\xi \dot{M}_{in}$, expressing the feeding rate in the re-accretion stream from the circumbinary disc onto the accretion disc.
	
	\textit{Disc wind temperature, $T_{wind}\,\mathrm{(K)}$}. Homogeneous equilibrium temperature for the disc wind assumed during RT. Higher temperatures correspond to higher electron population densities in the first excited hydrogen state, implying stronger $\mathrm{H_{\alpha}}$ absorption. This effect is limited by the wind's thermal blackbody emission, which will eventually dominate over the increase in absorption as $T_{wind}$ increases. Since the wind feature is seen in absorption, we are limited by $T_{wind}< T_{pAGB}$, with $T_{pAGB}$ the temperature of the primary (see Table \ref{table:target_star_params}), as the feature would appear in emission otherwise. 	
	
	\textit{Primary radius, $R_1\,(\mathrm{R_\odot})$}. The radius of the primary post-AGB star. Increasing it increases the $\phi_{orb}$ coverage of the wind absorption feature, since the primary is obscured by the wind for a longer duration. In order to limit computational time, we fixed it to the values found from parametric modelling by \citet{Bollen2020, Bollen2021, Bollen2022} ($R_{1, \mathrm{par}}$, see Table \ref{table:target_star_params}). $R_1$ is one of the more robust, reliably retrieved parameters found from the parametric modelling.
	
	\subsection{Omission of parameter errors}\label{sect:no_param_errors}
	Remarkably, parametric modelling of the $\mathrm{H_\alpha}$ time-series using Markov chain Monte Carlo (MCMC) resulted in very small formal errors on the fitting parameters \citep[e.g.][]{Bollen2022}. The inclination, for example, had formal errors typically only of the order $\sim 0.1^\circ$. Given the expected correlations, this already seems strongly underestimated. Moreover, runs with relatively minor changes in the input spectra or parameter exploration range resulted in best fit inclinations that could easily differ by $> 5^\circ$. In the case of BD+46\textdegree442, we found an equally well-fitting, degenerate minimum that differed by as much as $20^\circ$.
	
	The current MCMC implementation \citep[the \texttt{emcee} package by][]{Foreman-Mackey2013} seems both overconfident and badly conditioned, and does not properly take degeneracies into account. Preliminary error determination for the disc wind model fits in this work, using a simple $\chi^2$ criterion, was similarly overconfident. As a result, we refrained from giving formal errors on our fitted parameters. We suggest to use the maximum step size in the parameter grid as a lower limit for the parameter errors (Table \ref{table:grid_values}). In following works, parameter degeneracies can be significantly lifted by assuming inclinations from interferometric modelling of the circumbinary disc (assuming co-planarity with the accretion disc). Though outside the scope of this paper, the use of MCMC algorithms more suitable to high-dimensional and multi-modal distributions, such as Hamiltonian Monte Carlo \citep[e.g.][]{hoffman2014}, could lead to more robust error determination.
	
	\section{Target stars}\label{sect:target_stars}
	\begin{table*}
		\caption{Target system properties.}
		\label{table:target_star_params}
		\setlength\tabcolsep{0mm}
		\centering
		\begin{tabular}{L{0.16666666666\textwidth} C{0.16666666666\textwidth} C{0.16666666666\textwidth} C{0.16666666666\textwidth} C{0.16666666666\textwidth} C{0.16666666666\textwidth}}
			\hline\hline & \\[-1.7ex]
			Object & HD 52961 & BD+46\textdegree442 & TW Cam & IRAS 19135+3937 & HP Lyr\\
			Parameter & & & & & \\[1.0ex]
			\hline & \\[-1.6ex]
			$T_{pAGB}\,(\mathrm{K})$ & $6000$ & $6250$ & $5250$ & $6250$ & $6000$\\[0.5ex]
			$\mathrm{[Fe/H]}$ & $-4.8$ & $-0.8$ & $-0.5$ & $-1.0$ & $-1.0$\\[0.5ex]
			$\mathrm{Depletion}$ & Strong & None & None & None & Strong\\[0.5ex]
			$P\,(\mathrm{d})$ & $1288.6 \pm 0.3$ & $140.82\pm0.02$ & $662.2\pm5.3$ & $126.97\pm0.08$ & $1818\pm80$\\[0.5ex]
			$e$ & $0.23 \pm 0.01$ & $0.085\pm0.005$ & $0.25\pm0.04$ & $0.13\pm0.03$ & $0.20\pm0.04$\\[0.5ex]
			$L_{Gaia}\,\mathrm{(L_{\odot})}$ & $3300^{+1600}_{-400}$ & $2400^{+1100}_{-1100}$ & $1500^{+400}_{-500}$ & $3600^{+700}_{-1200}$ & $11600^{+12100}_{-6200}$\\[0.5ex]
			$i_{\mathrm{par}}\,\mathrm{(^\circ)}$ & $70.6\pm0.6$ & $49.6\pm0.6$ & $25.7\pm0.2$ & $71.8\pm0.5$ & $18.6\pm0.7$\\[0.5ex]
			$R_{1, \mathrm{par}}\,(\mathrm{R_\odot})$ & $131\pm4$ & $27.3\pm0.6$ & $83.0\pm0.6$ & $22.6\pm0.4$ & $75\pm1$ \\[0.8ex]
			\hline & \\[-1.7ex]
			$N_{obs}$ & $42$ & $40$ & $36$ & $32$ & $40$\\[0.5ex]
			$\mathrm{(S/N)_{avg}}$ & $110$ & $70$ & $70$ & $40$ & $60$\\[0.5ex]
			Refs.\ & 1,2,6,7 & 3,6,7,8 & 4,6,7,9 & 2,6,7,8 & 5,6,7,9\\[0.5ex]
			\hline
		\end{tabular}
		\tablefoot{Relevant properties of the target systems studied in this work, including the primary's effective temperature $T_{pAGB}$, metallicity $\mathrm{[Fe/H]}$ and refractory depletion strength. In addition, we list the binary period $P$ and eccentricity $e$, the \textit{Gaia} DR3 luminosity $L_{Gaia}$, and the inclination $i_{\mathrm{par}}$ and primary star radius $R_{1, \mathrm{par}}$ found from previous parametric modelling. Below that, the number of selected observed spectra $N_{obs}$, and their average signal-to-noise ratio $\mathrm{(S/N)_{avg}}$, as well as the references used. $T_{pAGB}$ and $\mathrm{[Fe/H]}$ values (formal errors of $250\,\mathrm{K}$ and $0.3\,\mathrm{dex}$, respectively) were based on (1) \citet{waelkens1991}, (2) \citet{rao2014}, (3) \citet{Gorlova2012}, (4) \citet{giridhar2000} and (5) \citet{Giridhar_2005}. The refractory depletion strength classification, as well as $P$ and $e$ were taken from (6) \citet{Oomen2018}. $L_{Gaia}$ was derived from (7) \citet{gaiaDR3_2023}, $i_{\mathrm{par}}$ and $R_{1, \mathrm{par}}$ from (8) \citet{Bollen2020}, (9) \citet{Bollen2021} and (10) \citet{Bollen2022}. 
			We note that the luminosities are not fully reliable, as none of these systems were flagged as astrometric binaries in \textit{Gaia} DR3.}
	\end{table*}
	
	This section describes our target binaries and their observed $\mathrm{H_\alpha}$ spectra. Table \ref{table:target_star_params} summarises the properties of these systems used for their selection and modelling. The observed dynamic $\mathrm{H_\alpha}$ spectra, after spectra selection from the full time-series, are shown in the left column of Fig. \ref{fig:spectra_main_results}. Most Galactic post-AGB binaries (see \citet{Kluska2022} for a full catalogue) for which HERMES time-series were assembled indeed show jet absorption. For this paper, we selected five targets covering a good range in orbital properties, metallicity, refractory depletion strength, inclination from previous parametric modelling and visual morphology of the $\mathrm{H_\alpha}$ spectra. As a result, these five systems form a small but representative sample of the diversity seen in jet-launching post-AGB binaries.
	
	\subsection{HD 52961}
	HD 52961 shows extreme refractory depletion \citep{waelkens1991}, as reflected by its remarkably low metallicity of $\mathrm{[Fe/H]}=-4.8$. It has a fairly long orbital period at $P\approx 1300\,\mathrm{d}$ (known values in the population range from $\approx 100-2500\,\mathrm{d}$) and a substantially eccentric orbit at $e =0.23$ (values range from $\approx 0.0-0.65$). Previous studies detected infrared emission from UV-excited polycyclic aromatic hydrocarbons (PAHs) and $\mathrm{C^{60}}$ fullerenes \citep{gielen2011}. Most importantly, it was the target of the parameter study performed using \#MDW 1 in Paper I. Its observed dynamic $\mathrm{H_\alpha}$ spectrum presents a deep double-peaked jet absorption feature slightly offset from conjunction, covering $\phi_{orb} \approx 0.10-0.45$. The feature reaches down to $\mathrm{RV}\approx -70\,\mathrm{km/s}$. In addition, it shows a small amount of redshifted jet absorption up to $\mathrm{RV}\approx 60\,\mathrm{km/s}$ at $\phi_{orb} \approx 0.3$, which suppresses the rightmost emission wing of the background spectrum. The background emission wings follow the RV curve of the primary.
	
	{{
			HD52961's position in the HR diagram is close to the region occupied by the RV Tauri stars, a group of radially pulsating post-AGBs in the population II Cepheid instability strip. RV Tauri lightcurves show successive shallow and deep minima due to the pulsations \citep[][]{preston1963, alcock1998}. If this is the only source of photometric variability in a system, it is called an RVa variable. Post-AGB binaries can, however, also show a long-term brightness modulation that follows the orbital period. This modulation is caused by high system inclinations, causing periodic obscuration of the primary by the circumbinary disc as the LOS grazes its edge \citep{manick2017}. Regardless of whether they show the characteristic brightness dips of RV Tauri stellar pulsations or not, these systems are called RVb variables. HD 52961 is an example of an RVb. It has a long term periodic brightness modulation but, due to it lying just outside the instability strip, its stellar pulsations do not follow the RV Tauri pattern of successive shallow and deep brightness dips \citep{VanWinckel1999}. Previous parametric jet modelling implied a high inclination of $i_{par}\approx70^\circ$, which is consistent with its RVb classification.
	}}
	
	\subsection{BD+46\textdegree442}
	BD+46\textdegree442 is a non-depleted system with moderate metallicity (typical values range from $\approx -0.3 - 0.0$) and an almost circularised, very short period orbit at $P\approx140\,\mathrm{d}$ and $e\approx 0.1$. Previous parametric modelling implied an intermediate inclination of $i_{par}\approx 50^\circ$. It does not show strong pulsations. Its dynamic $\mathrm{H_\alpha}$ spectrum shows a jet absorption feature down to $\mathrm{RV}\approx-250\,\mathrm{km/s}$, slightly offset from conjunction. The feature spans $\phi_{orb}\approx 0.4-0.7$. The background emission wings stay centred on the COM. It was the first of the HERMES-observed systems to which the hypothesis of a circumcompanion-launched jet was applied \citep{Gorlova2012}.
	
	\subsection{TW Cam}
	TW Cam is a non-depleted system with moderate metallicity. It has an intermediate orbital period at $P\approx660\,\mathrm{d}$ and substantial eccentricity at $e=0.25$. It is an RV Tauri pulsator, with strong pulsations of up to $20\,\mathrm{km/s}$ amplitude in radial velocity \citep[see Fig.\ 9 in][]{manick2017}. This results in strong variability of the observed $\mathrm{H_\alpha}$ spectrum. We refrained from trying to clean the observed spectra (e.g. by removing shock-affected spectra). {{It was tentatively classified as an RVb by \citet{manick2017} due to a photometric variability component of low amplitude ($\Delta V = 0.07 \, \mathrm{mag}$) following the orbital period, implying a high inclination under the circumbinary disc-grazing LOS hypothesis.}} \citet{giridhar2000}, however, identified it as an RVa variable (not disc-grazing), and parametric jet modelling also predicted a low inclination of $i_{par}\approx26^\circ$. This is due to the fact there is some jet absorption present throughout the entire orbit, which occurs when the inclination is smaller than the jet opening angle. {{This, together with the low amplitude of the long-term photometric variability, suggests it might not be due to periodic obscuration by the circumbinary disc.}} TW Cam's dynamic $\mathrm{H_\alpha}$ spectrum shows wide, double-peaked jet absorption centred at conjunction, reaching down to $\mathrm{RV}\approx -300\,\mathrm{km/s}$. The feature covers almost the entire orbit, but becomes more prominent at $\phi_{orb}\approx 0.35-1.0$. The background emission wings stay centred on the COM.
	
	\subsection{IRAS 19135+3937}
	IRAS 19135+3937 has only a weak signature of refractory depletion, and has a slightly sub-solar metallicity. It has the third shortest orbital period out of all known post-AGB binary orbit solutions, with $P\approx 130\,\mathrm{d}$ \citep{Oomen2018}, and a moderate eccentricity at $e=0.13$. It has been classified as a class d semi-regular variable. This variability has previously also been attributed to obscuration by the circumbinary disc \citep{gorlova2015}, implying a high inclination. Correspondingly, previous parametric modelling fit a high inclination at $i_{par}\approx 72^\circ$ . The primary's photosphere seems relatively stable \citep{gorlova2015}. It has an $\mathrm{H_\alpha}$ jet absorption feature slightly offset from conjunction, ranging down to $\mathrm{RV}\approx -360\,\mathrm{km/s}$ and spanning $\phi_{orb}\approx 0.3-0.7$. In addition, it shows a small amount of redshifted absorption around $\phi_{orb}\approx 0.40$, reaching up to $\mathrm{RV\approx 50\,\mathrm{km/s}}$. The background emission wings stay centred on the COM.
	
	\subsection{HP Lyr}
	HP Lyr shows a very strong depletion pattern, though not quite as strong as that of HD 52961 \citep{Oomen2019}. It has the second longest period out of all known post-AGB binary solutions at $P\approx 1800\,\mathrm{d}$ \citep{Oomen2018}, and a significant eccentricity at $e=0.20$. It is classified as an RVa variable due to the lack of long-term brightness variations \citep{manick2017}. This is consistent with the fact that parametric jet modelling resulted in an almost pole-on inclination of $i_{par}\approx 19^\circ$. The RV pulsations of the primary are on the order of $5\,\mathrm{km/s}$ \citep[see Fig.\ 11 in][]{manick2017}. The observed background $\mathrm{H_\alpha}$ spectrum is thus variable, though less severely than in the case of TW Cam. It has a deep $\mathrm{H_\alpha}$ jet absorption feature, reaching down to $\mathrm{RV}\approx -120\,\mathrm{km/s}$ and spanning $\phi_{orb}\approx 0.15-0.7$. Approximately from $\phi_{orb} \approx 0.42-0.70$, the jet absorption feature detaches from the photospheric absorption line, letting the leftmost background emission wing shine through. The background emission wings stay centred on the COM.
	
	\section{Results}\label{sect:results}
	\begin{table*}
		\caption{Fitting parameters and $\chi^2_\nu$ values for the best fitting disc wind models.}
		\label{table:pAGB_MDW_models_best_fit_params}
		\setlength\tabcolsep{0mm}
		\centering
		\begin{tabular}{L{0.18181818181 \textwidth} C{0.0809090909 \textwidth} C{0.0809090909 \textwidth} C{0.0809090909 \textwidth} C{0.0809090909 \textwidth} C{0.0909090909 \textwidth} C{0.0909090909 \textwidth} C{0.0909090909 \textwidth} C{0.1109090909 \textwidth} C{0.1109090909 \textwidth}}
			\hline\hline & \\[-1.7ex]
			Object & \#MDW & $i$ & $r_{in}$ & $r_{out}$ & $\dot{M}_{in}$ & $T_{wind}$ & $\chi^2_\nu$ & $\dot{M}_{out}$ & $\dot{M}_{wind}\tablefootmark{a}$\\
			& & $\mathrm{(^\circ)}$ & $(R_2)$ & $(R_{RL,2})$ & $(\mathrm{M_\odot /yr})$ & $(\mathrm{K})$ & & $(\mathrm{M_\odot /yr})$ & $(\mathrm{M_\odot /yr})$ \\[1.0ex]
			\hline & \\[-1.6ex]
			HD 52961 & 4 & 72 & 1 & 0.9 & $10^{-4}$ & 4200 & 5.32 & $3\cdot 10^{-4}$ & $1\cdot 10^{-4}$ \\[0.5ex]
			BD+46\textdegree442 & 3 & 56 & 5 & 0.8 & $10^{-5}$ & 5250 & 6.44 & $1.2\cdot 10^{-5}$ & $1\cdot 10^{-6}$\\ [0.5ex]
			TW Cam & 3 & 24 & 5 & 0.9 & $10^{-3}$ & 4850 & 5.42 & $1.4\cdot 10^{-3}$ & $2\cdot 10^{-4}$ \\ [0.5ex]
			IRAS 19135+3937& 4 & 48 & 1 & 0.9 & $10^{-5}$ & 5050 & 3.30 & $2.4\cdot 10^{-5}$ & $7\cdot 10^{-6}$ \\[0.5ex]
			HP Lyr & 4 & 29 & 20 & 0.3 & $10^{-4}$ & 5000 & 4.91 & $1.4\cdot 10^{-4}$ & $2\cdot 10^{-5}$  \\[0.8ex]
			\hline
		\end{tabular}\tablefoot{In addition to the fitted parameters, we also show the feeding rate at the outer disc rim and the wind mas-loss rate. \tablefoottext{a}{$\dot{M}_{wind}$ denotes the mass-loss rate of one of the wind lobes. Since the wind is bipolar, conservation of mass implies $\dot{M}_{out} = \dot{M}_{in} + 2\dot{M}_{wind}$.}}
	\end{table*}
	In this section, we describe the resulting $\mathrm{H_\alpha}$ profiles of our fitted models and compare them to the observations. The fitting routine of Sect.\ \ref{sect:fitting_routine} was run using each of the MDWs in Table \ref{table:MDW_solution_parameters}, resulting in five disc wind model fits per target, one per MDW solution. The parameters and dynamic $\mathrm{H_\alpha}$ spectra for the best fitting models are shown in Table \ref{table:pAGB_MDW_models_best_fit_params} and the right column Fig.\ \ref{fig:spectra_main_results}, respectively. For completeness, the results of the fitting procedure using all MDW solutions are summarised in Appendix \ref{sect:appendix_all_model_fits}.
	
	\subsection{HD 52961}
	\#MDW 4 provides the best fit for HD 52961's $\mathrm{H_\alpha}$ time-series at $\chi^2_\nu = 5.32$. It reproduces both the general absorption depth, RV extent and the phase positioning of the high $\phi_{orb}$ absorption peak. The lower $\phi_{orb}$ peak in the model, in contradiction with the observations, is redshifted. As discussed by Paper I, this redshift is induced by the prograde rotation of the wind material, causing the material's velocity to be pointed away from the observer as the LOS to the primary passes the wind's leading edge. Correspondingly, this also blueshifts the lower $\phi_{orb}$ absorption peak, as the trailing edge passes the LOS. This results in a strong RV asymmetry of the modelled wind absorption. Instead of being placed at its observed position of $\phi_{orb}\approx 0.20$, the modelled low $\phi_{orb}$ absorption peak is located at $\phi_{orb}\approx 0.30$, where it tries to fit the small redshifted absorption feature seen in the observations.
	
	Compared to the other targets, the rotation effect is remarkably strong in HD 52961's best fit. This is due to its high fitted inclination of $i=72^\circ$, causing the LOS to pass through regions close to the wind's base, where the rotation is strongest. Such a high fitted inclination is consistent with its RVb classification. None of the other models manage to suppress the wind rotation enough to blueshift the low $\phi_{orb}$ peak. \#MDW 1, 2 and 3 perform significantly worse at $\chi^2_\nu \leq 6.92$ due to the strong effect of rotation (see Fig.\ \ref{fig:all_fit_dynspec} and Table \ref{table:pAGB_MDW_models_all_fit_params}). Only \#MDW 5, the most efficient of the thicker disc solutions, can reduce the wind rotation more and reaches $\chi^2_\nu = 5.63$, but only manages to shift the low $\phi_{orb}$ peak into the RV position of the photospheric absorption line. We note that \#MDW 5 fits at $i=56^\circ$, which is significantly lower than for the other MDWs ($i>70^\circ$).	Given \#MDW 5's high efficiency, the density of the wind is too high close to the base. In order to compensate, the fit tends towards a low inclination in order to move the LOS to less dense regions. However, $i=56^\circ$ seems fairly low, given the system's robust RVb classification.
	
	Despite significant mismatches the $\chi^2_\nu$ values are still fairly low. This is a general feature of our fits due to the large selection boxes used, which include a lot of continuum points. These hold little statistical significance, but by construction of the background spectrum match up well to the observations. As a result, different MDW fits can thus sometimes fit the global $\mathrm{H_\alpha}$ pattern very differently while still achieving similar $\chi^2_\nu$ values (e.g.\ IRAS 19135+3937 \#MDW 3 and 4, see Fig.\ \ref{fig:all_fit_dynspec}). We thus stress that the choice of the best fitting MDW model was guided by the $\chi^2_\nu$ values, but ultimately depended on how well the specific MDW managed to recover the global $\mathrm{H_\alpha}$ pattern.
	
	\subsection{BD+46\textdegree442}
	BD+46\textdegree442 is best fitted by \#MDW 3 at $\chi^2_\nu = 6.44$. This model reproduces the absorption depth, $\phi_{orb}$ coverage and negative RV extent of the observed absorption feature. Two glaring mismatches with the observations remain, however. The first is the rotation-induced redshifted absorption feature centred at $\phi_{orb}\approx 0.41$, which is not seen in the observations. This feature is also present in the \#MDW 1 and 2 fits, which both fit more poorly at $\chi^2_\nu \leq 6.73$. Rotation also blueshifts the lower $\phi_{orb}$ peak. This detaches the wind absorption from the photospheric absorption at $\phi_{orb}\approx 0.55-0.70$. The high efficiency \#MDW 4 and 5 models do manage to suppress wind rotation to a significant degree, but produce poorer fits at $\chi^2_\nu \leq 6.85$. Their modelled absorption features are too broad in $\phi_{orb}$, and extend too much downwards into negative RV values. This is due to their lower fitted inclinations ($i < 45^\circ$ versus $i>55^\circ$ for the other fits), which are assumed because the wind regions close to the base are too dense. In addition, close to superior conjunction, the leftmost background emission wing shines through all fits, implying a general lack of low RV absorption in all the models.
	
	\subsection{TW Cam}
	TW Cam is best fitted by \#MDW 3 at $\chi^2_\nu = 5.42$. This model's wind absorption feature reproduces the extent in $\phi_{orb}$ and negative RV range of the main absorption peaks in the observations fairly well. The absorption depth is generally too shallow, most likely because the parameter grid is limited to $\dot{M}_{in} \leq 10^{-3} \, \mathrm{M_\odot /yr}$. The fit assumes a low inclination of $i = 24^\circ$. This is lower than the typical wind opening angle at the point where the outer wind edge and LOS rays cross, causing the primary to be obscured and the model to show wind absorption at all times. The low inclination implies we look through regions high above the wind's base, so the effect of rotation is a lot less pronounced. The observed blueshifted absorption peaks are fairly asymmetrical, where the lower $\phi_{orb}$ peak at $\phi_{orb}\approx 0.5$ is slightly more narrow in $\phi_{orb}$ coverage and is concentrated at more negative RV values than the other peak at $\phi_{orb}\approx 0.8$. In comparison, the modelled absorption peaks are fairly symmetrical. This is most likely due to a neglect of a wind tilt in our current models \citep[see e.g.][]{Bollen2020}, which can be dealt with by adding a jet tilt parameter. Since this is a second order effect, it is postponed for future work. The wind absorption away from superior conjunction (e.g.\ at $\phi_{orb}\approx 0.0-0.3$) detaches from the photospheric absorption due to high poloidal velocities, causing a mismatch with the observations. There is too much absorption in a broad RV band, and the background spectrum immediately blueward of the photospheric absorption shines through the model. The \#MDW 1 fit shows a similar effect, and correspondingly fits almost equally well at $\chi^2_\nu = 5.59$. The higher efficiency \#MDW 4 and 5 indeed manage to reattach the wind absorption outside of superior conjunction to the photospheric absorption due to their inherently lower poloidal velocities. In addition, \#MDW 5 reproduces the observed absorption depth much better. They however fail in reproducing the observed wind absorption's width in $\phi_{orb}$, and as a result fit significantly poorer at $\chi^2_\nu \leq 6.44$. The \#MDW 2 fit is also poorer at $\chi^2_\nu = 6.57$. We thus have been unable to find the right combination of kinematics and density profile to fully reproduce TW Cam’s observations.
	
	We note that the fitted inclinations are consistently on the lower end at $i < 44^\circ$. Parametric modelling also implied a low inclination at $i_{par} = 25.7^\circ$ (Table \ref{table:target_star_params}). Assuming a wind axis approximately perpendicular to the orbital plane, this strongly challenges the RVb classification of TW Cam by \citet{manick2017}, instead supporting the RVa classification by \citet{giridhar2000}.
	
	\subsection{IRAS 19135+3937}
	IRAS 19135+3937 is best fitted by \#MDW 4 at $\chi^2_\nu = 3.30$. This model matches the observations very well, reproducing both the general RV extent, $\phi_{orb}$ width, absorption depth and weak redshifted absorption indent at $\phi_{orb} \approx 0.40$. Similar to the BD+46\textdegree442 models, a lack of low RV absorption close to superior conjunction is indeed present. This is also the case for all other fits for this target. In the \#MDW 4 and 5 models this effect is however almost completely reduced due to the inherently lower poloidal velocities. \#MDW 1 and 2 are poorer fits at $\chi^2_\nu \leq 3.96$, but \#MDW 3 and 5 fit almost as well as \#MDW 4 at $\chi^2_\nu = 3.49 \, \& \, 3.53$, respectively. Nevertheless, the \#MDW 3 model's wind absorption is too shallow, and suffers the same mismatches of overestimated rotation and lack of low RV absorption as the BD+46\textdegree442 \#MDW 3 model. Aside from the \#MDW 2 model, fitted inclinations are relatively low at $i < 48^\circ$. \#MDW 5 fits an especially low inclination at $i = 28^\circ$, while parametric modelling fitted a high inclination at $i_{par} = 71.8^\circ$. The parametric prediction is more consistent with the identification of obscuration by the circumbinary disc as the cause for the observed photometric variability \citep{gorlova2015}.
	
	\subsection{HP Lyr}
	HP Lyr is best fitted by \#MDW 4 at $\chi^2_\nu = 4.91$. The model reproduces the observation very well, matching the RV extent and $\phi_{orb}$ width of the wind feature. The absorption depth is only slightly too shallow. The wind absorption from $\phi_{orb} \approx 0.42-0.70$ does not detach as much from the photospheric absorption line as in the observations due to excessive suppression of rotation, as the \#MDW 1 and 3 models do reproduce this feature. \#MDW 1, 2 and 3 however fit much worse overall with $\chi^2_\nu \leq 6.29$. The \#MDW 5 model is extremely similar to the \#MDW 4 one, reaching $\chi^2_\nu = 5.05$, though with a more shallow absorption depth due to its lower inclination of $i = 17^\circ$ ($i\geq 29^\circ$ for all other MDWs). As was the case for HD 52961, \#MDW 5 most likely assumes this lower inclination to avoid the densest regions close to the wind base. However, this implies a very large mass for the secondary star at $M_2 = 4.3\, \mathrm{M_\odot}$. In contrast, known SB1 orbital parameters imply $M_2$ follows a distribution with a mean of $1.09\, \mathrm{M_\odot}$ and standard deviation of $0.62\, \mathrm{M_\odot}$ (assuming random orbit orientations; \citealt{Oomen2018}). An upper limit on $M_2$ might serve as a good lower limit for $i$ during future fitting runs.
	
	\section{Discussion}\label{sect:discussion}
	This paper expands on Paper I in the pursuit to test MDW-based models in the context of post-AGB binary jets. We have successfully introduced our fitting routine and have made quantitative comparisons between the $\mathrm{H_\alpha}$ line predictions for five different MDWs and the observed HERMES time-series of our targets. In this section, we will proceed with an in-depth discussion on several aspect of our fitted models and the associated physics, using the results of Paper I as a comparative guideline.
	
	\subsection{Jet rotation}\label{sect:jet_rotation}
	The rotation of jets is a natural consequence of their formation, and is confirmed by direct observations of rotation and/or winding of the magnetic field lines in YSO jets \citep[e.g.][]{Launhardt2023, moscadelli2022}. From the lack of rotation-induced redshift in the observed HD 52961 wind absorption feature, Paper I concluded that wind rotation in this system must be strongly suppressed compared to what is predicted by the YSO-tailored \#MDW 1. As suggested by the authors, we considered more weakly magnetised, thicker and more efficient MDW solutions, which naturally lead to weaker jet rotation.
	
	For TW Cam and HP Lyr, both systems fitted under a low inclination ($i< 28^\circ \, \& \, 37^\circ$, respectively, excluding the poorly fitting \#MDW 2), the rotation-induced redshift is relatively weak for all considered MDW solutions (see Fig.\ \ref{fig:all_fit_dynspec}). As shown in Paper I, it are mostly the strongly rotating, sub-Alfv\'{e}nic regions close to the disc surface which are responsible for the redshifted absorption. Since the low inclinations cause the LOS to pass through higher up wind regions, the rotation is mostly avoided.
	
	The good fits for IRAS 19135+3937, using \#MDW 4 and 5, both reproduce a small but noticeable rotation-induced redshift in the absorption at $\phi_{orb}\approx 0.40$, matching a similar feature in the observations. The other, less efficient solutions predict redshifts that are too strong (see Fig.\ \ref{fig:all_fit_dynspec}). This implies that while rotation in the actual IRAS 19135+3937 jet is indeed relatively weak compared to typical YSO jets, it is still present to some degree. 
	
	Redshifted wind absorption is still strongly present in the best fit model for HD 52961 using \#MDW 4. The higher $\xi$ \#MDW 4 and 5 clearly suppress the rotation-induced redshift more compared to the lower $\xi$ \#MDW 2 and 3, and the lower $\xi$ and $\epsilon$ \#MDW 1. Due to the high fitted inclinations, rotation can even dominate the velocity field when using \#MDW 1, 2 and 3 (see Fig.\ \ref{fig:all_fit_dynspec}). Nevertheless, the rotation suppression is never sufficient to match the data.
	
	Also in the case of BD+46\textdegree442, the best fit using \#MDW 3 does not seem efficient enough to suppress the redshifted feature, which is completely absent in the observations (see Fig.\ \ref{fig:spectra_main_results}). While the \#MDW 4 and 5 models do manage to fully suppress the redshift, their absorption features are slightly too broad in $\phi_{orb}$, and reach RV values that are too negative (see Fig.\ \ref{fig:all_fit_dynspec}). These models fit lower inclinations of $i = 44^\circ \, \& \, 28^\circ$, respectively, compared to $i > 55^\circ$ for the other models. Since \#MDW 3 already fitted the absorption depth seen in the observations well at $i=56^\circ$, these more efficient solutions are too dense close to the base of the wind, and the inclination is correspondingly decreased so the LOS avoids these dense regions. As a result, the absorption feature becomes broader in phase, and since the angle between the LOS and the poloidal velocities decreases, absorption extends to more negative RV.
	
	For HD 52961 the rotation suppression is simply not strong enough using the currently considered MDW solutions, while for BD+46\textdegree442 they do not provide the correct combination of kinematics and density profile. New MDW solutions with thicker accretion discs at $\epsilon > 0.1$ must be considered to solve these discrepancies. As pointed out by Paper I, however, purely magnetically driven cold disc winds start failing as the disc becomes too thick. Discs that rotate too slowly simply cannot provide the rotational push to launch the material \citep{Casse2000}. As we will discuss in Sect.\ \ref{sect:MDW_solution_preference}, the inclusion of irradiation by the primary star might overcome this limitation.
	
	\subsection{Accretion rates}\label{sect:accretion_rates}
	{{Paper I raised the issue of lifetimes for the circumbinary discs, which feed the jet-launching accretion discs. Their \#MDW 1 model preferred a feeding rate of $\dot{M}_{out} \approx 10^{-4}\, \mathrm{M_\odot /yr}$. Unlike in some YSOs, there is no natal envelope which can replenish the circumbinary disc. Given that they have a typical total mass of up to $M_{cb} = 10^{-2}\,\mathrm{M_\odot}$ \citep[derived from radio observations, e.g.][]{Bujarrabal2013, Bujarrabal2015, Bujarrabal2016, Bujarrabal2017, Bujarrabal2018, GallardoCava2021}, the circumbinary discs should then dissipate on a timescale $\tau_{cb} = M_{cb}/\dot{M}_{out} \sim 100\, \mathrm{yr}$.}} The post-AGB phase lifetime for a lower-mass star can however easily be $\tau_{pAGB}\gtrsim 10\,000\,\mathrm{yr}$, and the re-accretion onto the primary might extend this by a factor of a few \citep{MillerBertolami2016, Oomen2019}. With a large fraction of $\approx 30\%$ of all observed post-AGBs being circumbinary disc-bearing \citep{Kamath2014, Kamath2015, Kluska2022}, they should not be able to dissipate so quickly, and we'd expect $\tau_{cb} \sim \tau_{pAGB}$. To push down the required feeding rate and increase the $\tau_{cb}$ estimate, Paper I suggested the use of more efficient MDW solutions, as such solutions can divert a larger fraction of the accretion stream toward the disc wind, requiring less material to be fed in order to obtain the same $\mathrm{H_\alpha}$ absorption depth.
	
	Across our targets, \#MDW 1 ($\xi=0.04$) fits in the range $\dot{M}_{out} \sim 10^{-4} - 10^{-3} \, \mathrm{M_\odot /yr}$ (see Table \ref{table:pAGB_MDW_models_all_fit_params}). \#MDW 2 and 3 ($\xi\approx 0.10$) manage to push the lower end down by an order of magnitude to $\dot{M}_{out} \sim 10^{-5} - 10^{-3} \, \mathrm{M_\odot /yr}$. \#MDW 4 ($\xi = 0.215$) fits $\dot{M}_{out} \sim 10^{-5} - 10^{-4} \, \mathrm{M_\odot /yr}$, while \#MDW 5 ($\xi = 0.305$) fits $\dot{M}_{out} \sim 10^{-6} - 10^{-4} \, \mathrm{M_\odot /yr}$. More efficient MDWs thus generally do predict lower feeding rates, but even for \#MDW 5, the most efficient of all known cold magnetocentrifugal solutions \citep{Jacquemin2019}, the high end of the range still predicts $\tau_{cb} \sim 100\, \mathrm{yr}$, which seems much too short. Evidently, the feeding rates must be pushed down even further, ideally by one to two orders of magnitude. We have, however, reached the limit of efficiency for purely magnetocentrifugal cold winds from an $\epsilon = 0.1$ disc with \#MDW 5 \citep{Jacquemin2019}. As was put forward as a possibility in the conclusions of Paper I, an additional driving force is needed. Irradiation from the bright post-AGB primaries, with luminosities easily over $1000\,\mathrm{L_\odot}$ (Table \ref{table:target_star_params}), is bound to deposit heat onto the accretion discs' surfaces, adding an extra thermal push to the wind material. Indeed, such 'warm' magnetothermal solutions have already been explored in the context of YSO jets with coronal heating \citep{Casse2000(2)}, which can readily reach huge efficiencies of $\xi > 0.5$. Such solutions will need to be explored in future work to push down the fitted values for $\dot{M}_{in}$ and $\dot{M}_{out}$, and solve the current circumbinary disc lifetime discrepancy. Lower $\dot{M}_{in}$ values will also lead to slightly colder accretion disc predictions (see Sect.\ \ref{sect:accretion_disc_temperature}) and lower secondary field strengths required to truncate the accretion discs through magnetospheric funnelling (see Sect.\ \ref{sect:disc_truncation_funneling}). It will also allow for thicker, more slowly rotating disc solutions at $\epsilon >0.1$ to be explored (see Sect.\ \ref{sect:jet_rotation}). 
	
	\subsection{Disc truncation and magnetospheric funnelling}\label{sect:disc_truncation_funneling}
	Considering the inefficient \#MDW 1 solution, Paper I discussed the necessity of large inner launching radii $\gtrsim 20\, R_2$ in order to match the HD 52961 observations. This would cut out the inner wind regions, which were too fast in their model. As there is no apparent physical reason for this inner launching radius to be different from the inner radius of the bulk accretion disc itself, this seemed to imply that the HD 52961 accretion disc had a large inner cavity and was heavily truncated. The required inner radii were, however, deemed too high for this truncation to be possible via magnetospheric funnelling, a mechanism often invoked to explain disc truncation around YSOs \citep[e.g.][]{bouvier1999, bouvier2003}, where the stellar magnetic dipole disrupts the accretion flow and funnels it along the magnetospheric field lines \citep[see e.g.][]{Bessolaz2008}. To relieve this tension, Paper I suggested the use of more efficient MDW solutions with inherently lower poloidal velocities, which we consider here.
	
	Both HD 52961 and IRAS 19135+3937 have a best fit with $r_{in} = 1\, R_2$, implying an inner disc rim that reaches the stellar surface. This is consistent with the fact that they are best fitted by \#MDW 4, the second-most efficient MDW solution at $\xi = 0.215$. Contrary to the solutions with lower $\xi$, the inner jet regions do not need to be cut out to suppress absorption at high negative RVs (see Table \ref{table:pAGB_MDW_models_all_fit_params}). BD+46\textdegree442 and TW Cam are best fitted by \#MDW 3, which is more efficient than \#MDW 1, but still less than \#MDW 4, and correspondingly fit an inner rim slightly further away at $r_{in} = 5\, R_2$. It is not clear if the discs are truly truncated in these two systems, since a more careful selection of the used MDW solution might result in best fits at $r_{in} = 1\, R_2$. In any case, these fits are consistent with the discs either reaching the stellar surface, or at least being truncated out to only several stellar radii.
	
	Following Paper I and considering magnetospheric funnelling as the truncation mechanism, the strength of the stellar dipole at the stellar equator required to truncate the disc can be written as \citep{Bessolaz2008}:
	\begin{equation}\label{eq:required_field_bessolaz}
		B_2 \sim 11 \,\mathrm{kG} \cdot \left( \frac{r_{in}}{R_2} \right)^{\frac{7}{4}} \left( \frac{\dot{M}_{in}}{10^{-4}\,\mathrm{M_\odot /yr}} \right)^{\frac{1}{2}} \left( \frac{M_2}{1\,\mathrm{M_\odot}} \right)^{\frac{1}{4}}\left( \frac{R_2}{1\,\mathrm{R_\odot}} \right)^{-\frac{5}{4}}.
	\end{equation}
	To start truncating the disc ($r_{in} > 1\, R_2$), the reasonably well-fitting models for the four targets discussed above imply magnetic field strengths greater than several kG and up to $\sim 15\, \mathrm{kG}$. For comparison, a typical heavily accreting YSO at say $\dot{M}_{in} \approx 10^{-7}\, \mathrm{M_\odot /yr}$ can achieve mean field strengths of up to a few kG \citep[e.g.][]{Johns-Krull_1999}, and at such accretion rates this is strong enough to truncate the accretion disc out to several stellar radii \citep{Bessolaz2008}. Our secondaries are mature main sequence stars, not entirely dissimilar to for example the Sun. The average equatorial solar dipole strength is only on the level of $\sim 1\,\mathrm{G}$, implying that our secondaries' inherent field strengths would be much too low for truncation at the current required accretion rates. However, if the secondary's magnetic field is driven by a stellar dynamo \cite[see e.g.][]{Vidotto2014, Wurster2018}, the strong accretion streams in our systems could be expected to increase stellar rotation and rejuvenate the stellar magnetic field. If, in addition, the required $\dot{M}_{in}$ rates are indeed much lower than our current models imply (see Sects.\ \ref{sect:accretion_rates} \& \ref{sect:MDW_solution_preference}), the accretion discs might be magnetically truncated, though in any case still close to the stellar surface. In future modelling, it will thus be reasonable to assume that the accretion disc either extends down to the stellar surface or is truncated out to several stellar radii at most, meaning $r_{in} \sim 1\, R_2$. Constraints on the secondaries' field strengths might provide more stringent limits on $r_{in}$, but the brightness contrast with the primaries unfortunately precludes direct determination of $B_2$ from spectra, using for instance Zeeman broadening.
	
	While the use of more efficient MDW solutions does lead to much smaller fitted $r_{in}$ values for four out of five of our targets (Table \ref{table:pAGB_MDW_models_all_fit_params}), one exception sticks out. The wind absorption feature in HP Lyr only extends down to $\mathrm{RV\approx -120\, \mathrm{km/s}}$, despite its low inclination (compare to TW Cam, where it reaches $\mathrm{RV\approx -320\, \mathrm{km/s}}$). Correspondingly, HP Lyr is fitted with $r_{in} \geq 20\, R_2$ for all MDW solutions (see Table \ref{table:pAGB_MDW_models_all_fit_params}). HP Lyr's best fit would imply a field strength of $B_2 \approx 1.6\, \mathrm{MG}$. The \#MDW 5 fit (Table \ref{table:pAGB_MDW_models_all_fit_params}) predicts the lowest field strength value for this target, at $B_2 \approx 800\, \mathrm{kG}$. These values are extraordinarily high, and impossible to reach via a MS stellar dynamo (e.g. because of saturation; \citealt{Vilhu1984}). Even if the accretion rate in our HP Lyr models is reduced to $\sim 10^{-8}\,\mathrm{M_\odot /yr}$, the required field strength would only be brought down to $\sim 16\, \mathrm{kG}$. Magnetospheric funnelling simply cannot be responsible for truncating HP Lyr's accretion disc out to this large a radius, and the accretion disc should extend further down towards the secondary.
	
	The consistently large required $r_{in}$ value for HP Lyr's models could be an artefact from the strongly simplified thermal structure in the RT routine, where we assumed a homogeneous wind temperature. This assumption forces all material included in the model to be set to a fairly high temperature, causing the entire wind structure to absorb. If HP Lyr's accretion disc wind has an inverse temperature gradient in reality, with colder inner regions and hotter outer regions (though still bellow $T_{pAGB}$), only the hydrogen in the slow outer regions would be hot enough to absorb significantly in $\mathrm{H_\alpha}$. In that case, our fitting routine accounts for this by simply cutting out the fast inner wind regions. An inverse temperature gradient might, for example, be caused by interaction between the outer disc wind regions and the fast stellar wind launched by the post-AGB primary (with mass loss rates up to $10^{-7}\,\mathrm{M_\odot /yr}$; \citealt{VanWinckel2003}), if the latter is strong enough and our accretion rate estimates are reduced. Future modelling of HP Lyr, and possibly similar objects that are fitting at high $r_{in}$ values despite the use of very efficient MDW solutions, warrants a more careful calculation and analysis of the thermal wind structure. This can be done either in detail by solving the energy equation along the field lines for specific disc wind models, or, at lesser computational expense, by including parametrised wind temperature profiles in the RT routine.
	
	\begin{figure*}
		\centering
		\includegraphics[width=\textwidth]{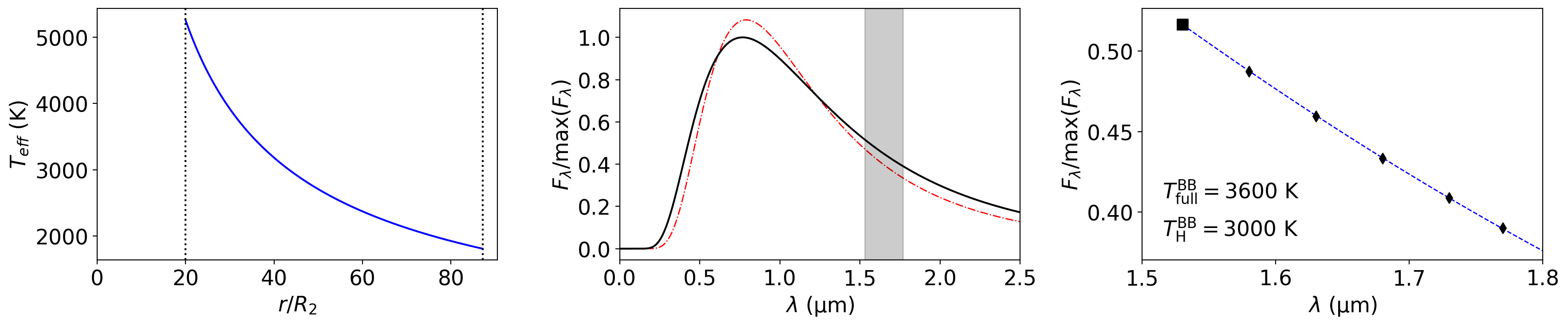}
		\includegraphics[width=\textwidth]{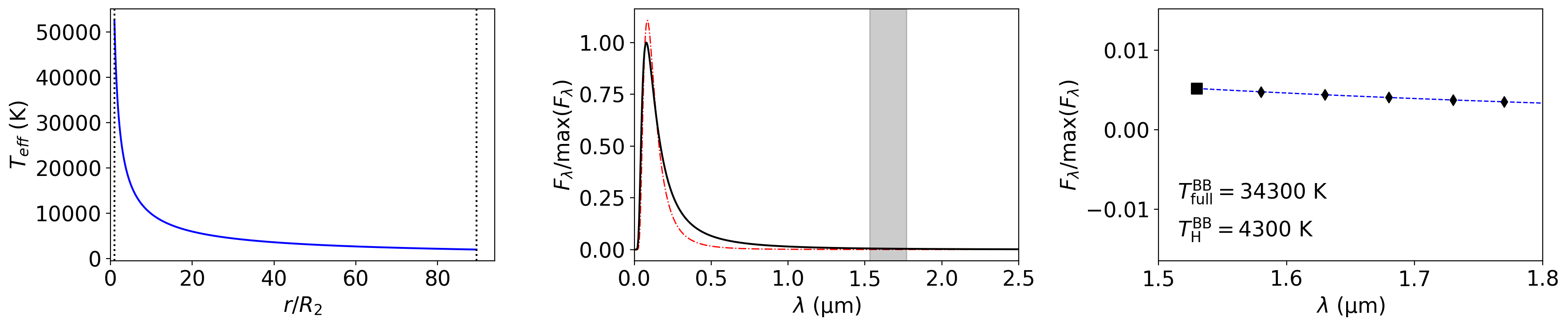}
		\caption{Example effective disc temperature profiles and spectra. \textit{Top:} HP Lyr's \#MDW 3 model. \textit{Bottom:} TW Cam's \#MDW 2 model. \textit{Left:} $T_{eff}(r)$ profile (Eq.\ (\ref{eq:disc_temperature_profile})). Dotted black lines indicate the disc boundaries. \textit{Middle:} the model disc spectrum (Eq.\ (\ref{eq:disc_spectrum_scaling_profile})), normalised to the peak flux, shown as a black line. A scaled blackbody fitted to the whole spectrum is drawn as a red dot-dashed line. The H band location is shown in gray. \textit{Right:} zoom-in of the spectrum in H band. The values at the six PIONIER wavelength channels are shown as black symbols. The blue dashed line shows a scaled blackbody fitted to these points only. Fitted blackbody temperatures for the full spectrum ($T^{\mathrm{BB}}_{\mathrm{full}}$) and H band only ($T^{\mathrm{BB}}_{\mathrm{H}}$) are shown.}
		\label{fig:example_temp_profiles}
	\end{figure*}
	
	\subsection{Missing low RV absorption}
	Paper I noticed a consistent lack of low RV absorption in their \#MDW 1 model compared to the HD 52961 observations, occurring close to superior conjunction. As a result, the $\mathrm{H_\alpha}$ emission component in the background spectrum shined through the models. The authors attributed this to the void cavity around the disc wind axis in the models, which is present by construction. Such cavities are indeed confirmed observationally in disc winds around YSOs \citep[e.g.][]{lopezvazquez2023}. As seen in Fig.\ \ref{fig:spectra_main_results}, the best fitting model for BD\textdegree46+442 shows this effect noticeably. As does the IRAS 19135+3937 best fit, to a lesser extent. This feature persists in the other fits for these systems as well (Fig.\ \ref{fig:all_fit_dynspec}). For BD\textdegree46+442 and IRAS 19135+3937, the current disc wind models alone are unable to provide enough low RV material to explain the observations, even with higher efficiency MDW solutions. 
	
	There are, however, physical reasons to expect extra outflow components inside the disc wind cavity \citep{Ferreira2006}, which might be able to provide the missing absorption. If magnetospheric funnelling onto the secondary star is indeed active close to the stellar surface, we would expect a thermally driven stellar jet to be launched from the secondary's poles and be collimated by the star’s open field lines \citep{Matt2005}. Even if the disc does extend down to the stellar surface, the collision of the accreting material with the equator might heat up the stellar atmosphere. Hence, a thermal wind from the equator, externally confined by the disc wind’s magnetic field, might be launched as well.
	
	In Appendix \ref{sect_append:extra_mass_loss} we derive Eq.\ (\ref{eq:missing_absorption_cavity_mass_loss}), a simple estimate of the required extra mass-loss in the cavity, $\dot{M}_{cav}$, needed to supply the missing absorption (assuming a homogeneous cylindrical flow at $T=T_{wind}$). If $\dot{M}_{cav} \ll \dot{M}_{in}$, we know that the extra outflow can be fed by material passing the inner disc rim in our models. If $\dot{M}_{cav} > \dot{M}_{in}$, our current models are inconsistent with the addition of an extra outflow component. Applying Eq.\ (\ref{eq:missing_absorption_cavity_mass_loss}) to BD+46\textdegree442's best fit, we derive $\dot{M}_{cav}/\dot{M}_{in} \sim 0.1\%$. Analyses of other model fits that show an analogous lack of low RV absorption (see Fig.\ \ref{fig:all_fit_dynspec}) result in similar values. While Eq.\ (\ref{eq:missing_absorption_cavity_mass_loss}) is derived under idealised conditions in terms of opacity and kinematics, this still shows that the inner rim accretion rate in our current models, in combination with an additional low-efficiency jet launching mechanism in the cavity, can provide the missing absorption. The implementation of such an additional outflow component in our models, however, currently falls outside the scope of this paper.

	\subsection{Accretion disc temperature}\label{sect:accretion_disc_temperature}
	
	\begin{figure}
		\resizebox{\hsize}{!}{\includegraphics{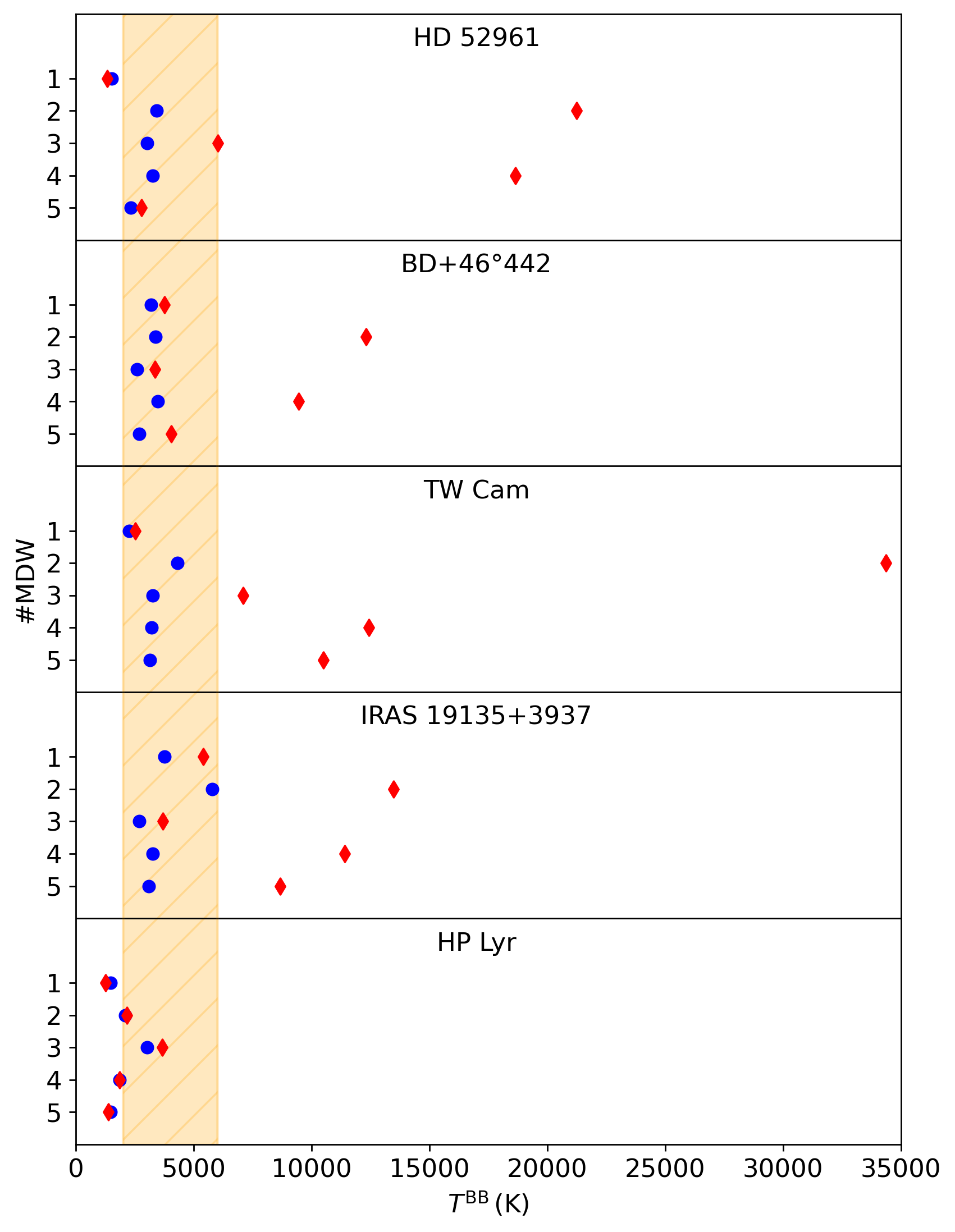}}
		\caption{Blackbody temperatures derived from the accretion disc's spectral slope, considering either the full spectrum ($T^{\mathrm{BB}}_{\mathrm{full}}$, red diamonds), or only the H band ($T^{\mathrm{BB}}_{\mathrm{H}}$, blue dots). The orange shaded area denotes the $\pm 1\sigma$ region of the $T_{\mathrm{disc}}$ estimate for IRAS 08544-4431 by \citet{Hillen2016}, around which the $T^{\mathrm{BB}}_{\mathrm{H}}$ values cluster.}
		\label{fig:all_temp_values}
	\end{figure}
	Following Paper I and assuming that the accretion disc is geometrically thin and radiates like an optically thick blackbody, its effective temperature profile follows \citep[generalised version of the expression in e.g.][]{Frank2002}:
	\begin{equation}\label{eq:disc_temperature_profile}
		\begin{aligned}
			T_{eff}(r) & = \left( \eta_{rad}\cdot \frac{GM_2 \dot{M}_{acc}(r)}{8\pi \sigma r^3} \right)^{\frac{1}{4}} \\
			& \approx 36.4\,\mathrm{kK}\cdot \left( \frac{\eta_{rad}\dot{M}_{acc}(r)}{10^{-4}\,\mathrm{M_\odot /yr}} \right)^{\frac{1}{4}} \left( \frac{M_2}{1 \, \mathrm{M_\odot}} \right)^{\frac{1}{4}} \left( \frac{r}{1 \, \mathrm{R_\odot}} \right)^{-\frac{3}{4}},
		\end{aligned}
	\end{equation}
	with $\sigma$ the Stefan-Boltzmann constant. We note how sensitive this is to the radii to which the disc reaches.
	
	We use two example model fits for our discussion: HP Lyr \#MDW 3 and TW Cam \#MDW 2 (see Table \ref{table:pAGB_MDW_models_all_fit_params}). Their $T_{eff}(r)$ profiles are shown in the left column of Fig.\ \ref{fig:example_temp_profiles}. HP Lyr's \#MDW 3 model has a high inner rim radius at $20\, R_2$. As a result, despite the high accretion rate of $10^{-3}\,\mathrm{M_\odot /yr}$, its profile only reaches up to $\approx 5000 \,\mathrm{K}$. TW Cam's \#MDW 2 model fits a disc that both reaches the stellar surface and has a high accretion rate of $10^{-3}\,\mathrm{M_\odot /yr}$. Correspondingly, it reaches very high temperatures in the inner disc regions of up to $\approx 50\,000\,\mathrm{K}$.
	
	\subsubsection{The IRAS 08544-4431 accretion disc temperature estimate}
	Using H band interferometric data from the Very Large Telescope Interferometer/Precision Integrated-Optics Near-infrared Imaging ExpeRiment instrument (VLTI/PIONIER), \citet{Hillen2016} singled out the spatially unresolved flux contribution from the pole-on accretion disc in IRAS 08544-4431. This is one of the most well-studied post-AGB binaries, in which all components have been spatially resolved. Fitting the spectral slope with a scaled blackbody, they derived an effective blackbody temperature $T_{\mathrm{disc}} = 4000 \pm2000\,\mathrm{K}$. This is currently the only direct observational temperature estimate known for a post-AGB binary accretion disc. Assuming this value as representative for the general population, Paper I concluded that the inner regions of their \#MDW 1 model were too hot, and that the accretion discs should be fairly cold in general. However, by applying a similar scaled blackbody fit to our models, we show below that the H band interferometric estimate is inherently biased, and unsuitable for comparison with disc wind models without broader wavelength coverage of the data. 
	
	\subsubsection{Spectral blackbody temperatures}
	Ignoring possible circum- or interstellar reddening, the relative flux scaling of the accretion disc with wavelength follows:
	\begin{equation}\label{eq:disc_spectrum_scaling_profile}
		L_\lambda \propto \int_{r_{in}}^{r_{out}} r \cdot B_\lambda(T_{eff}(r))\,\mathrm{d}r,
	\end{equation}
	with $B_\lambda$ the Planck function. The resulting spectra for our example models are shown in the middle column of Fig.\ \ref{fig:example_temp_profiles}. Scaled blackbodies were fitted to the full spectra to derive a blackbody temperature, $T^{\mathrm{BB}}_{\mathrm{full}}$. We note that this fits poorly to the multi-blackbody model spectra, which generally have a flattened peak and broader wings. $T^{\mathrm{BB}}_{\mathrm{full}}$ just serves as a rough but useful mean spectral temperature for the full disc spectrum. For the HP Lyr \#MDW 3 model $T^{\mathrm{BB}}_{\mathrm{full}} = 3600\,\mathrm{K}$, while for the TW Cam \#MDW 2 model $T^{\mathrm{BB}}_{\mathrm{full}} = 34\,300\,\mathrm{K}$. Indeed, due to the influence of the colder and geometrically larger outer disc regions, this is lower than the maximum achieved disc temperatures. Nevertheless, the TW Cam \#MDW 2 model seems too hot to comply with the IRAS 08544-4431 estimate of $T_{\mathrm{disc}} = 4000 \pm2000\,\mathrm{K}$.
	
	The analysis above, however, neglects a crucial aspect. The PIONIER data used by \citet{Hillen2016} only covers the H band. We again fitted a scaled blackbody, but now only to the PIONIER wavelength channels, deriving a new spectral temperature, $T^{\mathrm{BB}}_{\mathrm{H}}$. The resulting fits are shown in the right column of Fig.\ \ref{fig:example_temp_profiles}. For HP Lyr \#MDW 3, we derive $T^{\mathrm{BB}}_{\mathrm{H}} = 3000\,\mathrm{K}$, agreeing well with the $T^{\mathrm{BB}}_{\mathrm{full}}$ value. For TW Cam's \#MDW 2 model, however, we fit $T^{\mathrm{BB}}_{\mathrm{H}} = 4300\,\mathrm{K}$. This is much lower than $T^{\mathrm{BB}}_{\mathrm{full}}$, and suddenly agrees very well with the observational estimate for IRAS 08544-4431. As seen from Fig.\ \ref{fig:all_temp_values}, this sudden drop in derived spectral temperature is consistent for all hot disc models, which typically extend close to the stellar surface. In comparison, HP Lyr's models are all cold due to the high fitted inner radii ($\geq 20\, R_2$), and their spectral backbody temperature estimates are consistent between using the full spectrum or the H band only. Looking at the lower plot in the middle column of Fig.\ \ref{fig:example_temp_profiles}, the reason for this drop from $T^{\mathrm{BB}}_{\mathrm{full}}$ to $T^{\mathrm{BB}}_{\mathrm{H}}$ for the hot discs is elucidated. The peak disc flux for TW Cam's \#MDW 2 model is strongly blueshifted due to the hot inner disc regions. The H band itself lies far along the long-wavelength wing of the spectrum, and is dominated by the cold, large outer disc regions. The spectral slope is only weakly sensitive to the hottest parts of the disc in this narrow wavelength range. As a result, fitting the wavelength dependency of a hot disc's relative H band spectrum with a scaled blackbody results in a spectral blackbody temperature estimate biased to a few $1000\,\mathrm{K}$. The method presented in \citet{Hillen2016} is not a good probe of the disc's thermal structure. 
	
	Unlike what was posited by Paper I, hot accretion disc models thus do not contradict current near-infrared interferometric observations. If $T^{\mathrm{BB}}_{\mathrm{full}} 
	\gtrsim 10\,000\,\mathrm{K}$, the model disc spectrum sharply peaks in the UV at $\sim 200\,\mathrm{nm}$, blueward of the Balmer jump. If the total flux contribution of the accretion disc is significant, this should result in observable evidence for UV radiation. Interestingly, there are indeed other observational indicators for a hot component emitting in the UV. Several circumbinary disc-bearing post-AGB binaries show signatures of UV-excited PAH emission and/or a compact photodissociation region (PDR) \citep[e.g.][]{gielen2011, bujarrabal2023}. In the case of one system, the Red Rectangle, this has previously been attributed to a hot jet-launching accretion disc \citep{witt2009}. Our target HD 52961 also shows PAH emission \citep{gielen2011}, and our best fit does indeed predict a hot disc at $T^{\mathrm{BB}}_{\mathrm{full}} 
	\approx 18\,500\,\mathrm{K}$.
	
	We argue that it is very likely that several post-AGB binaries do host a hot accretion disc extending down close to the secondary's surface, which can act as a source of UV photons. The detection of PAH emission and/or PDRs can already prove suggestive. However, only combined multi-wavelength interferometric observations in the near-infrared and other wavelength regimes, for example combining VLTI/PIONIER and Center for High Angular Resolution Astronomy/Stellar Parameters and Images with a Cophased Array (CHARA/SPICA) observations, can directly and reliably probe the temperature structure of the accretion discs on a system-by-system basis. This, in turn, will place very strong constraints on our disc wind models.	
	
	\subsection{Trends in preferred MDW solution}\label{sect:MDW_solution_preference}
	From Table \ref{table:pAGB_MDW_models_best_fit_params} and our discussion of the results in Sect.\ \ref{sect:results}, it is clear that the targets show a clear preference for high efficiency MDW solutions, with the \#MDW 2, 3, 4 and 5 fits ($\xi > 0.10$), depending on the target, mitigating or solving multiple issues faced by \#MDW 1 ($\xi > 0.04$) in matching the observed $\mathrm{H_\alpha}$ profiles or the expected circumbinary disc lifetime. Such high efficiencies are only possible, at least for cold disc winds, at low magnetizations of $\mu \sim 10^{-3}-10^{-2}$. In addition, there is a strong preference for thicker discs at $\epsilon = 0.1$ in order to reduce the effect of jet rotation on the model spectra. Two targets imply the need for even thicker discs at say $\epsilon \gtrsim 0.3$ (Sect.\ \ref{sect:jet_rotation}). At such thicknesses, the driving of a cold, purely magnetocentrifugal wind starts to fail \citep{Casse2000}, and a thermal push from the primary's irradiation will indeed be necessary to create a bona fide disc wind. If the irradiation is also able to consistently heat the accretion disc midplane and puff up the disc, this would physically justify the high aspect ratios in the first place. In addition, such magnetothermal winds can reach $\xi > 0.5$ and might fully solve the persisting issue regarding the circumbinary disc lifetime. One might remark that our most efficient solution, \#MDW 5, already faced
	problems, as it systematically fitted unrealistically low inclinations for three of our targets (see Sect.\ \ref{sect:results}). The link between wind kinematics and density profiles will however be very different in an irradiated MDW solution.
	
	\section{Summary and conclusions}\label{sect:conclusions}
	We have successfully developed a fitting routine to fit the predictions of disc wind models, launched from a circumcompanion accretion disc and based on self-similar MHD disc wind solutions, to time-resolved observations of the $\mathrm{H_\alpha}$ jet features in post-AGB binaries. The five used MHD solutions cover the range of weakly magnetised and highly efficient cold disc winds, while the five target systems form a fairly representative sample of the known jet-launching post-AGB binaries. The models manage to reproduce many of the observed features fairly well, showing that MHD disc winds can indeed for a large part explain the $\mathrm{H_\alpha}$ features seen in post-AGB binaries. Several issues and points of interest remain, however. 
	
	Many of the models overestimate the wind's rotation. This manifests as a redshifted absorption feature which is not observed, implying the need for more slowly rotating solutions, such as those with thicker accretion discs. This issue is especially pertinent for two targets: HD 52961 and BD+46\textdegree442.
	
	The efficiency range covered by known cold magnetocentrifugal disc winds, topping out at $\xi \approx 0.30$ for discs of aspect ratio $\epsilon = 0.1$, is insufficient. Even at an efficiency of $\xi = 0.305$, predicted accretion rates may be as high as $10^{-4}\,\mathrm{M_\odot/yr}$, and correspondingly, predicted lifetimes for the dusty circumbinary disc as short as $100\, \mathrm{yr}$. Given the omnipresence of these circumbinary discs and a typical post-AGB lifetime on the order of $10\,000\,\mathrm{yr}$, these accretion rates need to be pushed down further by about two orders of magnitude. This would require disc wind solutions with $\xi \gtrsim 0.5$, which are unachievable using pure magnetocentrifugal driving, but can be reached when thermal effects help lift up disc material \citep{Casse2000(2)}.
	
	High efficiency MHD solutions predict inner rims closer to the secondary's stellar surface, and four out of five targets are indeed best fitted with an inner rim at a radius of at most five times the secondary's stellar radius. At the current high fitted accretion rates and assuming magnetospheric funnelling is responsible for truncating the accretion discs, the required magnetic field strengths of the secondaries should be greater than about several kG to $15\, \mathrm{kG}$. This is much higher than what can be generated by an isolated mature main sequence star. At reduced accretion rates and under possible rejuvenation of the stellar dynamo due to accretion, magnetospheric truncation close to the stellar surface remains plausible. HP Lyr stands out as an exception, consistently requiring a fitted inner rim radius equal to or greater than twenty times the stellar radius. This is impossible to achieve via magnetospheric funneling, even for lower accretion rates, requiring unreasonably high field strengths. This large inner radius could be an artefact due to the strong assumption of homogeneous wind temperature. Future modelling attempts should include more realistic and physically inspired temperature structures.
	
	All BD+46\textdegree442 models show a consistent lack of low RV absorption close to superior conjunction. This is shown to be, at least in principle, resolvable via the inclusion of an additional low efficiency outflow component inside the empty cavity currently present around the disc wind axis, such as a thermal stellar jet launched from the secondary's poles \citep{Matt2005}.
	
	A previous method of estimating the accretion disc's temperature from H band interferometry, performed on IRAS 08544-4431 by \citet{Hillen2016}, is shown to be biased to several $1000\,\mathrm{K}$. Hot accretion disc models that reach down to the stellar surface, and can achieve temperatures over $10\,000\,\mathrm{K}$ in the inner regions, are thus not inconsistent with current H band interferometric constraints. Instead, we argue that several post-AGB binaries that show evidence of UV-excited PAH emission and/or PDRs, such as our target HD 52961, indeed host such a hot accretion disc, providing a source of UV photons. Interferometric data covering a wider wavelength range, for example combining VLTI/PIONIER and CHARA/SPICA, will be required to reliably probe the thermal structure of the accretion discs. This would provide strong constraints on our disc wind models.
	
	There is a clear preference for all targets towards the efficient, thicker MHD solutions with $\xi > 0.10$ and $\epsilon = 0.10$, which are only achieved at a low magnetization level of $\mu \sim 10^{-3}-10^{-2}$. For future work, we advise the use of warm magnetothermal solutions, which include heat deposition via irradiation from the bright post-AGB primary at the disc surface. This should provide an extra push to the wind material, readily allowing the solutions to reach $\xi \gtrsim 0.5$ and for even thicker disc solutions at aspect ratios as high or even higher as $\epsilon = 0.3$ to be explored. This would provide a physically consistent solution to the problems currently faced in regards to overestimated wind rotation and the high required accretion rates.
	\begin{acknowledgements}
		TDP acknowledges support of the Research Foundation - Flanders (FWO) under grant 11P6I24N (Aspirant Fellowship). HVW acknowledges support of the FWO under grant G097619N. OV acknowledges support from the KU Leuven Research Council (grant C16/17/007: MAESTRO). DK acknowledges the support of the Australian Research Council (ARC) Discovery Early Career Research Award (DECRA) grant (DE190100813). DK is also supported in part by the ARC Centre of Excellence for All Sky Astrophysics in 3 Dimensions (ASTRO 3D), through project number CE170100013. JJ acknowledges support by the NSF AST-2009884 and NASA 80NSSC21K1746 grants. Based on observations obtained with the HERMES spectrograph, which is supported by the FWO, Belgium, the Research Council of KU Leuven, Belgium, the Fonds National de la Recherche Scientifique (F.R.S.-FNRS), Belgium, the Royal Observatory of Belgium, the Observatoire de Genève, Switzerland and the Thüringer Landessternwarte Tautenburg, Germany. The computational resources and services used in this work were provided by the VSC (Flemish Supercomputer Centre), funded by the FWO and the Flemish Government – department EWI. Finally, we thank the anonymous referee for helping improve the clarity of the paper.
	\end{acknowledgements}
	
	\begingroup
	\bibliographystyle{bibliography/aa.bst}
	\bibliography{bibliography/refs.bib}
	
	\begin{appendix}
		\section{MDW example structures}\label{sect:jed_example_structures}
		Fig.\ \ref{fig:all_jed_example_structures} shows five example disc wind structures, calculated entirely analogously except using the different MDW solutions as input. It showcases the general behaviour of the MDW solutions. 
		\begin{figure*}[b]
			\centering 
			\begin{subfigure}{0.283\textwidth}
				\includegraphics[width=\linewidth]{figs/jed1_density.png}
			\end{subfigure}\hfil 
			\begin{subfigure}{0.283\textwidth}
				\includegraphics[width=\linewidth]{figs/jed1_vp.png}
			\end{subfigure}\hfil
			\begin{subfigure}{0.283\textwidth}
				\includegraphics[width=\linewidth]{figs/jed1_vphi.png}
			\end{subfigure}
			
			\medskip
			\begin{subfigure}{0.283\textwidth}
				\includegraphics[width=\linewidth]{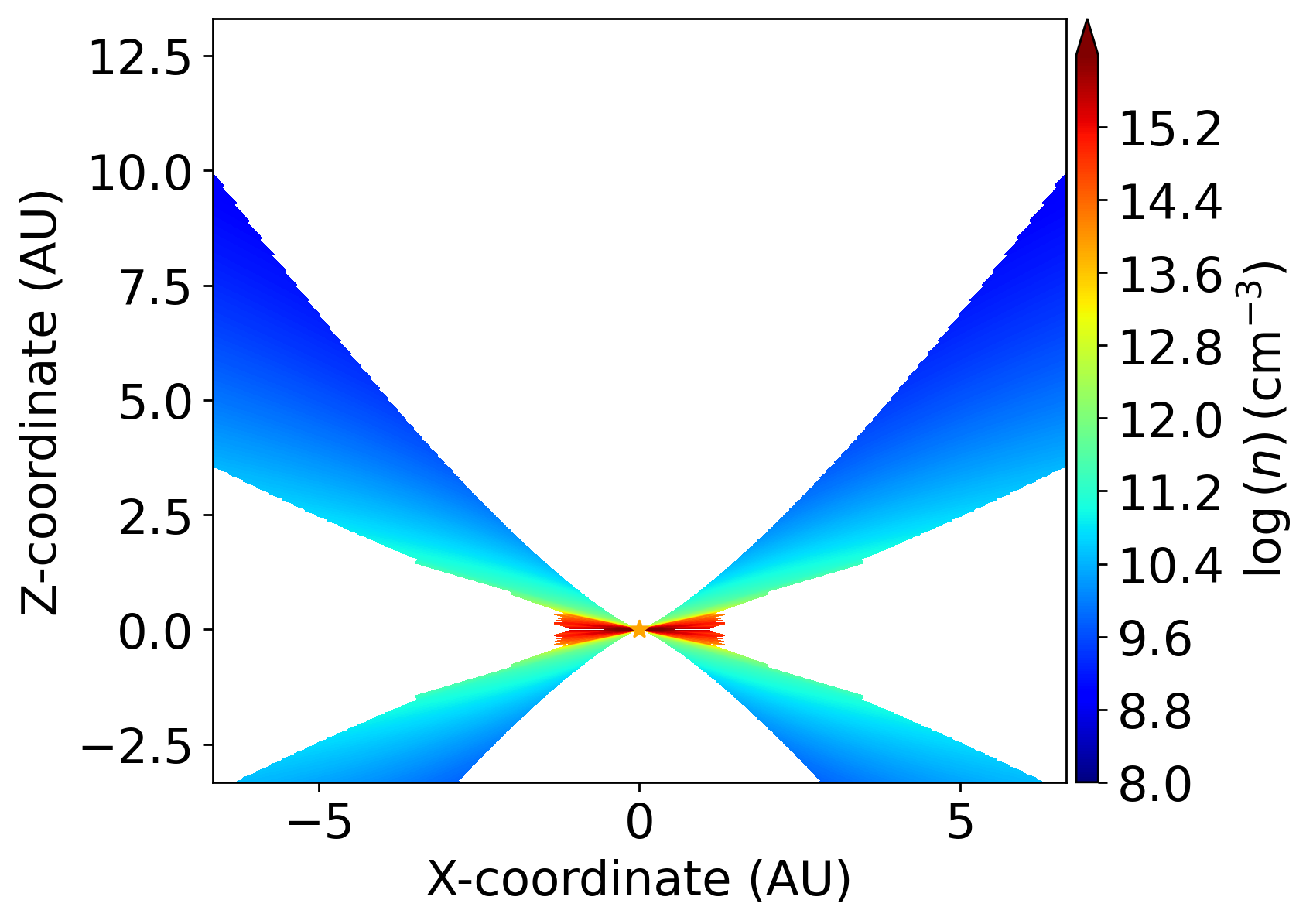}
			\end{subfigure}\hfil 
			\begin{subfigure}{0.283\textwidth}
				\includegraphics[width=\linewidth]{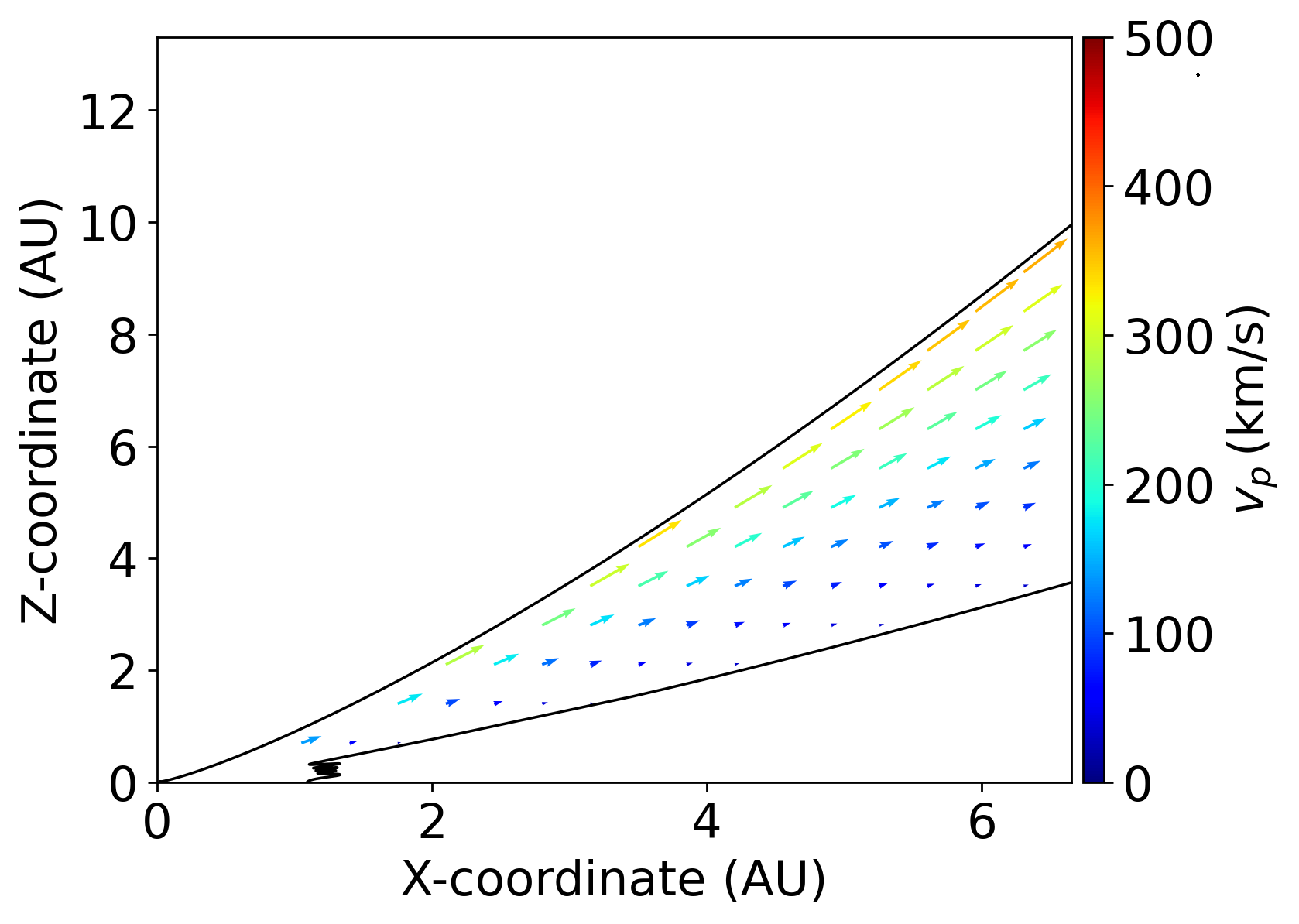}
			\end{subfigure}\hfil
			\begin{subfigure}{0.283\textwidth}
				\includegraphics[width=\linewidth]{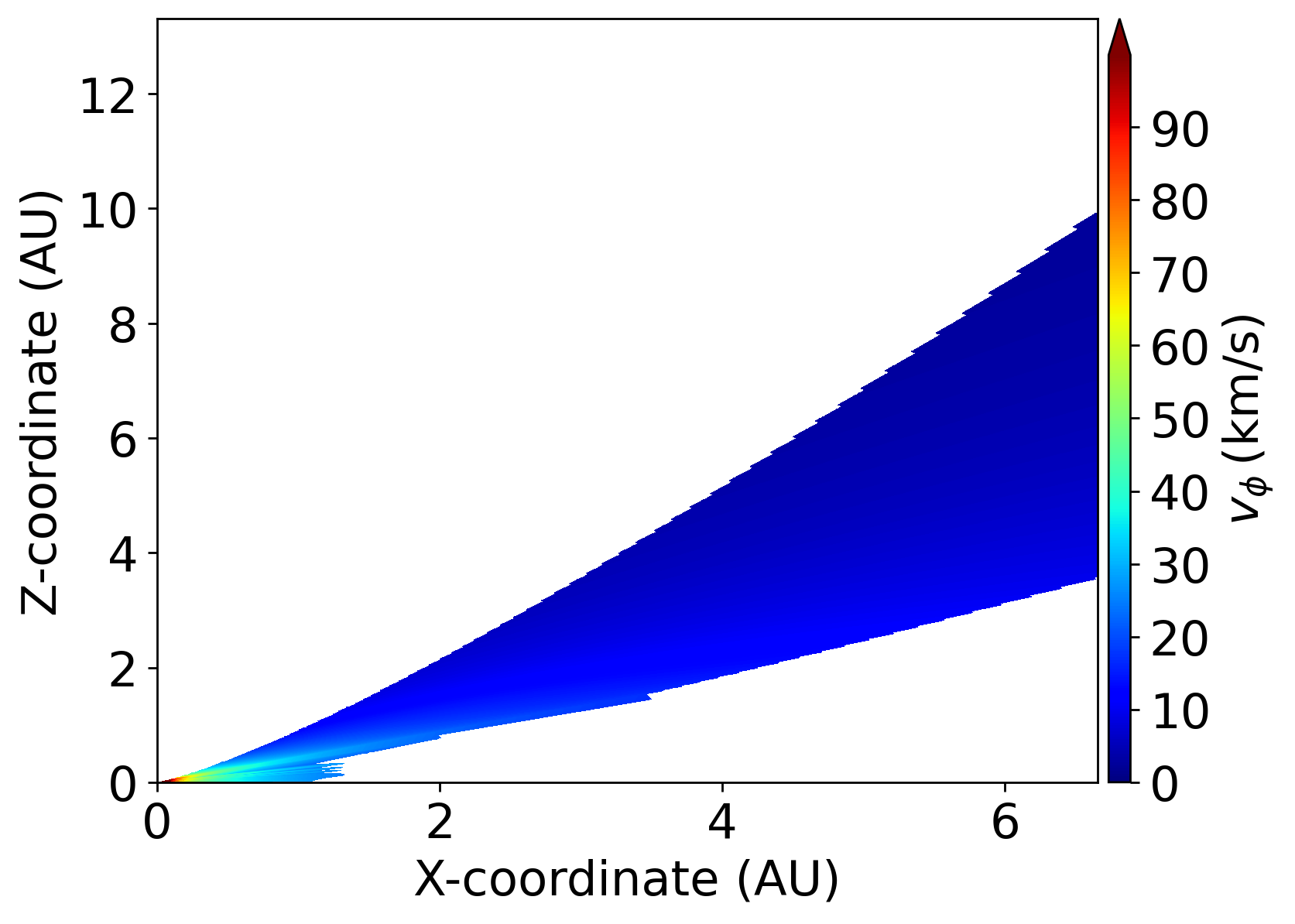}
			\end{subfigure}
			
			\medskip
			\begin{subfigure}{0.283\textwidth}
				\includegraphics[width=\linewidth]{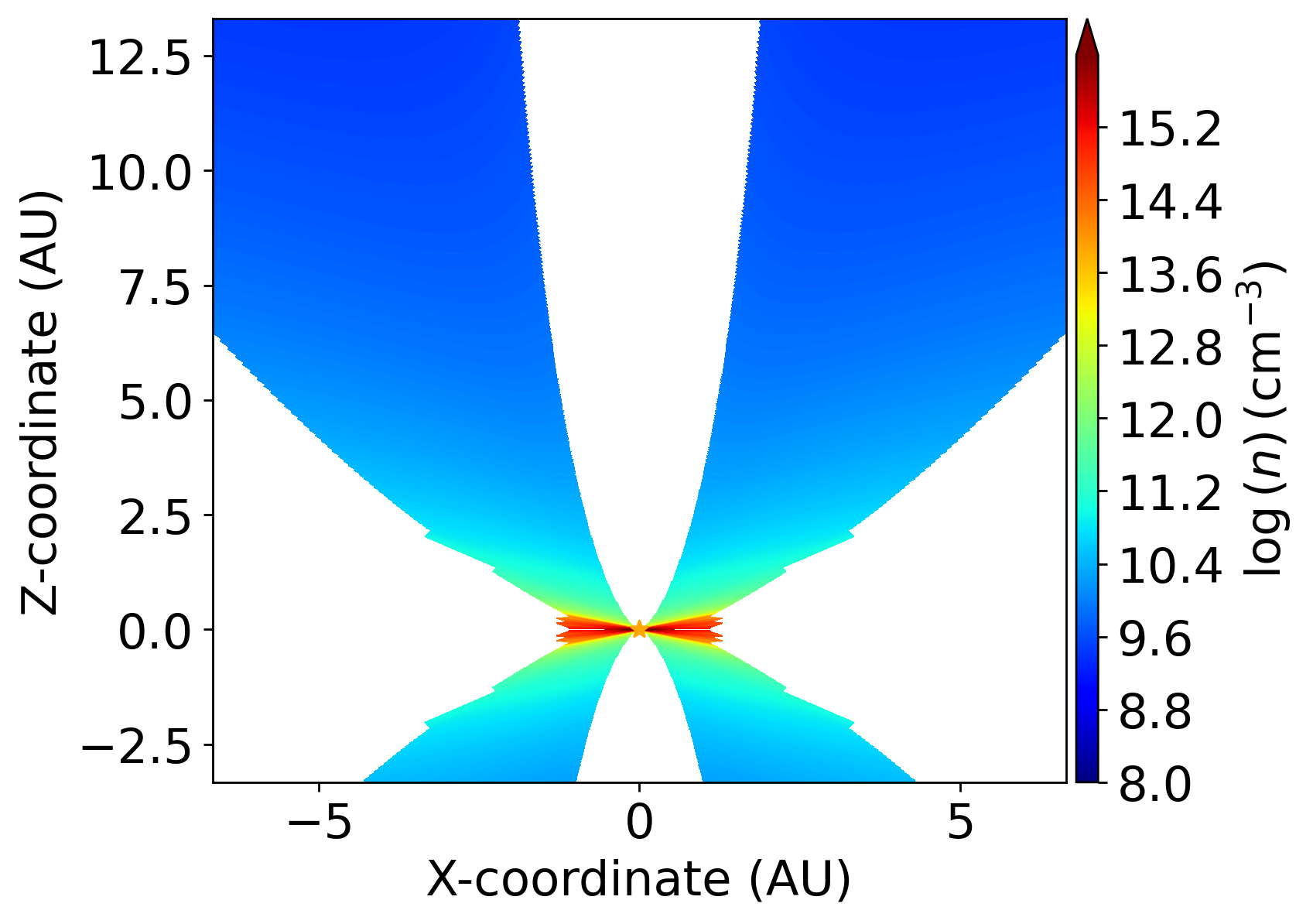}
			\end{subfigure}\hfil 
			\begin{subfigure}{0.283\textwidth}
				\includegraphics[width=\linewidth]{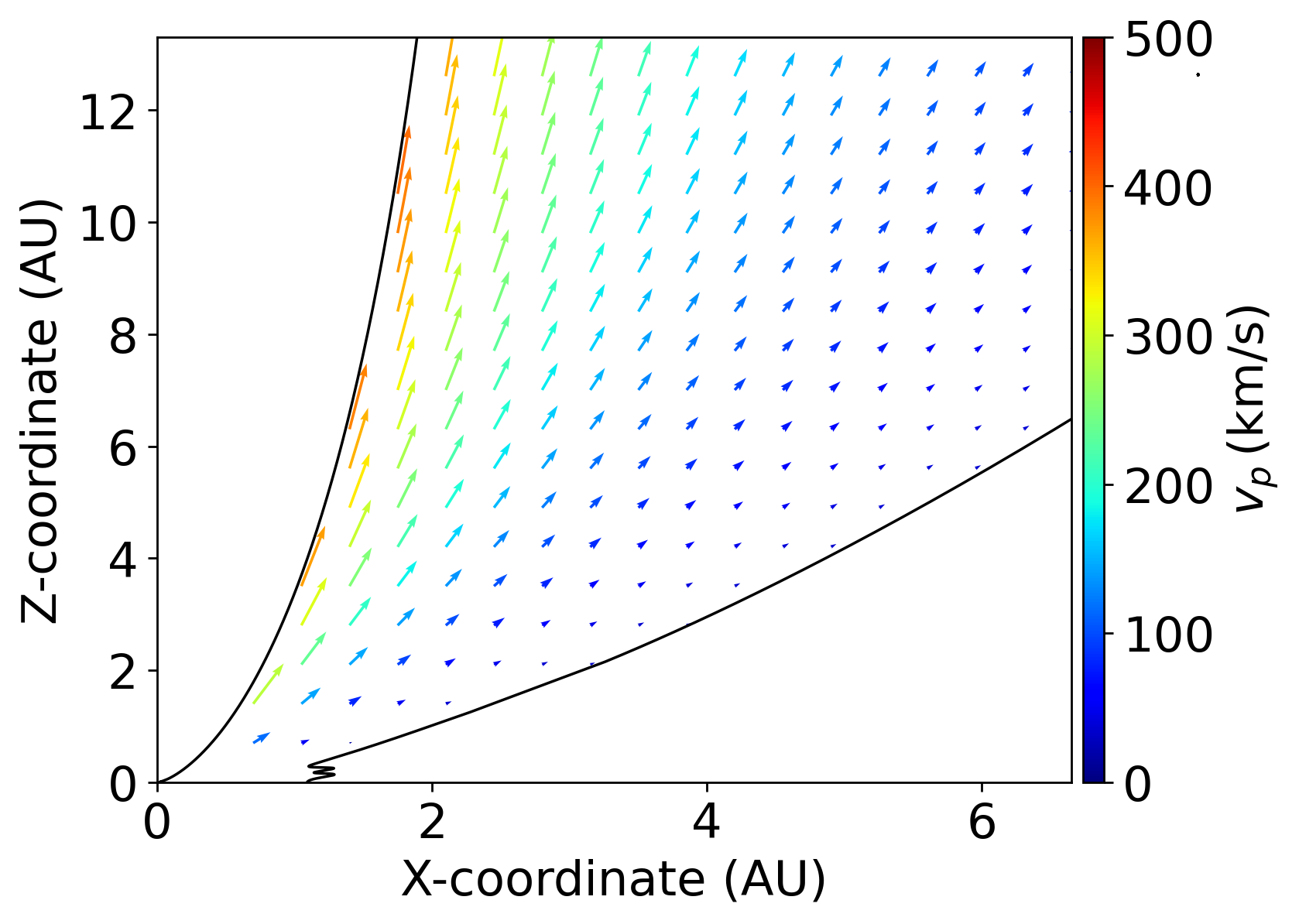}
			\end{subfigure}\hfil
			\begin{subfigure}{0.283\textwidth}
				\includegraphics[width=\linewidth]{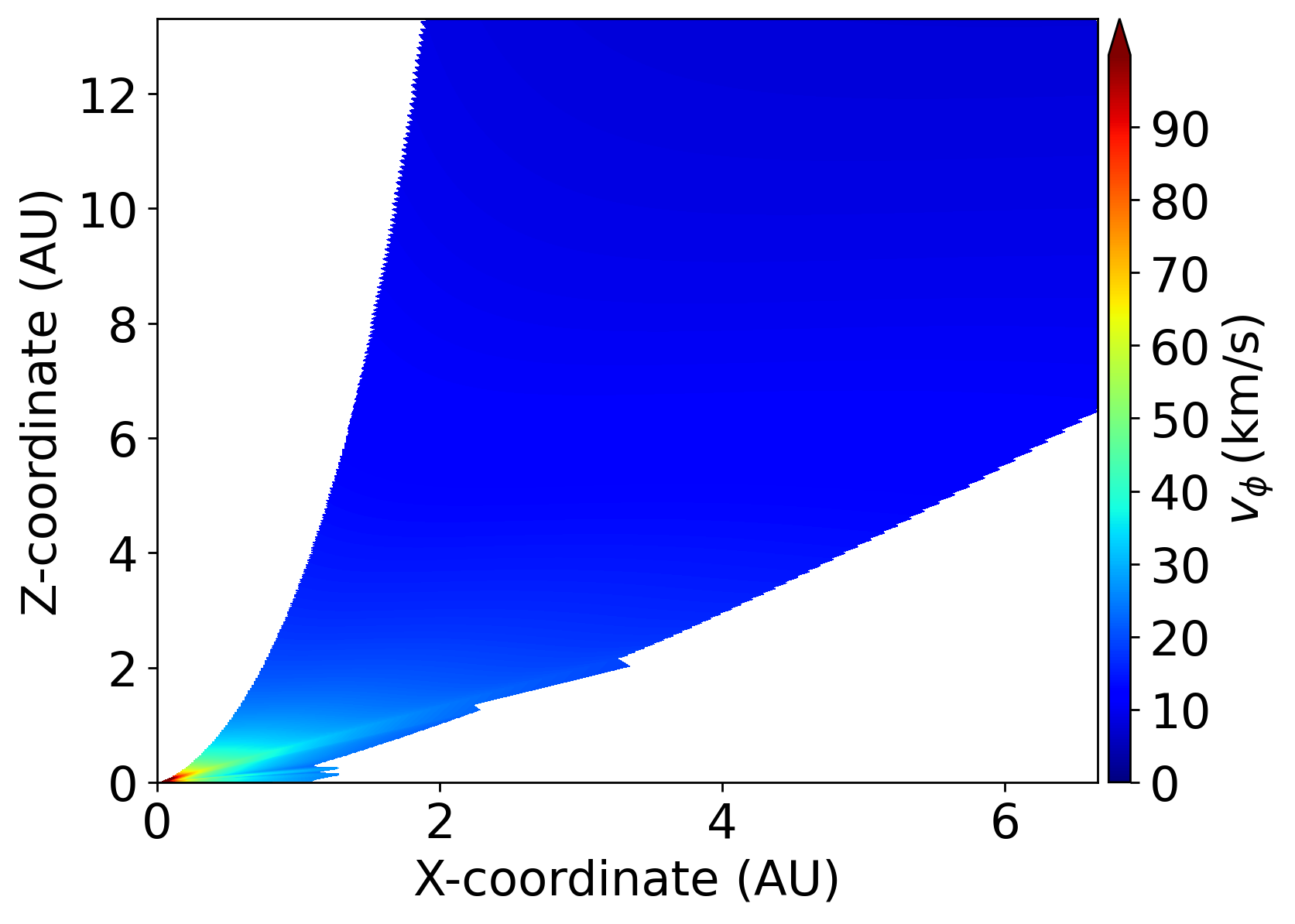}
			\end{subfigure}
			
			\medskip
			\begin{subfigure}{0.283\textwidth}
				\includegraphics[width=\linewidth]{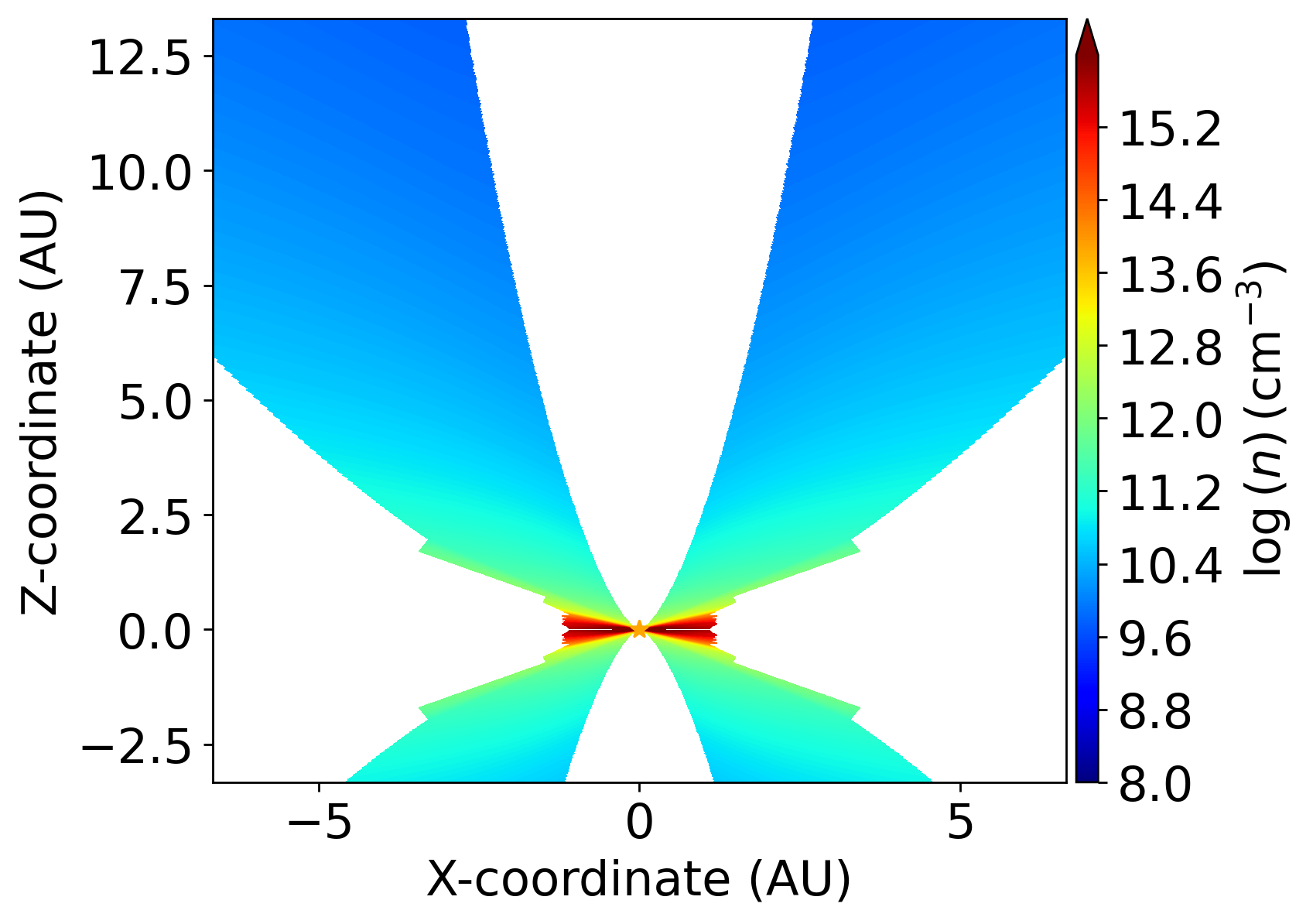}
			\end{subfigure}\hfil 
			\begin{subfigure}{0.283\textwidth}
				\includegraphics[width=\linewidth]{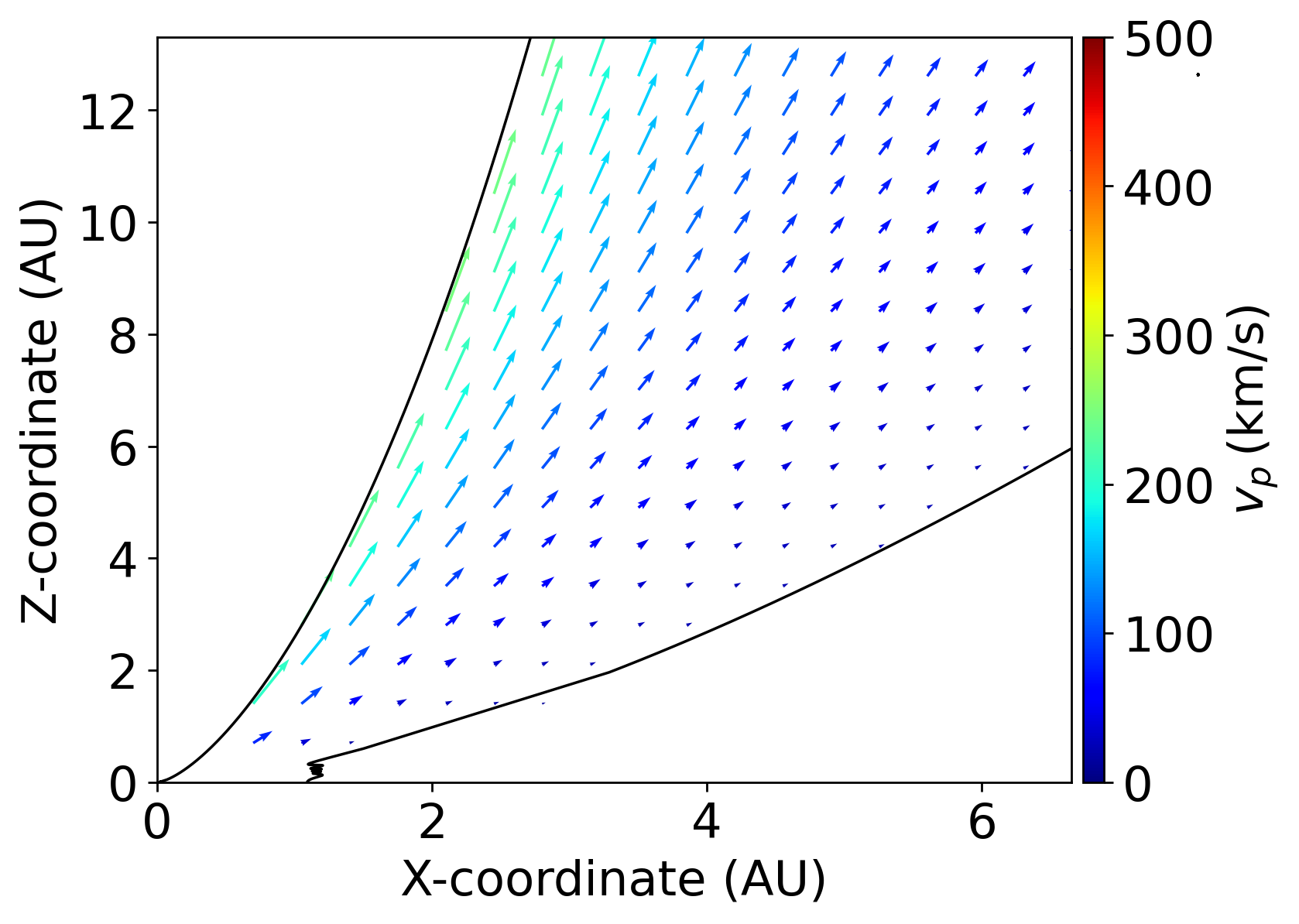}
			\end{subfigure}\hfil
			\begin{subfigure}{0.283\textwidth}
				\includegraphics[width=\linewidth]{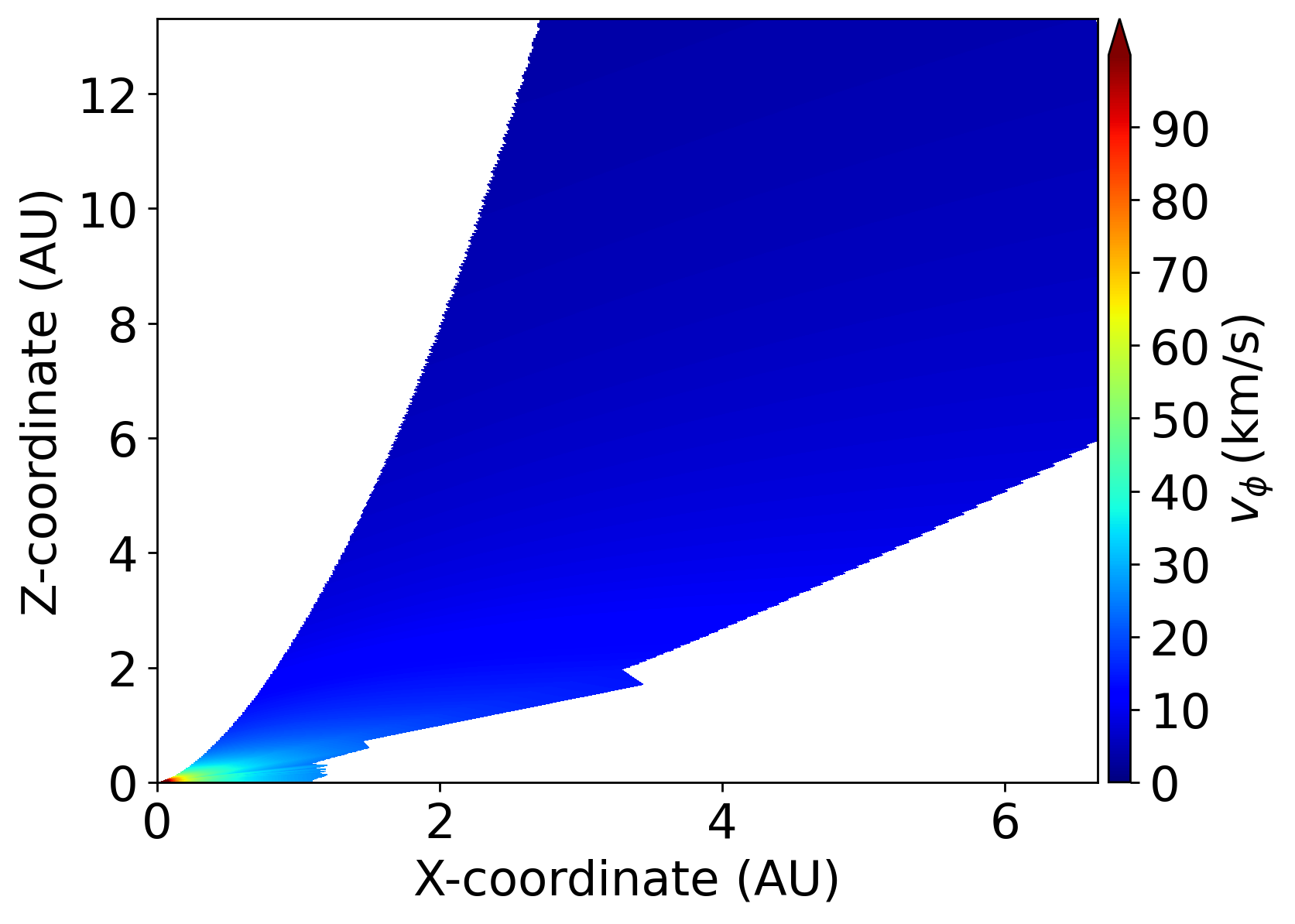}
			\end{subfigure}
			
			\medskip
			\begin{subfigure}{0.283\textwidth}
				\includegraphics[width=\linewidth]{figs/jed5_density.png}
			\end{subfigure}\hfil 
			\begin{subfigure}{0.283\textwidth}
				\includegraphics[width=\linewidth]{figs/jed5_vp.png}
			\end{subfigure}\hfil
			\begin{subfigure}{0.283\textwidth}
				\includegraphics[width=\linewidth]{figs/jed5_vphi.png}
			\end{subfigure}
			
			\caption{Model disc wind structures using different MDWs. \textit{Left:} hydrogen number density $n$. \textit{Middle:} poloidal velocity $v_p$. \textit{Right:} toroidal/rotational velocity $v_\phi$. From top to bottom, we show \#MDW 1, 2, 3, 4 and 5. Calculated assuming the HD 52961 system and $i=70^\circ$, $r_{in} = 5\, R_2$, $r_{out} = 0.99\, R_{RL,2}$ and $\dot{M}_{in} = 10^{-3}\,\mathrm{M_{\odot}/yr}$. We note the slight discontinuity at a fixed polar angle in the $n$ and $v_\phi$ profiles. This is an artefact from the leapfrog integration performed over the Alfv\'en surface in the MDW solutions \citep{Jacquemin2019}, in combination with our projection onto Cartesian coordinates.}
			\label{fig:all_jed_example_structures}
		\end{figure*}
		\clearpage
		
		\section{Spectral inputs from HERMES time-series}\label{sect:appendix_spectral_inputs}
		{{In this appendix, we describe in detail how the different required inputs for the modelling routine are derived from the observed HERMES time-series. We cover the pre-selection of observed spectra in Sect. \ref{sect:appendix_spec_selection}, the orbital parameters in Sect. \ref{sect:appendix_orb_params}, and the background spectrum and its associated uncertainty in Sects.\ \ref{sect:appendix_background_spec} \& \ref{sect:extra_background_error}, respectively.}}
		
		\subsection{Observed spectra selection}\label{sect:appendix_spec_selection}
		We require a sub-selection of spectra from the
		 full HERMES time-series for the object of interest. In order to limit both cycle-to-cycle variability and computational cost, we select several spectra lying within one to two orbital periods from each other and with a reasonable phase coverage across the orbit. If spectra lie too close to each other in orbital phase (i.e.\ within a difference of 0.01), only the spectrum with the highest signal-to-noise around $\mathrm{H_\alpha}$ is retained. Remaining gaps in the orbital phase, $\phi_{orb}$, are then filled by manually selecting additional spectra. The number of selected spectra for our targets, as well as their average signal-to-noise around $\mathrm{H_\alpha}$, are shown in Table \ref{table:target_star_params}. The resulting $\mathrm{H_\alpha}$ dynamic spectra for our targets are shown in the left column of Fig.\ \ref{fig:spectra_main_results}.
		
		\subsection{Orbital parameters}\label{sect:appendix_orb_params}
		{{The fitting routine makes extensive use of the system's SB1 orbital parameters in order to set te geometry of the binary}}. These parameters are derived from fitting the primary's RV curve with a Keplerian orbit while iteratively filtering out the signal due to stellar pulsations \citep{Oomen2018}. These parameters are summarised for our target systems in Table \ref{table:orbital_params}. This does not provide a full orbital solution for the binary. In addition, an assumption needs to be made for both the inclination and the primary's mass. The former is taken to be a fitting parameter (see Sect.\ \ref{sect:fitting_routine}), while the latter is set to $M_1 = 0.6\,\mathrm{M_\odot}$ \citep[typical for single white dwarfs in the Galactic field; e.g.][]{Tremblay2016}. This allows for the calculation of the secondary's mass, $M_2$, from the spectroscopic mass function, and solves the full 3D binary orbit.
		
		\subsection{Background spectrum}\label{sect:appendix_background_spec}
		A dynamic background spectrum for the $\mathrm{H_\alpha}$ line, valid for the entire orbit, needs to be constructed as input for the RT module. This background needs to represent the observed, unobscured spectra out of superior conjunction, which are unaffected by jet absorption. The construction procedure takes several observed unobscured spectra, and subtracts the RV shifted photospheric absorption component of the primary star. The photospheric component is represented using a suitable stellar spectrum from \citet{Coelho2014}. This leaves only the emission component, which is averaged over the spectra to create a template for the emission. This template is then added back on top of the photospheric absorption. If the observed emission component moves with the primary's RV curve in the observed time-series, the template is RV shifted according to $\phi_{orb}$ before addition. If the emission is centred on the systemic velocity, only the photospheric component is RV shifted before addition. This results in a suitable dynamic background spectrum, which is assumed to be fully localised on the primary's surface within the RT module. The background spectra constructed for our targets are shown in the middle column of Fig.\ \ref{fig:spectra_main_results}.

		\subsection{Additional background uncertainty}\label{sect:extra_background_error}
		The resulting dynamic background spectra are, by construction, smooth. However, looking at the observations in the left column of Fig.\ \ref{fig:spectra_main_results}, we see significant variability in spectra outside of superior conjunction, unrelated to variable jet obscuration. This variability is physical, as most post-AGB stars are photometrically variable \citep[e.g.][]{VanWinckel2003}. As such, there is an inherent uncertainty associated to the constructed background.
		
		In order to account for this inherent background variability, we include an additional error term, $\sigma_{\mathrm{back}}(\lambda)$. This term should be higher in the inherently more variable regions of the spectra. Correspondingly, the model predictions will be punished less severely there. We take several spectra outside of superior conjunction, and calculate the difference with the constructed background spectrum over the considered wavelengths. This difference profile is then averaged over the spectra to give $\sigma_{\mathrm{back}}(\lambda)$, which is added in quadrature to the S/N error to give the total spectral error $\sigma_{\mathrm{tot}}(\lambda) = ( \sigma_{\mathrm{S/N}}^2 + \sigma_{\mathrm{back}}^2(\lambda) )^{1/2}$. $\sigma_{\mathrm{tot}}$ is dominated by $\sigma_{\mathrm{back}}$ for all our targets. If the $\mathrm{H_\alpha}$ emission component follows the primary's RV, the $\sigma_{\mathrm{tot}}$ profile is shifted correspondingly. With this additional error, we typically get a reduced chi-squared $\chi^2_\nu \sim 1$ when comparing the background spectrum to the observed unobscured spectra. This shows that the combination of the background spectrum and $\sigma_{\mathrm{tot}}$ is a good statistical representation for the unobscured spectra. The total spectral error profiles for our targets are shown in Fig.\ \ref{fig:extra_background_error_profiles}.
		
		\begin{table*}
			\caption{Spectroscopic orbital parameters.}
			\label{table:orbital_params}
			\setlength\tabcolsep{0mm}
			\centering
			\begin{tabular}{L{0.16666666666\textwidth} C{0.16666666666\textwidth} C{0.16666666666\textwidth} C{0.16666666666\textwidth} C{0.16666666666\textwidth} C{0.16666666666\textwidth}}
				\hline\hline & \\[-1.7ex]
				Object & HD 52961 & BD+46\textdegree442 & TW Cam & IRAS 19135+3937 & HP Lyr\\
				Parameter & & & & & \\[1.0ex]
				\hline & \\[-1.6ex]
				$P\,\mathrm{(d)}$ & $1288.6\pm0.3$ & $140.82\pm0.02$ & $662.2\pm5.3$ & $126.97\pm0.08$ & $1818\pm80$\\[0.5ex]
				$e$ & $0.23\pm0.01$ & $0.085\pm0.005$ & $0.25\pm0.04$ & $0.13\pm0.03$ & $0.20\pm0.04$\\[0.5ex]
				$T_0-24\cdot10^5\,\mathrm{(d)}$ & $7308\pm24$ & $55233.9\pm1.3$ & $55111\pm18$ & $54997.7\pm1.0$ & $56175\pm61$\\[0.5ex]
				$\omega\,\mathrm{(^\circ)}$ & $297.4\pm5.9$ & $95.8\pm3.3$ & $144.4\pm10$ & $66.0\pm4.4$ & $14.2\pm13$\\[0.5ex]
				$K_1\,\mathrm{(km\,s^{-1})}$ & $13.1\pm0.3$ & $23.8\pm0.1$ & $14.1\pm0.6$ & $18.0\pm0.6$ & $7.8\pm0.2$\\[0.5ex]
				$\gamma\,\mathrm{(km\,s^{-1})}$ & $6.2 \pm 0.2$ & $-98.13 \pm 0.08$ & $-49.8 \pm 0.5$ & $2.1 \pm 0.4$ & $-115.6 \pm 0.2$\\[0.5ex]
				$a_1\sin{i}\,\mathrm{(AU)}$ & $1.507\pm0.034 $ & $0.3074\pm0.0014$ & $0.83\pm0.04$ & $0.209\pm0.008$ & $1.27\pm0.06$\\[0.5ex]
				$f(m)\,\mathrm{(M_{\odot})}$ & $0.274\pm0.019 $ & $0.195\pm0.003$ & $0.174\pm0.022$ & $0.075\pm0.008$ & $0.083\pm0.007$\\[0.8ex]
				\hline
			\end{tabular}
			\tablefoot{Values taken from \citet{Oomen2018}. Parameters include the orbital period $P$, eccentricity $e$, time of periastron passage $T_0$, argument of the periastron $\omega$, primary semi-amplitude $K_1$, systemic velocity $\gamma$, projected primary semi-major axis $a_1\sin{i}$ and the mass function $f(m)$.}
		\end{table*}
		
		\begin{figure*}
			\vspace{1.2cm}
			\centering 
			\begin{subfigure}{0.42\textwidth}
				\includegraphics[width=\linewidth]{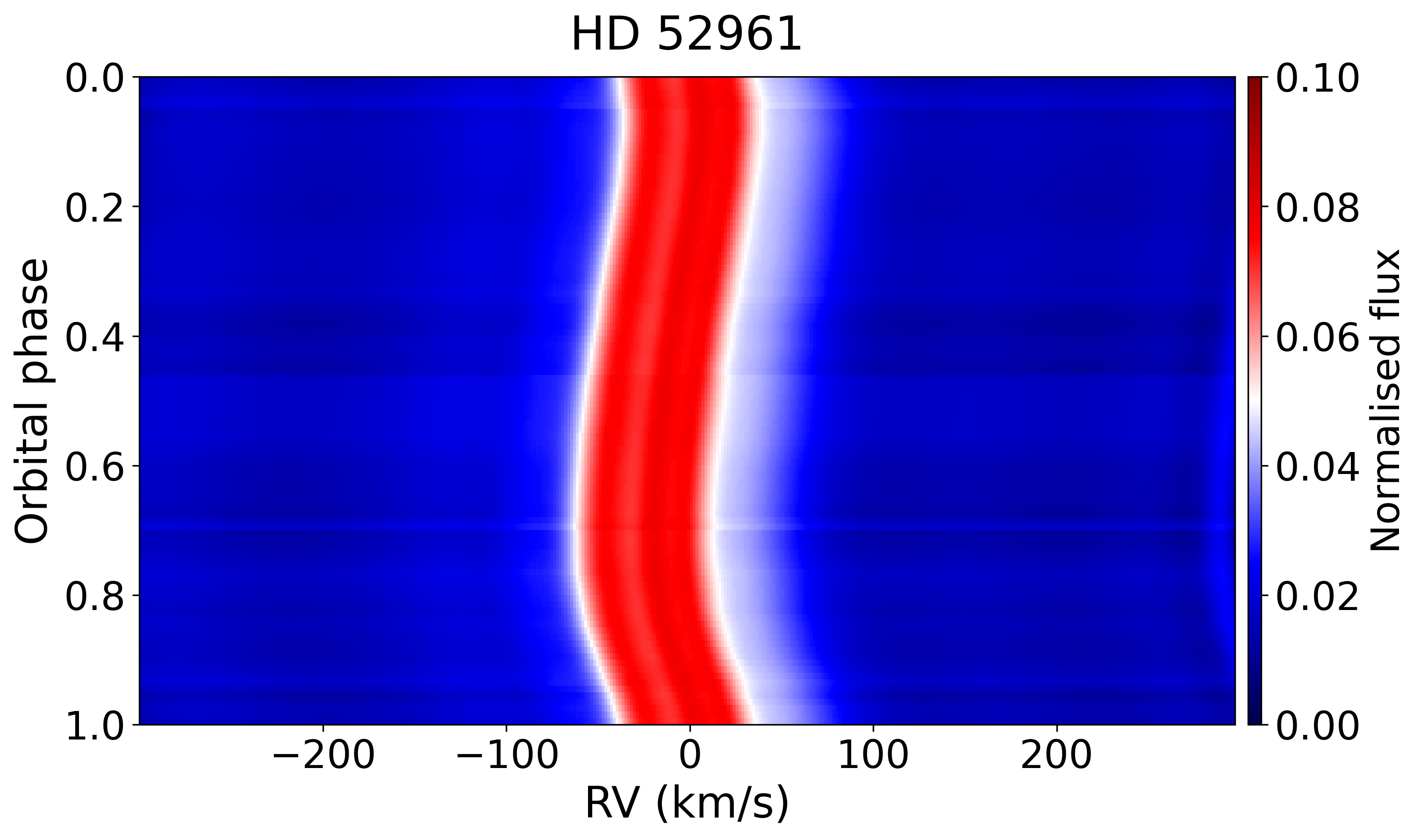}
			\end{subfigure}\hfil 
			\begin{subfigure}{0.42\textwidth}
				\includegraphics[width=\linewidth]{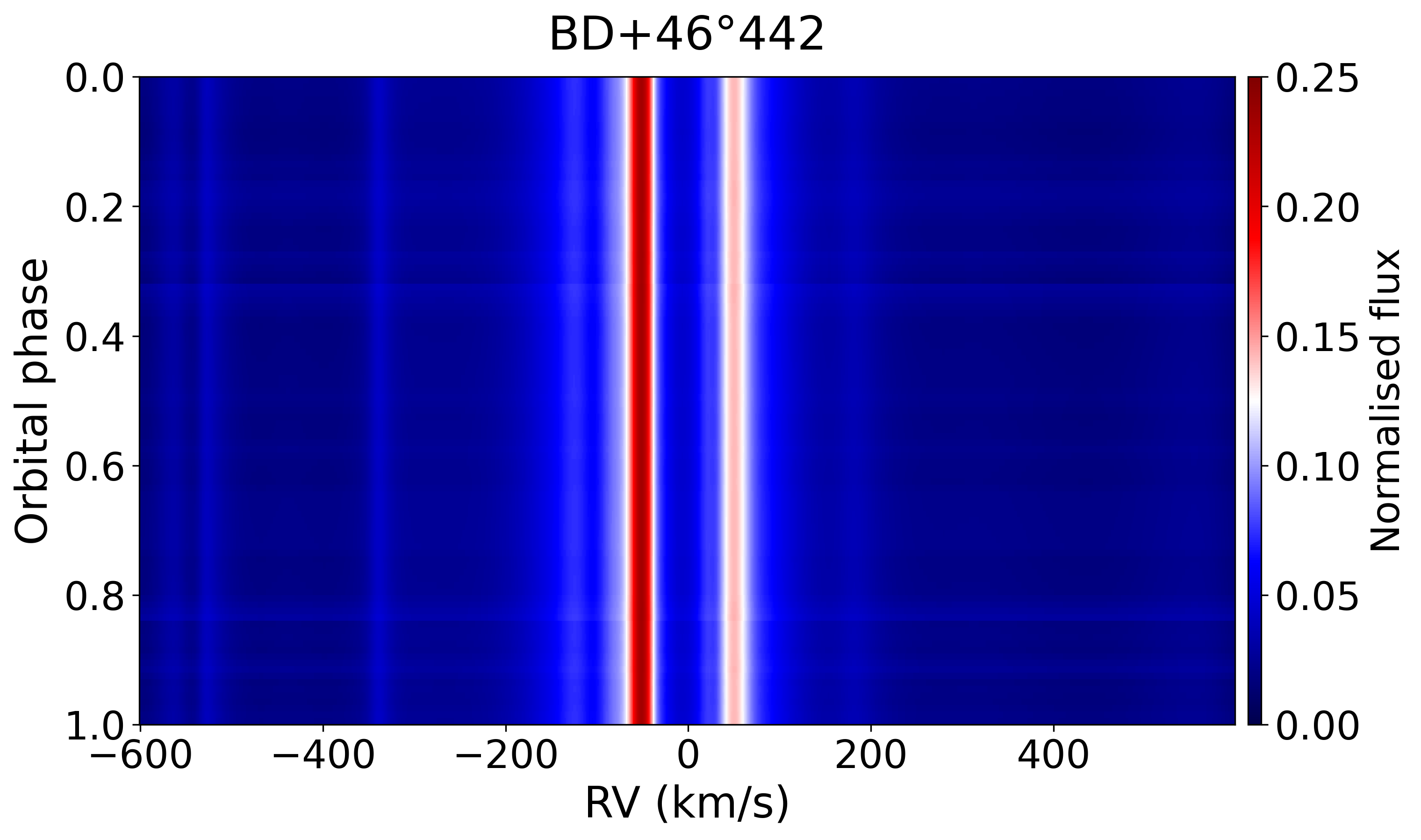}
			\end{subfigure}
			
			\medskip
			\begin{subfigure}{0.42\textwidth}
				\includegraphics[width=\linewidth]{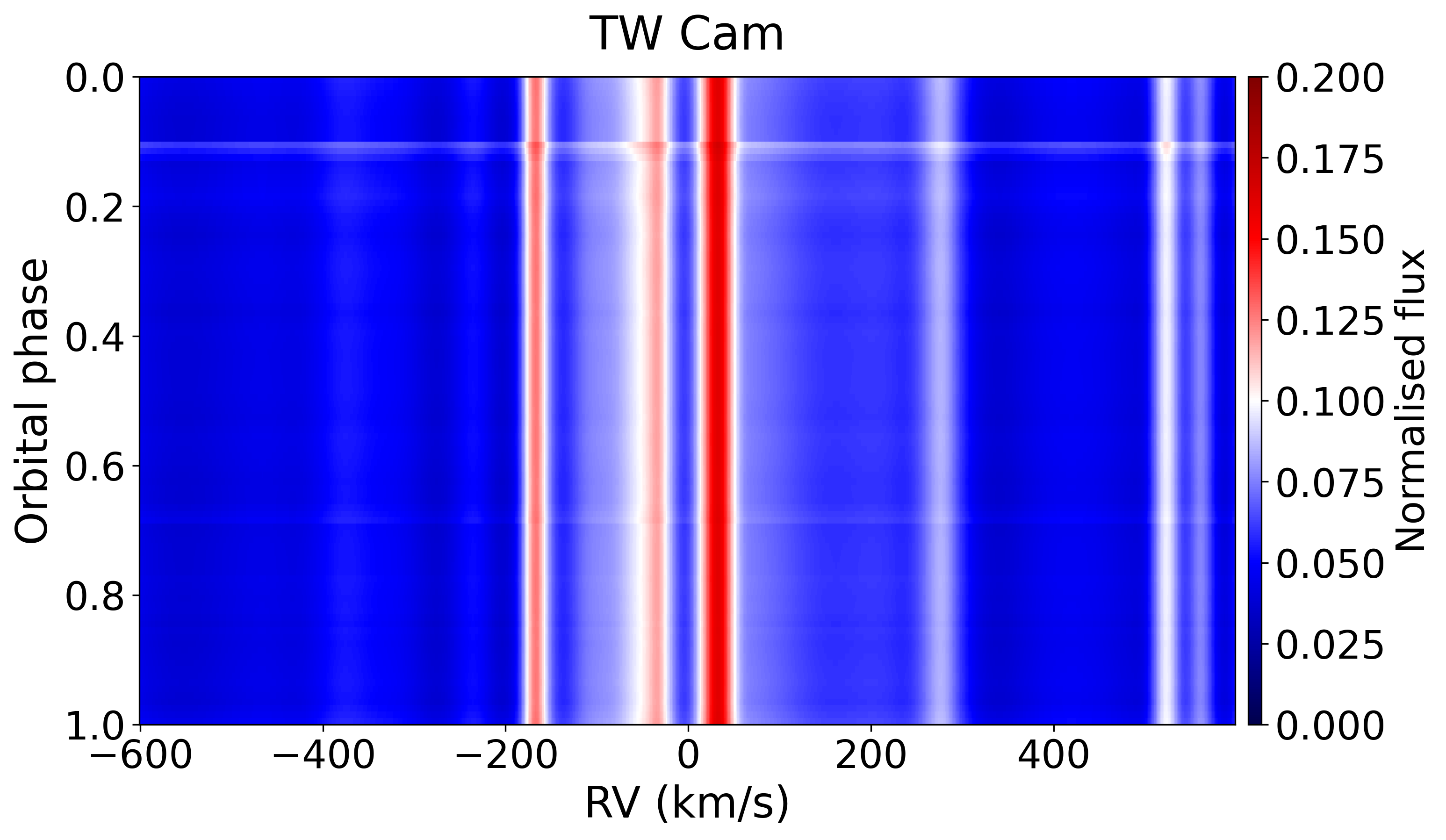}
			\end{subfigure}\hfil
			\begin{subfigure}{0.42\textwidth}
				\includegraphics[width=\linewidth]{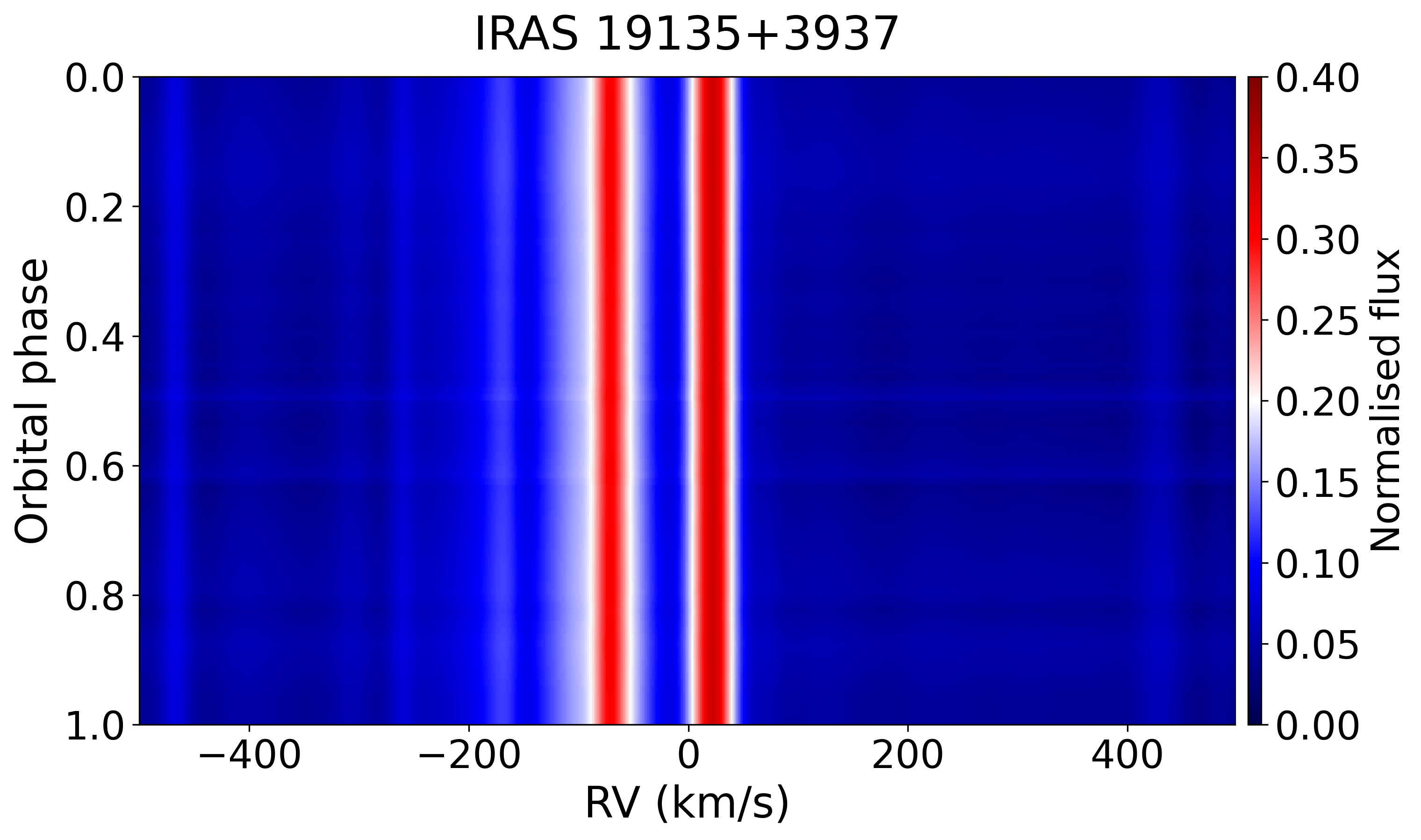}
			\end{subfigure}
			
			\medskip
			\begin{subfigure}{0.42\textwidth}
				\includegraphics[width=\linewidth]{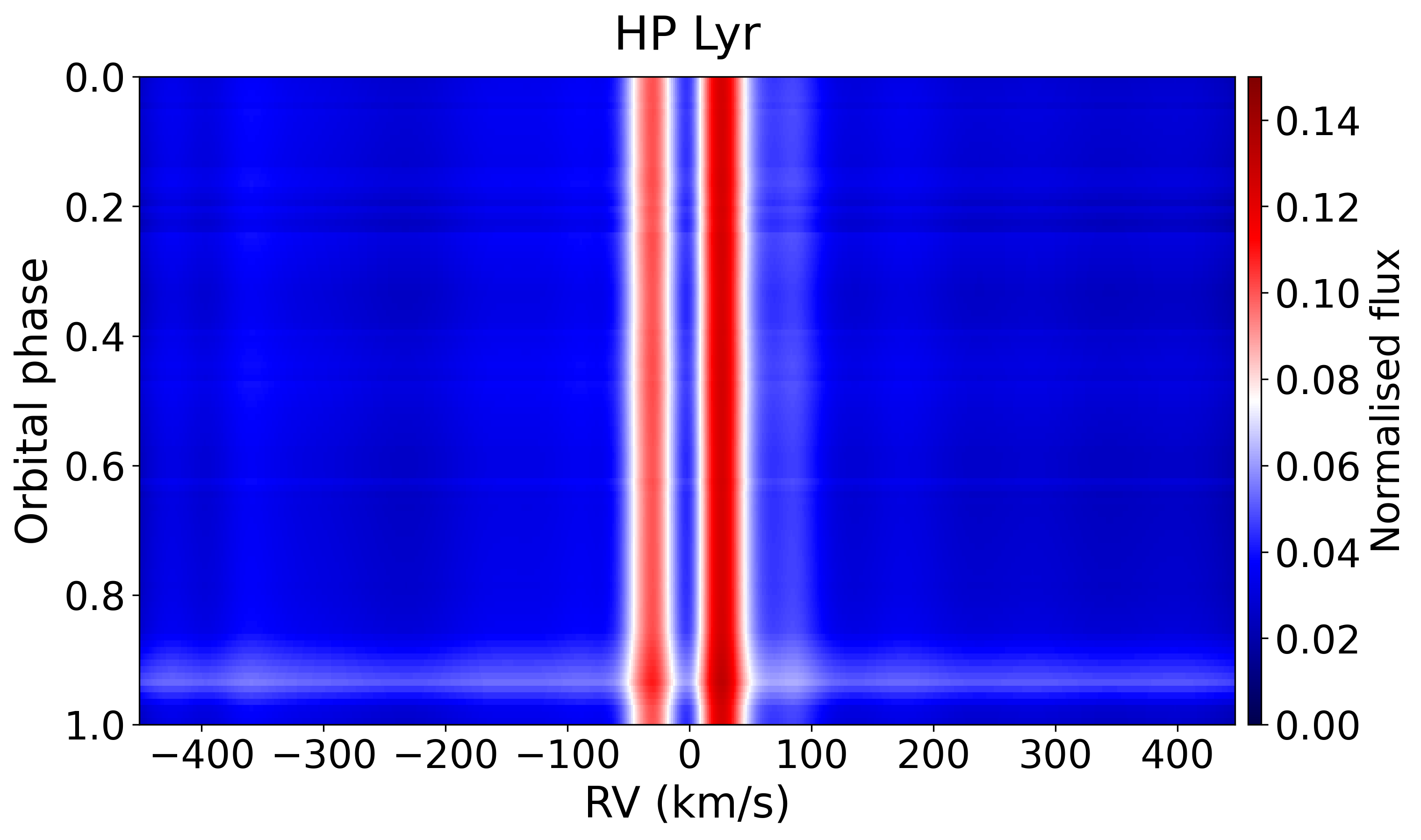}
			\end{subfigure}
			
			\caption{Total $\mathrm{H_\alpha}$ spectral error profiles, accounting for the variability of the observed unobscured spectra.}
			\label{fig:extra_background_error_profiles}
		\end{figure*}
		\clearpage
		\clearpage
		
		\section{All disc wind model fits}\label{sect:appendix_all_model_fits}
		The fitted parameters, for all five MDW solutions and for every one of our five targets, are summarised in Table \ref{table:pAGB_MDW_models_all_fit_params}. The corresponding model dynamic spectra are shown in Fig.\ \ref{fig:all_fit_dynspec}. These fits use the same $\chi^2_{\nu}$ calculation boxes that are shown in Fig.\ \ref{fig:spectra_main_results}. We note that the selection of the best fitting MDW model is guided by the $\chi^2_{\nu}$ values, but ultimately depends on the ability of the MDW to recover the global $\mathrm{H_\alpha}$ pattern.
		\begin{table*}[b]
			\caption{Parameters and $\chi^2_\nu$ values for all fitted disc wind models.}
			\label{table:pAGB_MDW_models_all_fit_params}
			\setlength\tabcolsep{0mm}
			\centering
			\begin{tabular}{L{0.15384615384 \textwidth} C{0.06692307692 \textwidth} C{0.06692307692 \textwidth} C{0.06692307692 \textwidth} C{0.06692307692 \textwidth} C{0.07692307692 \textwidth} C{0.07692307692 \textwidth} C{0.07692307692 \textwidth} C{0.10692307692 \textwidth} C{0.10692307692 \textwidth} C{0.06692307692 \textwidth} C{0.06692307692 \textwidth}}
				\hline\hline & \\[-1.7ex]
				Object & \#MDW & $i$ & $r_{in}$ & $r_{out}$ & $\dot{M}_{in}$ & $T_{wind}$ & $\chi^2_\nu$ & $\dot{M}_{out}$ & $\dot{M}_{wind}$ & $R_{RL,2}$ & $R_2$ \\
				& & $\mathrm{(^\circ)}$ & $(R_2)$ & $(R_{RL,2})$ & $(\mathrm{M_\odot /yr})$ & $\mathrm{(K)}$ & & $(\mathrm{M_\odot /yr})$ & $(\mathrm{M_\odot /yr})$ & $(R_2)$ & $(\mathrm{R_\odot})$ \\[1.0ex]
				\hline & \\[-1.6ex]
				HD 52961 & 1 & 80 & 25 & 0.9 & $10^{-4}$ & 4200 & 9.36 & $1.1\cdot10^{-4}$ & $5\cdot10^{-6}$ & 232 & 0.89 \\[0.5ex]
				& 2 & 76 & 1 & 0.9 & $10^{-4}$ & 4200 & 6.92 & $1.8\cdot10^{-4}$ & $4\cdot10^{-5}$ & 231 & 0.91 \\[0.5ex]
				& 3 & 72 & 5 & 0.9 & $10^{-4}$ & 4200 & 7.40 & $1.4\cdot10^{-4}$ & $2\cdot10^{-5}$ & 231 & 0.91 \\[0.5ex]
				& \textbf{4} & \textbf{72} & \textbf{1} & \textbf{0.9} & $\mathbf{10^{-4}}$ & \textbf{4200} & \textbf{5.32} & $\mathbf{3\cdot10^{-4}}$ & $\mathbf{1\cdot10^{-4}}$ & \textbf{228} & \textbf{0.93} \\[0.5ex]
				& 5 & 56 & 5 & 0.9 & $10^{-5}$ & 4400 & 5.63 & $3\cdot10^{-5}$ & $1\cdot10^{-5}$ & 212 & 1.10 \\[0.6ex]
				\hline & \\[-1.7ex]
				BD+46\textdegree442 & 1 & 64 & 15 & 0.9 & $10^{-3}$ & 5650 & 6.73 & $1.06\cdot10^{-3}$ & $3\cdot10^{-5}$ & 54 & 0.87 \\[0.5ex]
				& 2 & 68 & 1 & 0.9 & $10^{-5}$ & 5050 & 7.69 & $1.6\cdot10^{-5}$ & $3\cdot10^{-6}$ & 55 & 0.84 \\[0.5ex]
				& \textbf{3} & \textbf{56} & \textbf{5} & \textbf{0.8} & $\mathbf{10^{-5}}$ & \textbf{5250} & \textbf{6.44} & $\mathbf{1.2\cdot10^{-5}}$ & $\mathbf{1\cdot10^{-6}}$  & \textbf{46} & \textbf{0.96}\\[0.5ex]
				& 4 & 44 & 1 & 0.5 & $10^{-5}$ & 5050 & 6.85 & $2.0\cdot10^{-5}$ & $5\cdot10^{-6}$ & 26 & 1.20 \\[0.5ex]
				& 5 & 28 & 1 & 0.4 & $10^{-6}$ & 5450 & 7.18 & $2.4\cdot10^{-6}$ & $7\cdot10^{-7}$ & 17 & 2.12 \\[0.6ex]
				\hline & \\[-1.7ex]
				TW Cam & 1 & 28 & 15 & 0.9 & $10^{-3}$ & 4650 & 5.59 & $1.08\cdot10^{-3}$ & $4\cdot10^{-5}$  & 107 & 2.00 \\[0.5ex]
				& 2 & 44 & 1 & 0.6 & $10^{-3}$ & 4650 & 6.57 & $1.6\cdot10^{-3}$ & $3\cdot10^{-4}$  & 89 & 1.14 \\[0.5ex]
				& \textbf{3} & \textbf{24} & \textbf{5} & \textbf{0.9} & $\mathbf{10^{-3}}$ & \textbf{4850} & \textbf{5.42} & $\mathbf{1.4\cdot10^{-3}}$ & $\mathbf{2\cdot10^{-4}}$ & \textbf{98} & \textbf{2.52}\\[0.5ex]
				& 4 & 24 & 1 & 0.9 & $10^{-4}$ & 4850 & 6.44 & $2.6\cdot10^{-4}$ & $8\cdot10^{-5}$  & 98 & 2.52\\[0.5ex]
				& 5 & 20 & 1 & 0.6 & $10^{-4}$ & 4450 & 6.68 & $3\cdot10^{-4}$ & $1\cdot10^{-4}$  & 58 & 3.41\\[0.6ex]
				\hline & \\[-1.7ex]
				IRAS 19135+3937& 1 & 56 & 10 & 0.9 & $10^{-3}$ & 5850 & 4.28 & $1.08\cdot10^{-3}$ & $4\cdot10^{-5}$ & 57 & 0.67 \\[0.5ex]
				& 2 & 68 & 5 & 0.9 & $10^{-3}$ & 5650 & 3.96 & $1.2\cdot10^{-3}$ & $1\cdot10^{-4}$ & 60 & 0.59 \\[0.5ex]
				& 3 & 48 & 5 & 0.9 & $10^{-5}$ & 5450 & 3.49 & $1.2\cdot10^{-5}$ & $1\cdot10^{-6}$  & 54 & 0.75 \\[0.5ex]
				& \textbf{4} & \textbf{48} & \textbf{1} & \textbf{0.9} & $\mathbf{10^{-5}}$ & \textbf{5050} & \textbf{3.30} & $\mathbf{2.4\cdot10^{-5}}$ & $\mathbf{7\cdot10^{-6}}$ & \textbf{54} & \textbf{0.75} \\[0.5ex]
				& 5 & 28 & 1 & 0.9 & $10^{-5}$ & 4650 & 3.53 & $3\cdot10^{-5}$ & $1\cdot10^{-5}$ & 42 & 1.32 \\[0.6ex]
				\hline & \\[-1.7ex]
				HP Lyr \hspace{0.1mm} \tablefootmark{a} & 1 & 37 & 25 & 0.8 & $10^{-4}$ & 4800 & 7.17 & $1.1\cdot10^{-4}$ & $5\cdot10^{-6}$ & 247 & 1.00 \\[0.5ex]
				& 2 & 61 & 25 & 0.8 & $10^{-4}$ & 5000 & 8.19 & $1.2\cdot10^{-4}$ & $1\cdot10^{-5}$ & 298 & 0.65 \\[0.5ex]
				& 3 & 33 & 20 & 0.3 & $10^{-3}$ & 4400 & 6.29 & $1.16\cdot10^{-3}$ & $8\cdot10^{-5}$ & 87 & 1.15 \\[0.5ex]
				& \textbf{4} & \textbf{29} & \textbf{20} & \textbf{0.3} & $\mathbf{10^{-4}}$ & \textbf{5000} & \textbf{4.91} & $\mathbf{1.4\cdot10^{-4}}$ & $\mathbf{2\cdot10^{-5}}$  & \textbf{82} & \textbf{1.32} \\[0.5ex]
				& 5 & 17 & 20 & 0.3 & $10^{-4}$ & 5200 & 5.05 & $1.4\cdot10^{-4}$ & $2\cdot10^{-5}$ & 61 & 2.91 \\[0.8ex]
				\hline
			\end{tabular}
			\tablefoot{The best fitting model per target is marked in bold. In addition to the fitted parameters, we show the feeding rate at the outer disc rim and the wind mas-loss rate, and also provide $R_{RL,2}$ in units of $R_2$ and $R_2$ in solar radii so the extent of the disc can be calculated for future reference.\tablefoottext{a}{The values for HP Lyr \#MDW 2 correspond to a secondary, high inclination minimum in the parameter space. The first minimum represented an unphysical solution, implying an unrealistically large secondary mass of $M_2 \approx 23\,\mathrm{M_\odot}$ and, unlike what is observed, a wind absorption feature spanning the entire orbit.}}
			\vspace{2.5cm}
		\end{table*}
		
		\begin{figure*}
			\vspace{3.2cm}
			\centering
			\includegraphics[width=\textwidth]{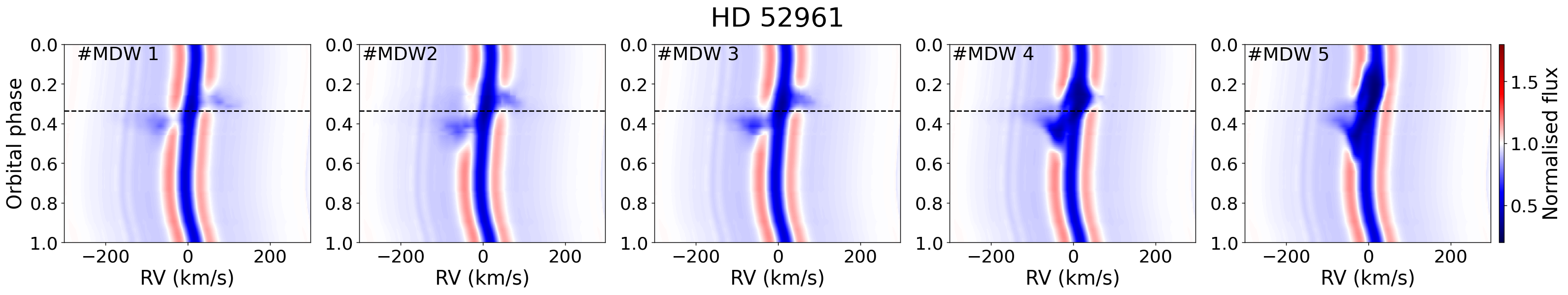}
			\includegraphics[width=\textwidth]{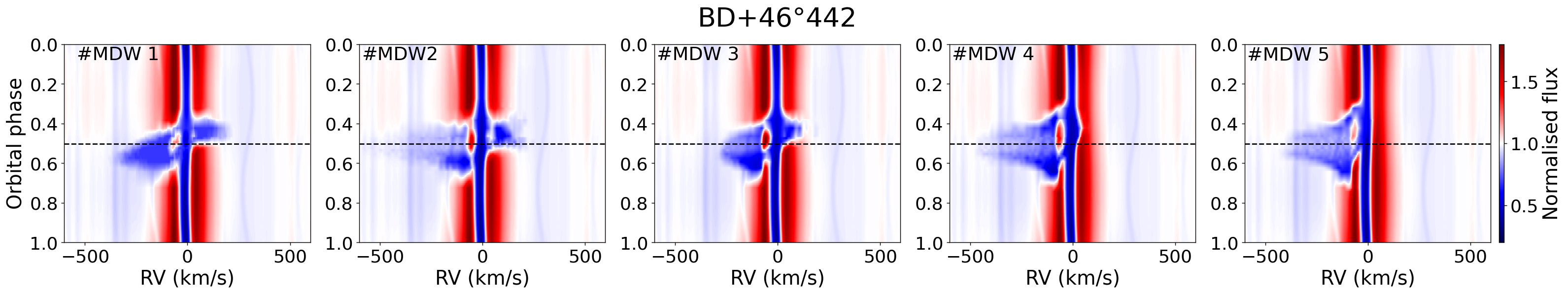}
			\includegraphics[width=\textwidth]{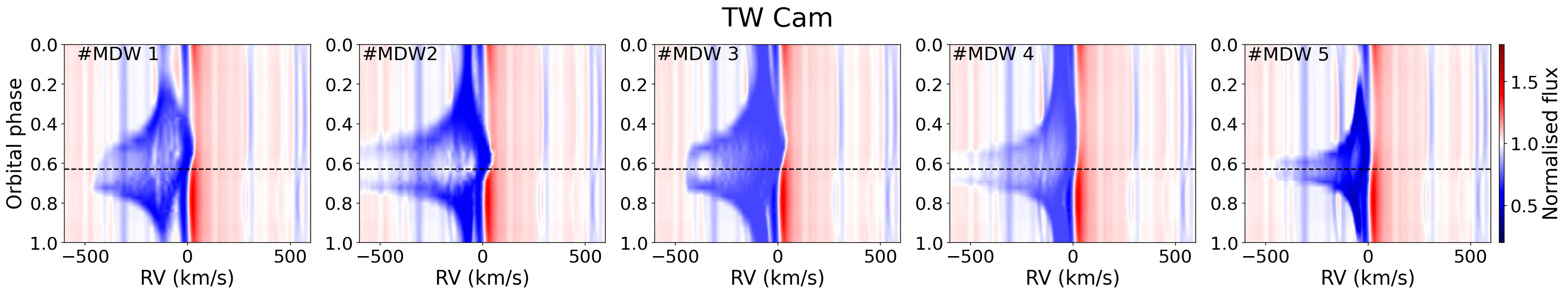}
			\includegraphics[width=\textwidth]{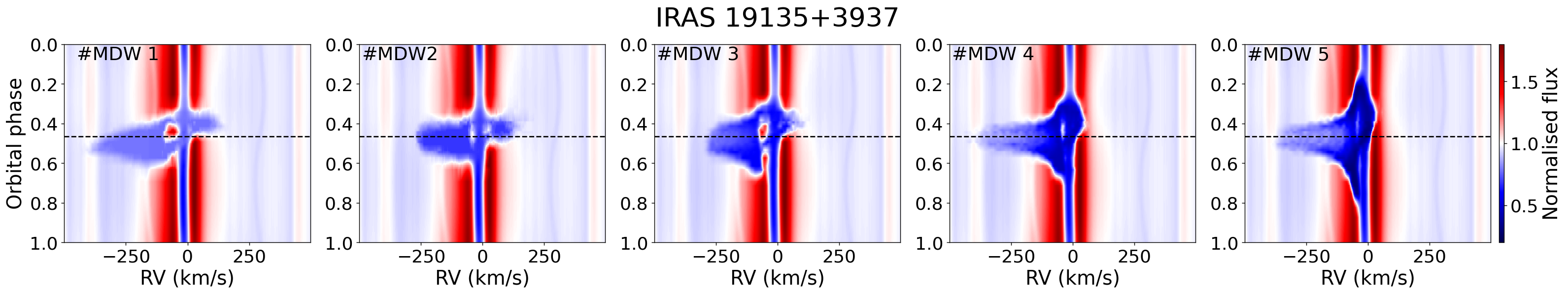}
			\includegraphics[width=\textwidth]{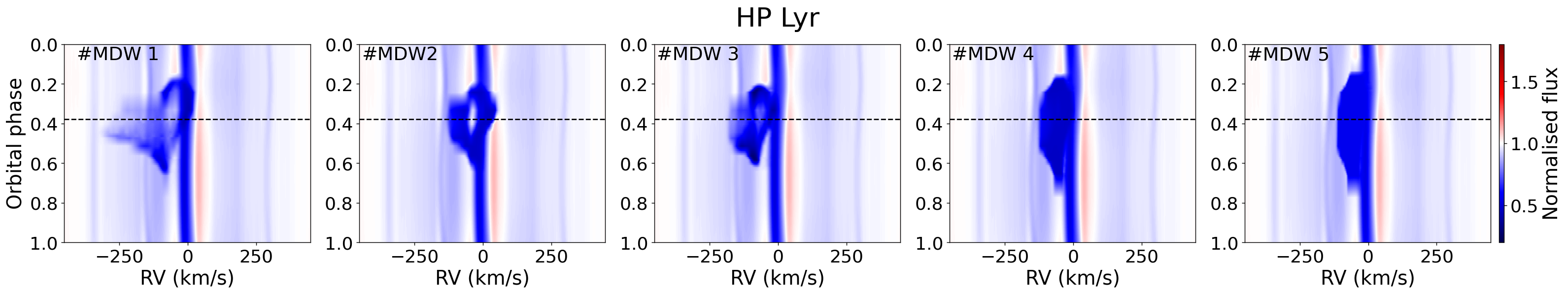}
			\caption{Fitted model dynamic spectra for all targets (Table \ref{table:target_star_params}) and considered MDW solutions, ordered left to right by their \#MDW number. The corresponding model parameters are summarised in Table \ref{table:pAGB_MDW_models_all_fit_params}.}
			\label{fig:all_fit_dynspec}
		\end{figure*}
		\clearpage
		
		\section{Mass-loss rate for extra outflow inside the disc wind cavity}\label{sect_append:extra_mass_loss}
		In this appendix, we derive a simple expression for the mass-loss rate that an extra outflow inside the disc wind cavity would need in order to supply the missing low RV absorption occurring close to superior conjunction in some of our models. We assumed a simple cylindrical geometry for this extra outflow.
		
		Consider a disc wind model close to superior conjunction, where a chosen LOS ray passes through both the left and right wing of the wind lobe facing the observer (Fig.\ \ref{fig:extra_mass_loss_estimate_cartoon}). In addition, assume that there is some extra component of material flow inside the wind cavity. Neglecting the wind emission, the intensity at wavelength $\lambda$ leaving the disc wind is then given by:
		\begin{equation}\label{eq:intensity_components_separate}
			I^{out}_\lambda = I^{0}_\lambda \cdot e^{-\Delta \tau_{\lambda,tot}} = I^{0}_\lambda \cdot e^{-(\Delta \tau_{\lambda,left} + \Delta \tau_{\lambda,cav} + \Delta \tau_{\lambda,right})},
		\end{equation}
		with $I^{0}_\lambda$ the background spectrum. We have split the total optical depth along the LOS ray into the separate contributions of the two jet wings and the hypothetical material in the cavity. Identifying $I^{out}_\lambda$ with the observed spectrum, $I_{\lambda,obs}$, and $I^{0}_\lambda \cdot e^{-(\Delta \tau_{\lambda,left} + \Delta \tau_{\lambda,right})}$ as the spectrum modelled by our MDW fits, $I_{\lambda, mod}$, this gives:
		\begin{equation}\label{eq:delta_tau_cav_non_averaged}
			I_{\lambda, obs} = I_{\lambda, mod}\cdot e^{-\Delta \tau_{\lambda, cav}} \implies \Delta \tau_{\lambda, cav} = \ln{\left( \frac{1}{1-\frac{\Delta I_\lambda}{I_{\lambda, mod}}} \right)},
		\end{equation}
		where $\Delta I_\lambda = I_{\lambda, mod}-I_{\lambda, obs}$. Eq.\ (\ref{eq:delta_tau_cav_non_averaged}) thus identifies how much optical depth is missing at a certain wavelength in order to bring the emission peak shining through our models in line with the observations.
		Fig.\ \ref{fig:emission_peak_single_spec} visualises the different quantities involved in the calculation of Eq.\ (\ref{eq:delta_tau_cav_non_averaged}).
		
		With the estimate for the extra required optical depth at hand, we can construct a simple geometrical model for our extra outflow component. This model is visualised in Fig.\ \ref{fig:extra_mass_loss_estimate_cartoon}. Consider the length $s$ of an LOS ray inside the cavity. The contribution of the cavity material to the optical depth is given by $\Delta \tau_{\lambda, cav} = \kappa_{\lambda, cav} \cdot \rho_{cav} \cdot s$,
		with $\kappa_{\lambda, cav}$ the material's $\mathrm{H_\alpha}$ line opacity profile and $\rho_{cav}$ its mass density, assuming homogeneity. For $\kappa_{\lambda, cav}$, we choose the $\mathrm{H_\alpha}$ line opacity of the disc wind, at the point where the LOS ray enters the cavity. This mainly assumes that the material in the cavity has the same temperature as the disc wind ($T = T_{wind}$), since the $\mathrm{H_\alpha}$ opacity is strongly dependent on temperature, but only weakly pressure-dependent at the considered densities. Choosing a range of wavelengths $\Delta \lambda$, spanning the background emission peak that shines through the model, this can then be averaged over $\Delta \lambda$:
		\begin{eqnarray}\label{eq:relation_delta_tau_cav_and_rho_cav_averaged}
			\overline{\Delta \tau}_{cav} & = & \overline{\kappa}_{cav} \cdot \rho_{cav} \cdot s \implies \rho_{cav} = \frac{\overline{\Delta \tau}_{cav}}{\overline{\kappa}_{cav} \cdot s},\\
			\overline{\kappa}_{cav} & = & \frac{\int_{\Delta \lambda} \kappa_{\lambda, cav}\,\mathrm{d}\lambda}{\Delta \lambda},\\
			\overline{\Delta \tau}_{cav} & = & \frac{\int_{\Delta \lambda} \ln{\left(\frac{1}{1-\Delta I_\lambda/I_{\lambda, mod}}\right)}\,\mathrm{d}\lambda}{\Delta \lambda}.
		\end{eqnarray}
		Eq.\ (\ref{eq:relation_delta_tau_cav_and_rho_cav_averaged}) can thus be used to obtain a single estimate for the density of the cavity material needed to supply the missing absorption. A representative LOS ray needs to be chosen, for which we chose the ray leaving the centre of the post-AGB primary.
		
		Assume that the extra component consists of an outflow along a cylinder of radius $r = (s/2)\cdot \sin{i}$, with a uniform velocity $v$ along the jet axis (Fig.\ \ref{fig:extra_mass_loss_estimate_cartoon}). The velocity $v$ is related to the observed central RV value of the missing absorption via $v = \mathrm{|RV|}/\cos{i}$ (i.e.\ we assume all the missing absorption to be concentrated at the central RV value). Using Eq.\ (\ref{eq:relation_delta_tau_cav_and_rho_cav_averaged}) to calculate $\rho_{cav}$, we retrieve an estimate of the extra component's mass-loss rate:
		\begin{equation}\label{eq:missing_absorption_cavity_mass_loss}
			\dot{M}_{cav} \sim \pi r^2 \cdot v \cdot \rho_{cav} = \frac{\pi \sin^2{i}}{4\cos{i}}\cdot \frac{\mathrm{RV \cdot  s}}{\overline{\kappa}_{cav}} \cdot \overline{\Delta \tau}_{cav}.
		\end{equation}
		
		Typical values for models that show a clear lack of wind absorption near superior conjunction, for example \#MDW 3 for BD+46\textdegree442 and IRAS 19135+3937, are  $\overline{\kappa}_{cav} = 10 - 100 \,\mathrm{cm^2 / g}$, $\rho_{cav} \approx 10^{-15}\,\mathrm{g/cm^3}$, $v\approx 100\,\mathrm{km/s}$ and $\dot{M}_{cav} \approx 10^{-8}\, \mathrm{M_\odot/yr}$. These are the physical conditions that an additional outflow inside the cavity, for example a thermal stellar jet, would need to achieve in order to provide the missing absorption. 
		
		\begin{figure}
			\resizebox{\hsize}{!}{\includegraphics{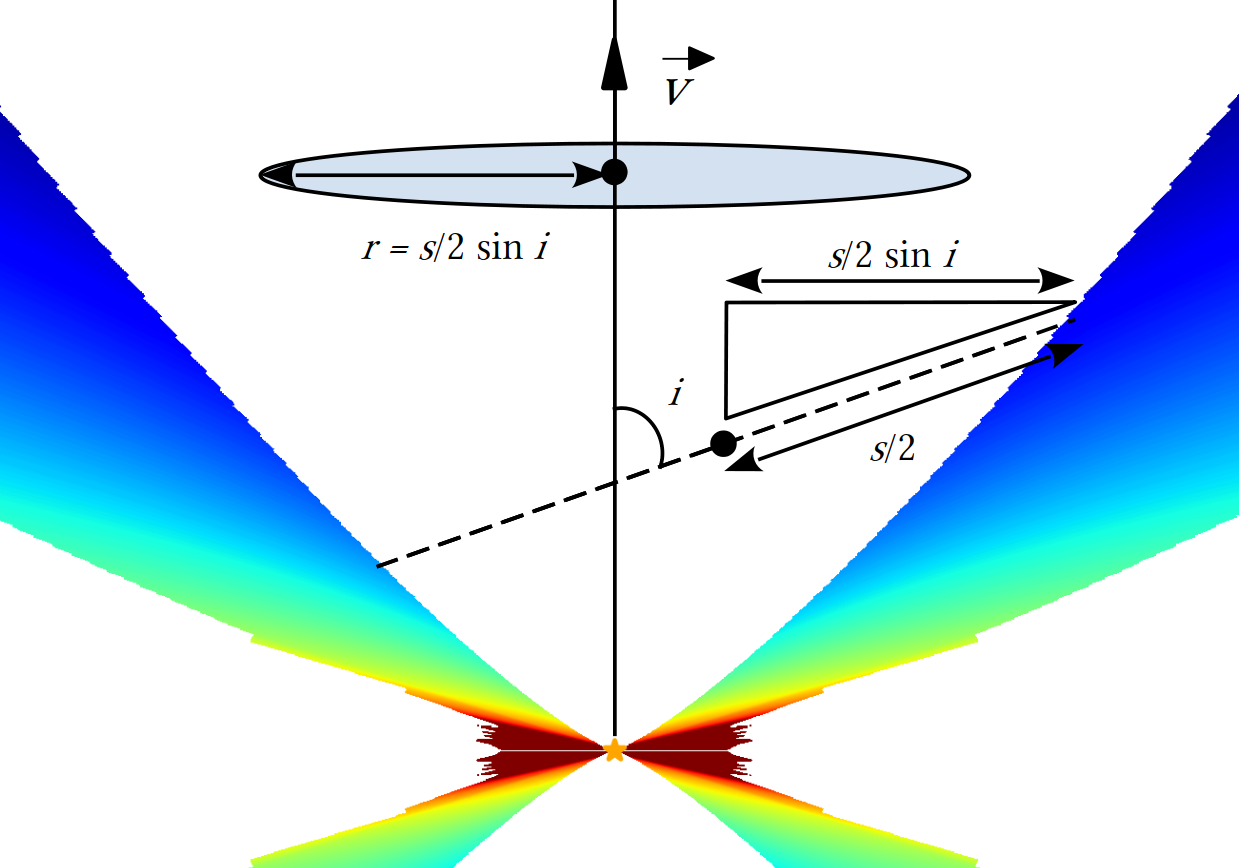}}
			\caption{Simple geometrical model for an extra outflow component in the disc wind cavity. The dashed line indicates the stretch of the LOS inside the cavity. It has length $s$. The extra component is taken to be a homogeneous flow of uniform velocity $v$, pointed along the jet axis and flowing along a cylinder of radius $r = (s/2)\cdot \sin i$.}
			\label{fig:extra_mass_loss_estimate_cartoon}
		\end{figure}
		
		\begin{figure}
			\resizebox{\hsize}{!}{\includegraphics{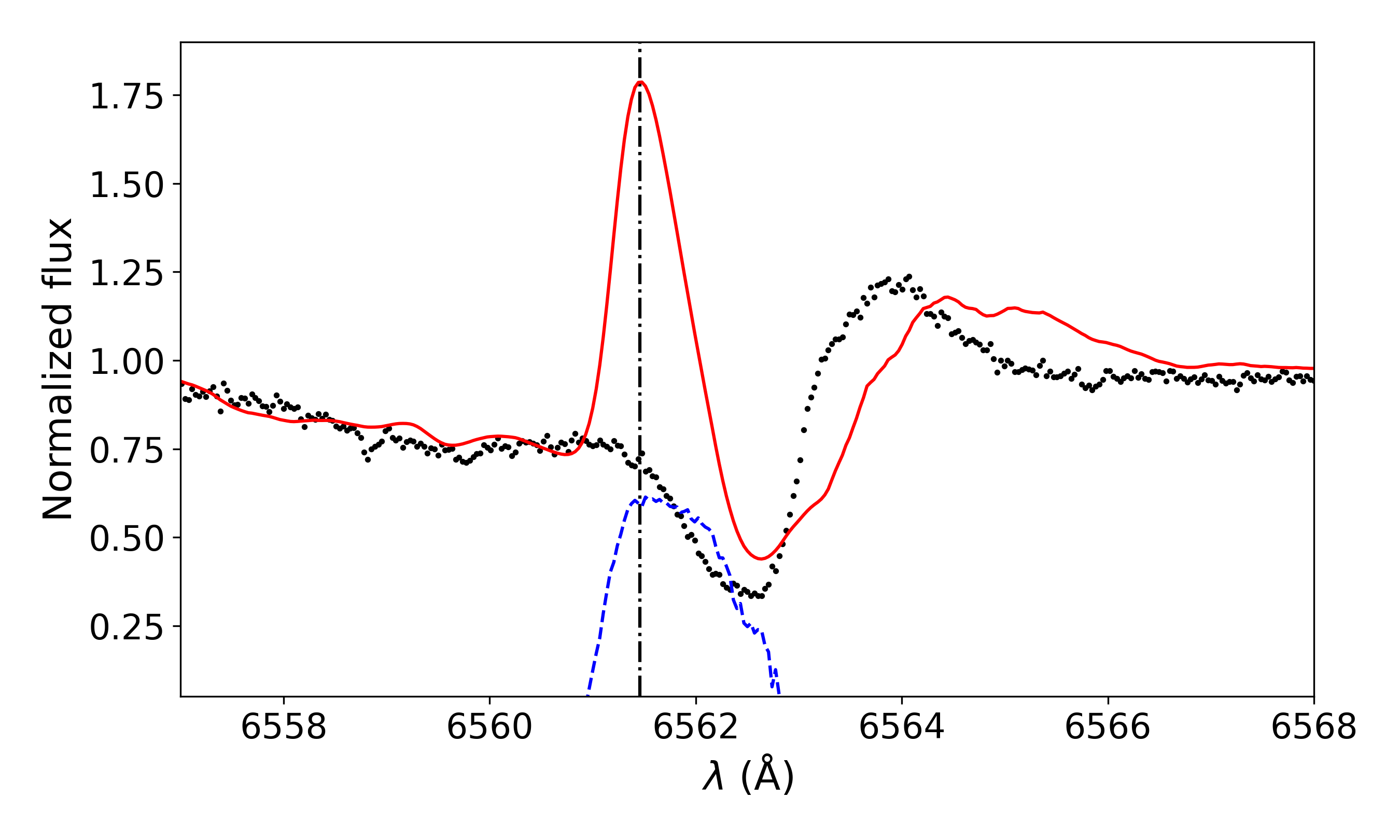}}
			\caption{Example spectrum at $\phi_{orb} = 0.49$ of BD+46$^\circ$442's \#MDW 3 fit. The model is shown as a solid red line, while the observations are shown as black dots. The curve of $\Delta I_\lambda/I_{\lambda, mod}$ is shown as a dashed blue line. The central wavelength value at which absorption is missing ($\mathrm{RV} \approx -60 \,\mathrm{km/s}$) is indicated by the black dot-dashed line.}
			\label{fig:emission_peak_single_spec}
		\end{figure}
		
	\end{appendix}
	\endgroup
	
\end{document}